\documentclass[twocolumn,prx,superscriptaddress,longbibliography]{revtex4-1}

\usepackage{titlesec}%
\usepackage{subfigure}%
\usepackage{amsmath}%
\allowdisplaybreaks
\usepackage{amsfonts}%
\usepackage{amssymb}%
\usepackage{feynmp}%
\usepackage{graphicx}%
\usepackage{setspace}%
\usepackage{comment}%
\usepackage{dsfont}%
\usepackage{color}%
\usepackage{mathtools}
\usepackage{multirow}
\usepackage[pdftex,colorlinks=true,linkcolor=blue,citecolor=blue]{hyperref}

\titleformat{\paragraph}
{\normalfont\normalsize\bfseries}{\theparagraph}{1em}{}
\titlespacing*{\paragraph}
{0pt}{3.25ex plus 1ex minus .2ex}{1.5ex plus .2ex}

\begin{document}

\title{Quantization of fractional corner charge \\in $C_n$-symmetric higher-order topological crystalline insulators}

\author{Wladimir A. Benalcazar}
\affiliation{Department of Physics, The Pennsylvania State University, University Park, PA 16802, USA}
\affiliation{Department of Physics and Institute for Condensed Matter Theory, University of Illinois at Urbana-Champaign, IL 61801, USA}
\author{Tianhe Li}
\author{Taylor L. Hughes}
\affiliation{Department of Physics and Institute for Condensed Matter Theory, University of Illinois at Urbana-Champaign, IL 61801, USA}
\date{\today}

\begin{abstract}
In the presence of crystalline symmetries, certain topological insulators present a \emph{filling anomaly}: a mismatch between the number of electrons in an energy band and the number of electrons required for charge neutrality. In this paper, we show that a filling anomaly can arise when corners are introduced in $C_n$-symmetric crystalline insulators with vanishing polarization, having as consequence the existence of corner-localized charges quantized in multiples of $\frac{e}{n}$. We characterize the existence of  this charge systematically and build topological indices that relate the symmetry representations of the occupied energy bands of a crystal to the quanta of fractional charge robustly localized at its corners. When an additional chiral symmetry is present, $\frac{e}{2}$ corner charges are accompanied by zero-energy corner-localized states. We show the application of our indices in a number of atomic and fragile topological insulators and discuss the role of fractional charges bound to disclinations as bulk probes for these crystalline phases.
\end{abstract}

\maketitle

Topological crystalline insulators (TCIs) \cite{teo2008,fu2011,hsieh2012,slager2013,shiozaki2014,fang2015,watanabe2017} are known to exhibit a variety of quantized electromagnetic phenomena. They host bulk dipole moments that lead to surface charge densities quantized in fractions of the electronic charge $e$~\cite{zak1989,king-smith1993, vanderbilt1993, resta1994,hughes2011inversion,turner2012}. Recently, it was found that TCIs can also host higher bulk multipole moments that manifest lower-order moments bound to their boundaries \cite{benalcazar2017quad,benalcazar2017quadPRB}. For example, a quadrupole insulator in  two dimensions has edge-bound dipole moments and corner-bound charges, while an octupole insulator in three dimensions has surface-bound quadrupole moments, hinge-bound dipole moments, and corner-bound charges. Just as in the case of insulators with symmetry-protected dipole moments, crystalline symmetries quantize the boundary signatures in quadrupole or octupole TCIs. Indeed, TCIs with quantized multipole moments are symmetry protected topological phases of matter; their quantization is robust and can change only in discrete jumps at phase transitions~\cite{benalcazar2017quad,benalcazar2017quadPRB}, unless the protecting symmetries are broken. 

A salient property of TCIs with quantized higher multipole moments is that some of their protected observables at the boundary are at least two dimensions less than the protecting bulk. This property has now been extended to a broader family of TCIs, broadly referred to as higher-order topological insulators~\cite{benalcazar2017quadPRB,fu2017,song2017,langbehn2017,schindler2018,brouwer2018,noh2018,ezawa2018,ortix2018,xie2018,khanikaev2018,xue2018,ahn2018,ezawa2018b,ezawa2018c,ezawa2018d,kunst2018,fulga2018,wang2018,watanabe2018,schindler2018a,khalaf2018,khalaf2018b,you2018,vanderbilt2018,xue2018,wang2018,wieder2018,wieder2018b}. In this paper, we focus on two-dimensional (2D) higher-order TCIs having zero-dimensional topological signatures. A number of studies have recently shown examples of such TCIs which exhibit in-gap corner-localized states~\cite{benalcazar2017quadPRB,noh2018,ezawa2018,ortix2018,xie2018,khanikaev2018, xue2018,ahn2018}, some of which have been related to fractionally quantized corner charges~\cite{benalcazar2017quadPRB,ezawa2018,ortix2018,ahn2018}. Interestingly, many such TCIs have these corner signatures in spite of vanishing quadrupole moments, and their mechanisms of protection and associated topological invariants are still not completely elucidated.

In this article we systematically study 2D second-order TCIs in class AI (spinless and time-reversal symmetric insulators) protected by $C_n$ symmetry and find the topological indices that connect the bulk topology of these TCIs with corner or defect-bound fractional charges. We show that the fractional quantization of corner charge arises from a \emph{filling anomaly}: a topological property of the occupied energy bands of a TCI that keeps track of the mismatch between the number of electrons required to simultaneously satisfy charge neutrality and the crystal symmetry. This mismatch exists even in first-order TCIs with quantized dipole moments --giving rise to quantized fractional charge at edges~\cite{kivelson1982,vanderbilt1993,hughes2011inversion,miert2017,rhim2017}-- and we discuss this type of filling anomaly to introduce the concept. Our focus, however, is on a refined form of a filling anomaly that originates \emph{only} when corners are created in a lattice. Such corner-induced filling anomalies are particular of higher-order topological phases. We build topological indices that allow us to identify the cases in which the filling anomaly arising from edges is avoided, but the filling anomaly due to corners is not. Given the set of rotation topological invariants for a particular $C_n$ symmetry (extracted from the representations of the little groups of the occupied bands at the high symmetry points of the Brillouin zone), the topological indices we derive relate the set of rotation topological invariants to the quanta of the corner-bound charge. We show that in obstructed atomic insulators, i.e., insulators that admit a Wannier representation~\cite{bradlyn2017}, the filling anomaly is intimately related to the locations of the Wannier centers of the electrons in the bulk of the crystal. However, the index theorems apply even for crystalline insulators that are \emph{not} Wannier-representable, as we will show by providing examples for the quantization of charge fractionalization at the corners of fragile topological crystalline phases~\cite{vishwanath2018,cano2018a,bohuon2018,bradlyn2018}.

The paper is organized as follows. In Section~\ref{sec:Classification}, we first classify $C_n$-symmetric TCIs in terms of rotation topological invariants which we define (see Refs. \onlinecite{fang2012,fang2013,teo2013,benalcazar2014,ortix2018} for other related invariants and classifications). In Section~\ref{sec:PrimitiveGenerators}, we present model Hamiltonians that constitute \emph{primitive generators} of these classifications. All of our generators are Wannier-representable~\cite{resta1998,yu2011,alexandradinata2014,bradlyn2017,cano2018} and have the property that they can be combined to span the entire set of phases in these classifications. The Wannier-representablility of our generators is advantageous because it transparently connects the Wannier centers of the electrons in the bulk and boundaries of a lattice to the filling anomaly, and consequently to the edge and corner charge. Additionally, because of their simple structure and our choice of the AI symmetry class, all of our generator models can be straightforwardly constructed in metamaterial contexts through the evanescent coupling of wave-guide or resonator modes. This will allow for the immediate realization of our predictions in experiments.

\begin{figure}[t]
\centering
\includegraphics[width=\columnwidth]{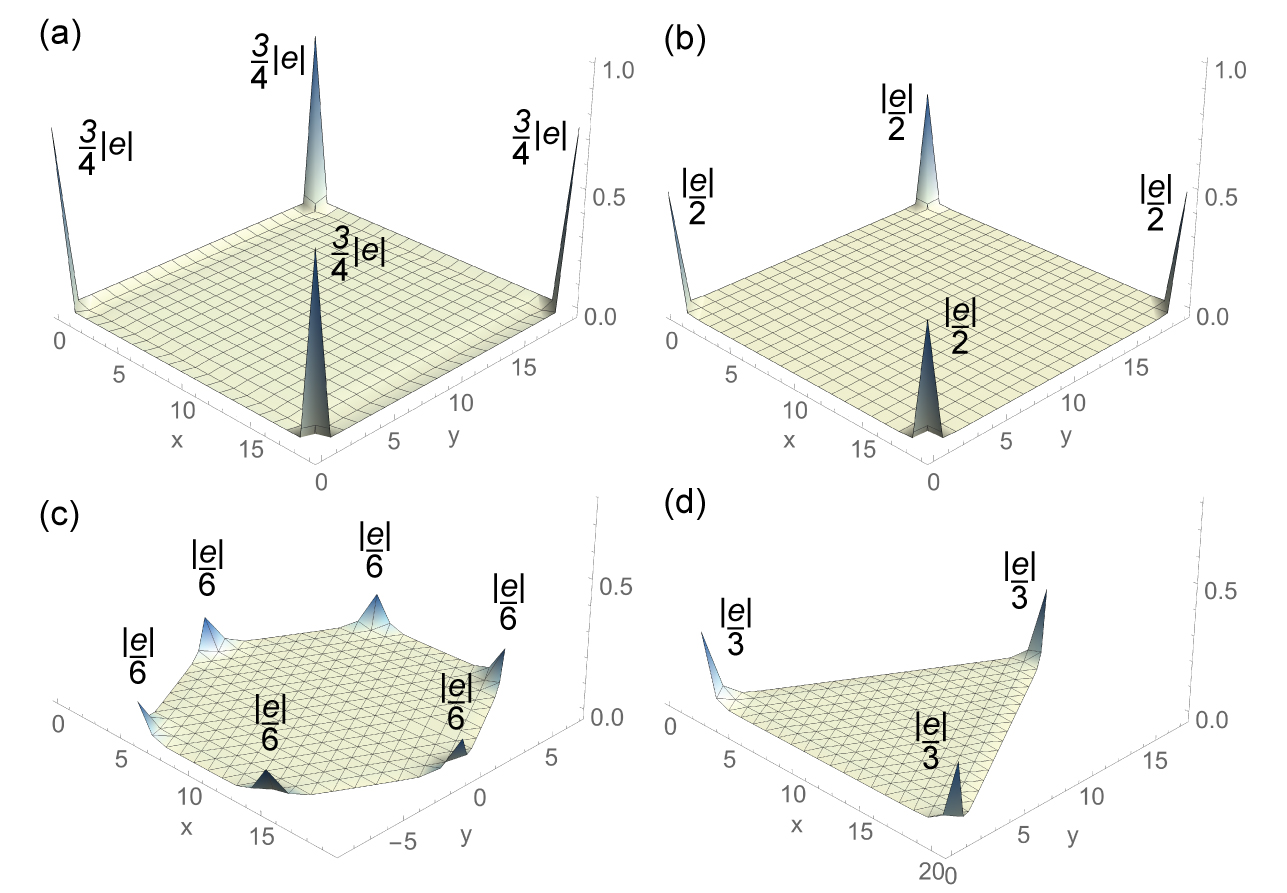}
\caption{Quantized fractional corner charge in $C_n$-symmetric TCIs. The plots show the total (electronic and ionic) charge density of two-dimensional TCIs. (a) a $C_4$-symmetric TCI with corner charge $\frac{3|e|}{4}$, (b) a $C_4$-symmetric TCI with corner charge $\frac{|e|}{2}$, (c) a $C_6$-symmetric TCI with corner charge $\frac{|e|}{6}$, and (d) a $C_3$-symmetric TCI with corner charge $\frac{|e|}{3}$. In all cases, the bulk \emph{and} edges are neutral. These charge patterns are obtained by stacking the primitive generator models as described in Section~\ref{sec:IndexConstruction}.}
\label{fig:CornerCharge}
\end{figure}

After introducing the generators, in Sections~\ref{sec:AnomalyEdges} and~\ref{sec:AnomalyCorners} we describe the mechanism that gives rise to the edge and corner filling anomalies, respectively, and how they relate to edge and corner fractional charges. In Section~\ref{sec:IndexConstruction} we apply our insights from the previous sections to identify the boundary charges in the primitive generators, which we then use to build index theorems for the filling anomaly and corner fractional charge of any $C_n$-symmetric higher-order TCI in class AI. We find that TCIs in a lattice with a global $C_n$-symmetry host corner-localized charges that, when added within a spatial sector subtended by an angle of $\frac{2\pi}{n}$ from the center of the lattice, are quantized in multiples of $\frac{e}{n}$. In particular, if each $\frac{2\pi}{n}$ sector has only one corner, the charge is fractionally quantized at each corner of the lattice, as shown in Fig.~\ref{fig:CornerCharge}.

After the construction of the topological indices, in Section~\ref{sec:FragilePhases} we provide examples of the fractional quantization of charge in TCIs without a Wannier representation, and in Section~\ref{sec:Defects} we apply our theory to show that higher-order TCIs bind fractional charge at the core of certain topological defects. Finally, we present a discussion and our conclusions in Section~\ref{sec:Conclusions}.

\section{Classification} 
\label{sec:Classification}
Two-dimensional TCIs in class AI \cite{altland1997,schnyder2008,kitaev2009} preserve time-reversal symmetry (TRS), having Bloch Hamiltonians satisfying $h({\bf k}) = h^*(-{\bf k}).$  These systems have a vanishing Hall conductance, indicated by a zero Chern number. The presence of additional $C_n$ symmetry, however, allows for a finer classification of topological phases in these insulators~\cite{fu2011,fang2012,fang2013,teo2013,benalcazar2014,ortix2018} (see Appendix~A for the detailed construction of the classification). These classes can be most directly distinguished by the value of their polarization~\cite{zak1989,resta1992, resta1993, resta1994, king-smith1993, vanderbilt1993, resta2007} 
\begin{align}
{\bf P}^{(n)}=p_1 {\bf a_1}+p_2 {\bf a_2},\label{eq:polarization}
\end{align} where the superindex $n$ labels the $C_n$-symmetry of the classification, ${\bf a_1}$ and ${\bf a_2}$ are primitive unit lattice vectors, and the components $p_1$ and $p_2$ are topological indices that correspond to quantized Berry phases along the non-contractible loops of the Brillouin zone (BZ) \cite{zak1989, hughes2011inversion, turner2012, fang2012}. We take ${\bf a_1}$ and ${\bf a_2}$ to be ${\bf a_1=\hat{x}}$, ${\bf a_2=\hat{y}}$ in $C_4$ and $C_2$-symmetric lattices, and ${\bf a_1=\hat{x}}$, ${\bf a_2}=\frac{1}{2}{\bf \hat{x}}+\frac{\sqrt{3}}{2}{\bf \hat{y}}$ in $C_6$ and $C_3$-symmetric lattices (note that we have set all lattice constants to unity). As reviewed in Appendix~B, the values of the polarization ${\bf P}$ form a $\mathbb{Z}_2$ index in $C_4$-symmetric TCIs as it can only take the values $p_1=p_2 \in \{0,\frac{e}{2}\}$; a $\mathbb{Z}_2 \times \mathbb{Z}_2$ index in $C_2$-symmetric TCIs with values $p_1, p_2 \in \{0,\frac{e}{2}\}$; and a $\mathbb{Z}_3$ index in $C_3$-symmetric TCIs with values $p_1=p_2 \in \{0,\frac{e}{3},\frac{2e}{3}\}$; while in $C_6$-symmetric TCIs the polarization always vanishes.
 
More generically, we can distinguish nontrivial topological classes arising from the $C_n$ symmetry through the symmetry representations that the occupied energy bands take at the high symmetry points of the BZ (HSPs)\cite{fu2011,fang2012,teo2013,benalcazar2014,kruthoff2017,bradlyn2017,po2017}. 
Consider $C_n$-symmetric Bloch Hamiltonians, which obey $\hat{r}_n h({\bf k}) \hat{r}_n^\dagger = h(R_n {\bf k})$, where $\hat{r}_n$ is the $n$-fold rotation operator obeying $\hat{r}_n^n=1,$ and $R_n$ is the $n$-fold rotation matrix acting on the crystal momentum ${\bf k}$. We denote the HSPs as ${\bf \Pi}^{(n)}$. These are defined as the special points in the BZ which obey $R_n {\bf \Pi}^{(n)} = {\bf \Pi}^{(n)}$ modulo a reciprocal lattice vector. Rotation symmetry then implies that $[\hat{r}_n,h({\bf \Pi}^{(n)})]=0$. Thus, the energy eigenstates of the Bloch Hamiltonian at HSPs are also eigenstates of the rotation operator. Let us denote the eigenvalues of $\hat{r}_n$ at HSP ${\bf \Pi}^{(n)}$ as
\begin{align}
\Pi^{(n)}_p = e^{2\pi i(p-1)/n}, \quad \mbox{for } p=1,2,\ldots n,
\label{eq:RotationEigenvalues}
\end{align}
(see a complete list of HSPs in Appendix~A). Given a subspace of energy bands, we can compare these rotation eigenvalues at the various HSPs. If the eigenvalues change at different HSPs, the energy bands have nontrivial topology. Accordingly, we use the rotation eigenvalues at ${\bf \Pi}^{(n)}$ compared to a reference point ${\bf \Gamma}=(0,0)$ to define the integer topological invariants
\begin{align}
[\Pi^{(n)}_p] \equiv \# \Pi^{(n)}_p - \# \Gamma^{(n)}_p,
\end{align} 
where $\# \Pi^{(n)}_p$ is the number of energy bands below the (in-gap) Fermi level with eigenvalue $\Pi^{(n)}_p$. Not all these invariants are independent, however. First, rotation symmetry can force representations at certain HSPs to be the same. $C_4$ symmetry forces the representations at ${\bf X}$ and ${\bf X'}$ in the BZ to be equal, while $C_6$ symmetry forces equal representations at ${\bf M}$, ${\bf M'}$, and ${\bf M''}$, as well as at ${\bf K}$ and ${\bf K'}$. Furthermore, there are redundancies in the invariants due to: (i) the fact that the number of bands in consideration is constant across the BZ, from which it follows that  $\sum_p \# \Pi^{(n)}_p = \sum_p \# \Gamma^{(n)}_p$, or $\sum_p [\Pi^{(n)}_p]=0$, and (ii) the existence of TRS, which implies that rotation eigenvalues at ${\bf \Pi}^{(n)}$ and $-{\bf \Pi}^{(n)}$ are related by complex conjugation, from which it follows that $[M^{(4)}_2] = [M^{(4)}_4]$, $[K^{(3)}_2]=[K^{'(3)}_3]$, and $[K^{(3)}_3]=[K^{'(3)}_2]$. Dropping the redundant invariants due to these constraints, the resulting topological classes of TCIs with TRS and $C_n$ symmetry are given by the indices $\chi^{(n)}$, as follows,
\begin{align}
\chi^{(4)}&= ( [X^{(2)}_1],[M^{(4)}_1],[M^{(4)}_2])\nonumber\\
\chi^{(2)}&= ( [X^{(2)}_1],[Y^{(2)}_1],[M^{(2)}_1])\nonumber\\
\chi^{(6)}&=( [M^{(2)}_1],[K^{(3)}_1] )\nonumber\\
\chi^{(3)}&=( [K^{(3)}_1], [K^{(3)}_2] ).
\label{eq:ClassificationIndices}
\end{align}
The $C_2$ invariants of a $C_4$-symmetric insulator obey $[X^{(2)}_1]=[Y^{(2)}_1]$ and $[M^{(2)}_1]=-2[M^{(4)}_2]$, and the $C_3$ invariants of a $C_6$-symmetric insulator obey $[K^{(3)}_1]=[K^{(3)}_2]$.  
$C_n$-symmetric TCIs with different $\chi^{(n)}$ belong to different topological classes, as they cannot be deformed into one another without closing the bulk energy gap or breaking the symmetry~\cite{teo2010,Kitaevtable08,teo2013,benalcazar2014}.

Having identified the rotation invariants that distinguish the $C_n$ protected topological phases, we can apply the algebraic method developed in Refs.~\onlinecite{teo2013,benalcazar2014} to connect these invariants to physical properties. The topological classification $\chi^{(n)}$ forms a free Abelian additive structure. Two $C_n$-symmetric TCIs with Hamiltonians $h^{(n)}_1$ and $h^{(n)}_2$, in classes $\chi^{(n)}_1$ and $\chi^{(n)}_2$, and having rotation operators $\hat{r}_n$ and $\hat{r}_n'$, respectively, can be stacked leading to a third $C_n$-symmetric insulator with Hamiltonian $h^{(n)}_3 = h^{(n)}_1 \oplus h^{(n)}_2,$ and with rotation operator $\hat{r}_n''=\hat{r}_n \oplus \hat{r}_n'$. The resulting insulator is in class $\chi^{(n)}_3=\chi^{(n)}_1+\chi^{(n)}_2$. Thus, given a $C_n$ symmetry which classifies TCIs using $N$ topological invariants, all topological classes - and their topological observables - can be accessed by a set of $N$ \emph{primitive generators}: a set of $C_n$-symmetric TCIs having invariants represented by vectors $\chi^{(n)}$ which are linearly independent to one another. From the classifications in Eq.~\ref{eq:ClassificationIndices}, it follows that all of our topological classes can be accessed by combinations of 3 primitive generators for each of $C_4$ and $C_2$-symmetric TCIs, and by 2 primitive generators for each of $C_6$ and $C_3$-symmetric TCIs.

\section{Primitive generators}
\label{sec:PrimitiveGenerators}

\begin{figure}[t]
\centering
\includegraphics[width=\columnwidth]{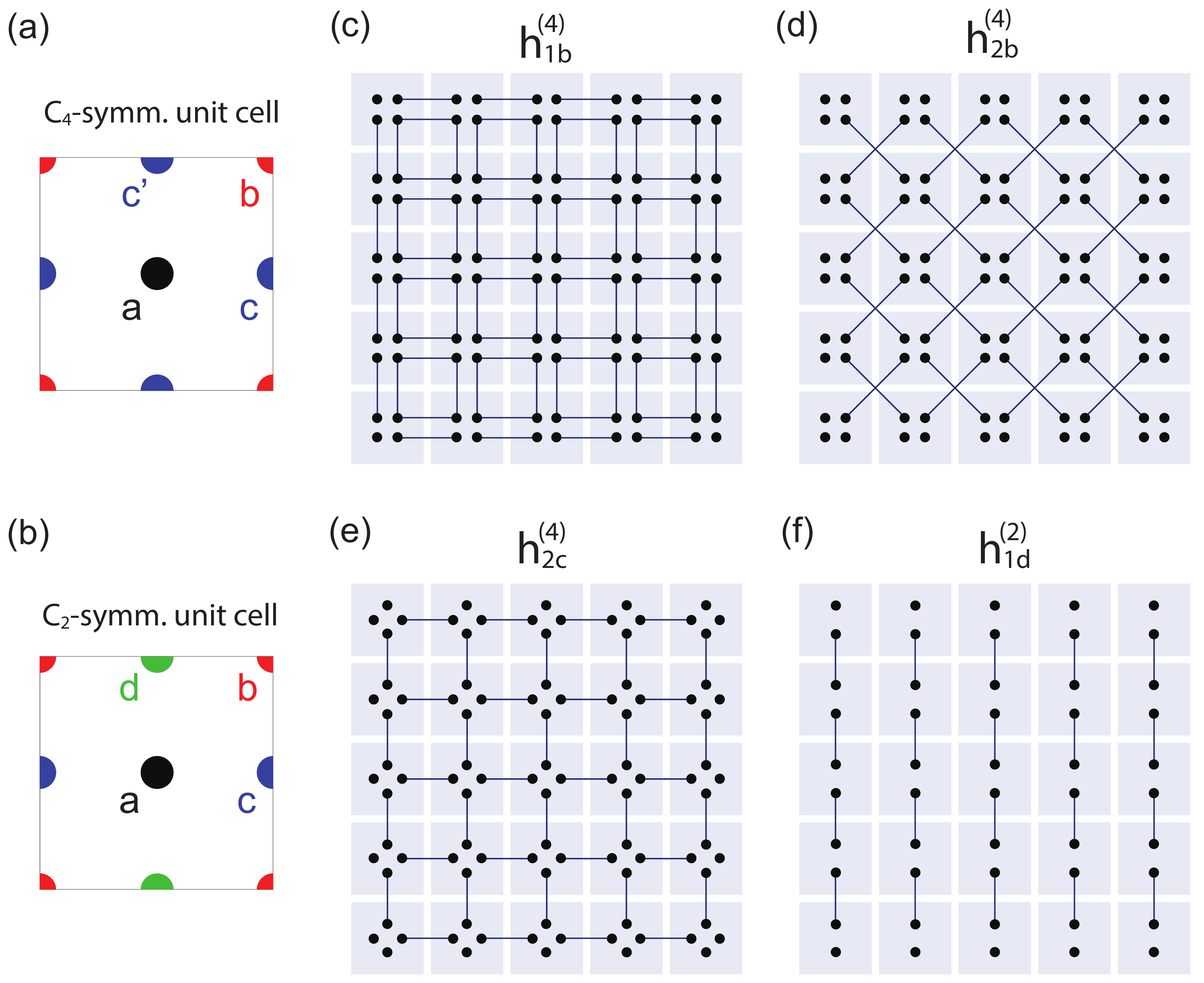}
\caption{(a,b) Maximal Wyckoff positions for (a)  $C_4$- and (b) $C_2$-symmetric unit cells. (c-e) Lattices for the three primitive generators that span the classification of $C_4$-symmetric TCIs. The lattices for the primitive generators for the classification of $C_2$-symmetric TCIs are those in (c), (e), and (f).}
\label{fig:PrimitiveGenerators1}
\end{figure}

\begin{figure}[t!]
\centering
\includegraphics[width=\columnwidth]{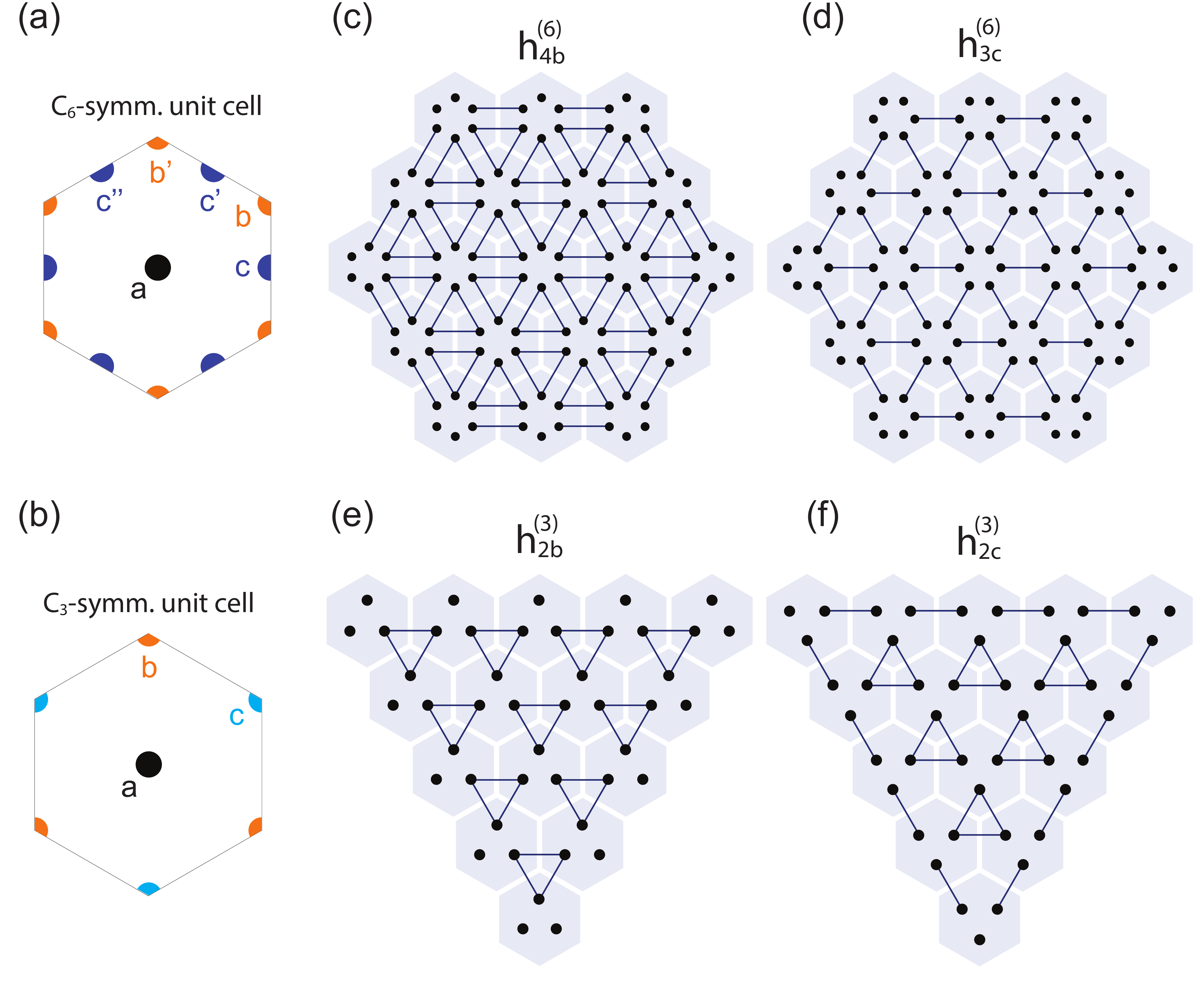}
\caption{(a,b) Maximal Wyckoff positions for (a) $C_6$- and (b) $C_3$-symmetric unit cells. (c,d) Primitive generators that span the classification of $C_6$-symmetric TCIs. (e,f) Primitive generators for the classification of $C_3$-symmetric TCIs.}
\label{fig:PrimitiveGenerators2}
\end{figure}

The primitive generators we consider are illustrated in Figs.~\ref{fig:PrimitiveGenerators1}(c-f) and \ref{fig:PrimitiveGenerators2}(c-f). The shaded squares and hexagons delimit the unit cells. Within each unit cell, the black dots represent its degrees of freedom; for example, they could represent different ions -- each hosting an electronic orbital -- or different orbitals generated by a single ion~\cite{bradlyn2017}. Although the ionic charges do not enter the tight binding Hamiltonians represented in this lattice, our formulation requires that each unit cell contains an integer ionic charge. In all our models, we assume the center of all the positive ionic charge is localized at the maximal Wyckoff position~$a$ of the unit cell (see Appendix~C for a description of ionic positions and choices of unit cells) [black dots in Figs.~\ref{fig:PrimitiveGenerators1}(a,b) and Figs.~\ref{fig:PrimitiveGenerators2}(a,b)]. All the generators are TCIs that admit a Wannier representation~\cite{marzari1997,vanderbilt2012} of their occupied bands.

The $\chi^{(n)}$ invariants of these generators are indicated in Table~\ref{tab:PrimitiveGenerators}. In the bulk, they are Wannier-representable~\cite{marzari1997,vanderbilt2012}, with Wannier centers pinned, by symmetry, to maximal Wyckoff positions other than at the center of the unit cell. In contrast, trivial bands, in class $\chi^{(n)}={\bf 0}$, will necessarily have Wannier centers at the center of the unit cell;  coinciding with the position of the ionic centers. Our primitive generators are in \emph{obstructed atomic limits}~\cite{bradlyn2017,cano2018}, because a connection to the \emph{trivial atomic limit} $\chi^{(n)}={\bf 0}$ is not allowed unless a gap-closing phase transition occurs or the symmetry is broken. 

We present the generators in Figs. \ref{fig:PrimitiveGenerators1} and \ref{fig:PrimitiveGenerators2} in a simple limit without hopping terms within unit cells to allow a pictorial identification of the Wannier centers. Their topological classes are stable to the addition of intra-cell hopping terms or any other symmetry-preserving terms that do not close the bulk gap. In Appendix~D, we detail how adding intra-cell hopping terms can transition our models into a variety of classes in their $\chi^{(n)}$ classifications.  Since our generators are spinless and only require real-valued hoppings (i.e., without any phase factors), they are easy to fabricate in a variety of metamaterials. Indeed, the lattices presented in Refs. ~\onlinecite{noh2018}, \onlinecite{xie2018}  and \onlinecite{xue2018} coincide with the generators shown in Fig.~\ref{fig:PrimitiveGenerators1}(c), Fig.~\ref{fig:PrimitiveGenerators2}(d) and Fig.~\ref{fig:PrimitiveGenerators2}(f), respectively. A first instance of a possible solid state material realization of one of these primitive generators is detailed in Ref.~\onlinecite{ezawa2018} for the generator shown in Fig.~\ref{fig:PrimitiveGenerators2}(f).

We use the notation that a generator $h^{(n)}_{mW}$ is $C_n$-symmetric, has $m$ filled bands, and has Wannier centers at the maximal Wyckoff positions $W$ shown in Figs. \ref{fig:PrimitiveGenerators1} (a,b) and \ref{fig:PrimitiveGenerators2} (a,b).
The classification of $C_4$-symmetric TCIs has three generators: $h^{(4)}_{1b}$, $h^{(4)}_{2b}$, and $h^{(4)}_{2c}$ [Fig.~\ref{fig:PrimitiveGenerators1}(c,d,e)]. All of them have four energy bands. The lattice model in Fig.~\ref{fig:PrimitiveGenerators1}(c) has a gap that separates the first and the second bands, and another gap that separates the third and fourth bands. We take the first generator $h^{(4)}_{1b}$ to occupy only the lowest band, i.e., $\tfrac{1}{4}$-filling.  The generators $h^{(4)}_{2b}$ and $h^{(4)}_{2c}$ are gapped at half filling; hence, we take both of these generators to occupy the lowest two bands. As indicated by their labels, the first two generators have one and two Wannier centers at position $b$, respectively [red dot in Fig.~\ref{fig:PrimitiveGenerators1}(a)], while the third generator has Wannier centers at the two inequivalent positions $c$ and $c'$ [blue dots in Fig.~\ref{fig:PrimitiveGenerators1}(a)].

The classification of $C_2$-symmetric TCIs also requires three generators. We choose the first two of them to be $h^{(4)}_{1b}$ and $h^{(4)}_{2c}$ [Fig.~\ref{fig:PrimitiveGenerators1}(c,e)]. The generator $h^{(4)}_{2b}$ is not independent because its $C_2$ invariants are given by the vector $\chi^{(2)}=(2,2,0)$, which is linearly proportional to the invariant vector of $h^{(4)}_{1b}$, $\chi^{(2)}=(-1,-1,0)$. 
The third generator is a two-dimensional version of the Su-Schrieffer-Heeger (SSH) model \cite{SSH1979}, labeled as $h^{(2)}_{1d}$ and shown in Fig. \ref{fig:PrimitiveGenerators1}(f) in its extremely dimerized limit. As the label indicates, it is an obstructed atomic limit with one Wannier center at position $d$. 

The classification of $C_6$-symmetric TCIs requires two generators. We take them to be $h^{(6)}_{4b}$ and $h^{(6)}_{3c}$ [Fig.~\ref{fig:PrimitiveGenerators2}(c,d)]. Both of them have six energy bands. $h^{(6)}_{4b}$ is taken to occupy the lowest four bands, and has a pair of Wannier centers at each of the Wyckoff positions $b$ and $b'$ [orange dots in Fig.~\ref{fig:PrimitiveGenerators2}(a)], while $h^{(6)}_{3c}$ is taken to occupy the lowest three bands, and has its three Wannier centers at positions $c$, $c'$, and $c''$ [blue dots in Fig.~\ref{fig:PrimitiveGenerators2}(a)].

The classification of $C_3$-symmetric TCIs requires two generators. We take them to be $h^{(3)}_{2b}$ and $h^{(3)}_{2c}$ [Fig.~\ref{fig:PrimitiveGenerators2}(e,f)], which are related to each other by a $\pi$-rotation. Each of these generators has three energy bands with a degeneracy in the lowest two bands protected by $C_3$ symmetry and TRS at the $\bf \Gamma$ point. We therefore take these two models to occupy the lowest two energy bands. $h^{(3)}_{2b}$ has its two Wannier centers at the Wyckoff position $b$ [orange dot in Fig.~\ref{fig:PrimitiveGenerators2}(b)], while $h^{(3)}_{2c}$ has them at position $c$ [cyan dot in Fig.~\ref{fig:PrimitiveGenerators2}(b)]. 

In Appendix~E we induce the representations for Wannier orbitals at all maximal Wyckoff positions for all the $C_n$-symmetric configurations, and by comparing these representations with those of our primitive generators, show that they have the Wannier centers described in this Section.

\section{Filling Anomaly and Charge Fractionalization: Polarization}
\label{sec:AnomalyEdges}
\begin{figure}[!t]
\centering
\includegraphics[width=\columnwidth]{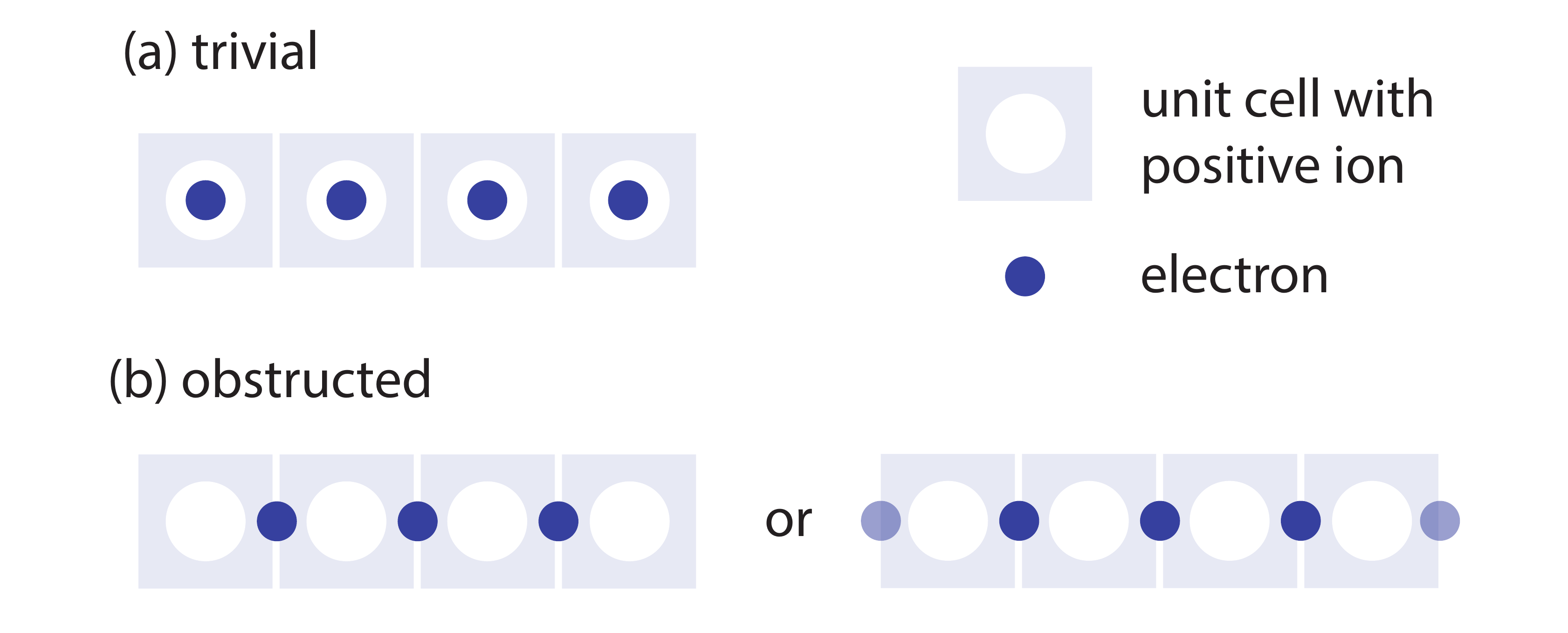}
\caption{Filling anomaly in the reflection symmetric Su-Schrieffer-Hegger model with the Bloch Hamiltonian of Eq.~\ref{eq:SSH_model} and open boundaries. (a) Trivial atomic limit. Charges are balanced. (b) Obstructed atomic limit. Positive and negative charges are unbalanced. For $N$ positive ions, there are $N-1$ electrons (left) or $N+1$ electrons (right). Solid (dimmer) circles represent bulk (boundary) Wannier centers.}
\label{fig:FillingAnomalySSH}
\end{figure}

Due to the crystalline symmetry of a TCI, it may be impossible to maintain the number of electrons required for charge neutrality. To illustrate the simplest case in which this happens, consider the SSH model~\cite{SSH1979}, which has the Bloch Hamiltonian
\begin{align}
h^{SSH}(k) = \left(\begin{array}{cc}
0 & t_0 + t_1 e^{ik}\\
t_0 + t_1 e^{-ik} & 0
\end{array}\right).
\label{eq:SSH_model}
\end{align}
This model has a reflection symmetry,
\begin{align}
\hat{M} h^{SSH}(k) \hat{M}^{-1} = h^{SSH}(-k), \quad \hat{M}=\left(\begin{array}{cc}0&1\\1&0\end{array}\right),\nonumber
\end{align}
that protects two gapped phases separated by a gapless point at $t_0=t_1$. We consider this insulator with electrons occupying only the lowest energy band. At this filling, and with periodic boundary conditions, each unit cell has only one electron. To have a neutral insulator, each unit cell in the crystal has one positive ion with charge $|e|$. When we open the boundaries (with edge terminations that do not cut inside unit cells), on the other hand, the number of electrons is different at each phase. When $t_0>t_1$, $h^{SSH}(k)$ is in the trivial atomic limit and its Wannier centers are as shown in Fig.~\ref{fig:FillingAnomalySSH}(a). In the other phase, $t_0<t_1$, $h^{SSH}(k)$ is in an obstructed atomic limit with Wannier centers as shown in Fig.~\ref{fig:FillingAnomalySSH}(b). Notice that in the trivial phase there is charge neutrality: for $N$ ions in the crystal (one per unit cell), there are $N$ electrons and the configuration is reflection symmetric. On the other hand, in the obstructed atomic limit, charge neutrality is lost:  for $N$ ions, there are either $N-1$ or $N+1$ electrons. Reflection symmetry in $h^{SSH}(k)$ guarantees pairwise degeneracies in the energies of the electronic states at the boundaries. Thus, raising the Fermi level can transition from $N-1$ to $N+1$ electrons, but not from $N-1$ to $N$ which would be needed for neutrality. 

More generically, for a preserved crystalline symmetry that divides a lattice into $n$ symmetry-related sectors, we can define a \emph{filling anomaly} to be
\begin{align}
\eta =  \# \mbox{ions} - \# \mbox{electrons} \quad \mbox{mod }n.
\label{eq:FillingAnomaly}
\end{align}
Thus, in the case of reflection symmetry, which divides the lattice into left and right halves, the filling anomaly (defined modulo 2) captures the parity of charge imbalance. Reflection symmetry guarantees that any extra charge due to charge imbalance in the obstructed atomic limit is distributed equally among the two halves of the lattice. Thus, when the charge imbalance is odd we will have fractional charge $\frac{e}{2}$ modulo $|e|$ in each sector. This happens for the obstructed atomic limit, which has a dipole moment of $p=\frac{e}{2}$; hence, the filling anomaly due to edges is a manifestation of the bulk-boundary correspondence for polarization.

We now extend the formulation of the filling anomaly to TCIs with dipole moments in two dimensions. Let us consider vertically aligned SSH chains having $N_y$ unit cells along $y$. We stack $N_x$ such chains along the $x$-direction as in $h^{(2)}_{1d}$ [Fig.~\ref{fig:PrimitiveGenerators1}(f)] to form a two-dimensional lattice with open boundaries along $y$. To avoid introducing corners, we impose periodic boundary conditions along $x$. The charge imbalance in the obstructed atomic limit will be $N_x$. Following the analysis for the one-dimensional case, we can define the charge density at each of the (upper or lower) halves of the lattice \emph{per unit cell along $x$} by 
\begin{align}
\rho = \frac{\# \mbox{ions} - \# \mbox{electrons}}{2 N_x} |e| \quad\mbox{mod }|e|,
\label{eq:filling_anomaly_charge_density}
\end{align}
where the denominator has a factor of $2$ due to the two symmetry-related halves, and a factor of $N_x$ to determine the charge per unit length. 
The charge density in Eq.~\ref{eq:filling_anomaly_charge_density} captures the usual fractionalization of edge charge density due to a bulk polarization that is quantized under symmetries~\cite{zak1989, vanderbilt1993, hughes2011inversion,turner2012}. It is useful to note that the filling mismatch associated with polarization scales with the system size along $x$, $N_x$. The definition of charge density in Eq.~\ref{eq:filling_anomaly_charge_density} also provides us with a microscopic picture of charge fractionalization; in the extremely dimerized limits we are considering, the fractional boundary charge can be pictorially determined by counting the fraction of bulk Wannier orbitals that fall into the boundary unit cells modulo $|e|$ (e.g., only half of a bulk Wannier orbital falls into the boundary unit cell in Fig.~\ref{fig:FillingAnomalySSH}(b), right).

In previous work, the polarization components $p_{i=1,2}$ (Eq.~\ref{eq:polarization}) of reflection or inversion symmetric TCIs were related to the inversion or reflection symmetry eigenvalues that the occupied states take at the HSPs \cite{hughes2011inversion, turner2012}.  Extending this approach to $C_n$ symmetries~\cite{fang2012}, the values of polarization in terms of the invariants of Eq.~\ref{eq:ClassificationIndices} (detailed in Appendix~B) are
\begin{align}
{\bf P}^{(4)}&=\frac{e}{2}[X^{(2)}_1]({\bf a_1+a_2})\nonumber\\
{\bf P}^{(2)}&=\frac{e}{2}([Y^{(2)}_1]+[M^{(2)}_1]) {\bf a_1}+\frac{e}{2}([X^{(2)}_1]+[M^{(2)}_1]) {\bf a_2}\nonumber\\
{\bf P}^{(6)}&={\bf 0}\nonumber\\
{\bf P}^{(3)}&=\frac{2e}{3}([K^{(3)}_1]+2[K^{(3)}_2])({\bf a_1 + a_2}),
\label{eq:WeakInvariantsFromRotationInvariants}
\end{align}
all of which are defined modulo $e$. These indices can be directly applied to our primitive generators to determine their polarizations. Furthermore, the surface charge theorem immediately relates the bulk polarization to a surface charge density and, for our $C_n$ protected TCIs, yields a quantized fractional charge per edge unit cell. The values of polarization for our primitive generators are indicated in Table~\ref{tab:PrimitiveGenerators}.

\section{Filling Anomaly and Charge Fractionalization: Corner Charge}
\label{sec:AnomalyCorners}

When a TCI has two open edges that intersect to form a corner, a filling anomaly arising from the corner itself may occur. This filling anomaly lies at the heart of higher-order topological insulators in two dimensions. In the initial study of topological quadrupole insulators, for example, there was a recognition that an overall charge imbalance exits in the subspace of occupied bands~\cite{benalcazar2017quad, benalcazar2017quadPRB}, which has latter been found in other higher-order topological phases~\cite{wieder2018,ahn2018,wieder2018b}. The work by Song et al.~\cite{song2017} additionally identified that, in higher-order TCIs that allow a Wannier representation, a mismatch exists between the Wannier centers of the occupied bands and the atomic positions in the crystal. Here, we connect the notion of Wannier center mismatch with the overall deficit of charge in energy bands by considering the bulk and edge electrons in real space representations of higher-order topological bands. This will allow us to put forward a formal definition of the filling anomaly in two dimensions, and to relate all the possible filling anomalies to topological indices (Eq.~\ref{eq:CornerIndices}) written in terms of the topological invariants defined in Eq.~\ref{eq:ClassificationIndices}. 

For this purpose, we will use of our primitive generators defined in Section~\ref{sec:PrimitiveGenerators}. The connection between real space and crystal momentum space via our Wannier-representable primitive generators will make evident the connection between this higher-order filling anomaly and the quantization of fractional corner charge. Our topological indices, however, are more general than the primitive generators they are derived from, and are valid also for TCIs that are not Wannier representable, as we will show for fragile phases in Section~\ref{sec:FragilePhases}.  

\begin{figure}[!t]
\centering
\includegraphics[width=\columnwidth]{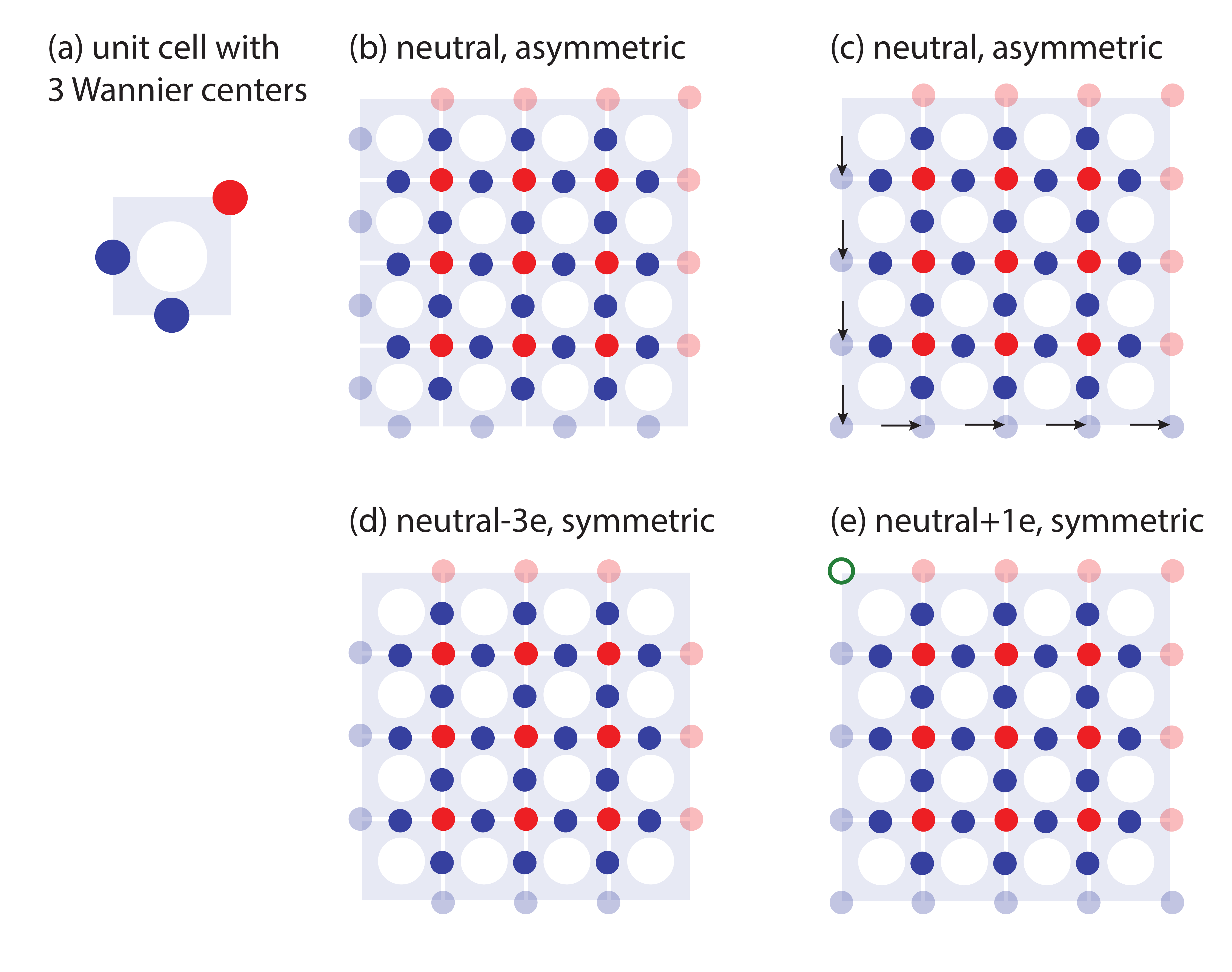}
\caption{Filling anomaly in the $C_4$-symmetric insulator of Eq.~\ref{eq:Simulation}. (a) A unit cell with charge 3$|e|$ at position $1a$ and three electrons with Wannier centers at positions $b$ (red circle) and $c,c'$ (blue circles). (b) A $4 \times 4$ lattice formed by tiling the unit cell shown in (a) along $x$ and $y$. The configuration is neutral but breaks $C_4$-symmetry. (c) A deformation of (b) as an attempt to restore $C_4$ symmetry along the edges; symmetry is still broken at corners. (d,e) Two choices that restore full $C_4$ symmetry in the lattice by either removing 3 corner electrons (d) or adding one (e); in either case, charge neutrality is lost.}
\label{fig:FillingAnomaly2D}
\end{figure}

Let us first illustrate the existence of a corner-induced filling anomaly with an example. Consider Fig.~\ref{fig:FillingAnomaly2D}, which shows the Wannier centers of the $C_4$-symmetric crystalline insulator with the Bloch Hamiltonian
\begin{align}
h^{(4)} = \begin{pmatrix}
h^{(4)}_{1b} & \gamma_c\\
\gamma_c^\dagger & h^{(4)}_{2c}
\end{pmatrix}.
\label{eq:Simulation}
\end{align}
This is an 8 band TCI formed by stacking the primitive generators $h^{(4)}_{1b}$ and $h^{(4)}_{2c}$. $\gamma_c$ represents any $C_4$ symmetry-preserving couplings between the generators that do not close the energy band gap. We will enforce a global $C_4$ symmetry in the lattice of Fig.~\ref{fig:FillingAnomaly2D}, and consider the 4 quadrants --each having one corner -- as our 4 symmetry-related sectors.  At $\frac{3}{8}$-filling, each unit cell has a positive ionic charge of $3|e|$, and its electrons have Wannier centers at the three maximal Wyckoff positions $b$, $c$, and $c'$. For the choice of Wannier centers at each unit cell shown in Fig.~\ref{fig:FillingAnomaly2D}(a), a lattice of $4 \times 4$ unit cells is shown in Fig.~\ref{fig:FillingAnomaly2D}(b). Now we show that this TCI must have a charge imbalance caused by the presence of corners if it is to preserve $C_4$ symmetry: the configuration in Fig.~\ref{fig:FillingAnomaly2D}(b) preserves $C_4$ symmetry in the bulk, but not at the edges. This configuration, of course, is incompatible. Hence, we deform the edge electrons to procure the preservation of the overall $C_4$-symmetry, as in Fig.~\ref{fig:FillingAnomaly2D}(c). We find, however, that $C_4$-symmetry at the edges can be achieved only at the expense of breaking $C_4$ symmetry at the corners. To restore the overall $C_4$ symmetry, we need to cause a charge imbalance by either removing 3 electrons [Fig.~\ref{fig:FillingAnomaly2D}(d)] or adding one [open green circle in Fig.~\ref{fig:FillingAnomaly2D}(e)]. This argument holds for any other choice of deformation of the edges. We conclude that it is not possible to have any choice of Wannier center assignment that preserves charge neutrality and $C_4$ symmetry simultaneously. The filling anomaly in this case is $\eta=3$ or $\eta=-1$, and only $\eta~\mbox{mod 4}=3$ is well-defined (c.f. Eq.~\ref{eq:FillingAnomaly}). Since by symmetry the charge has to be equally distributed over each of the 4 sectors, there has to be a total charge per sector modulo $|e|$ of $Q_{corner}=\frac{3|e|}{4}$. A plot of the charge density for this insulator (with added intra-cell hopping terms as detailed in Appendix~F) is shown in Fig.~\ref{fig:CornerCharge}(a). There, we verify that each quadrant has a charge of $\frac{3|e|}{4}$, and that the charges in each quadrant exponentially localize at the corners of the lattice. A more rigorous demonstration of the exponential localization of the corner charge can be found in Appendix~H.

TCIs with ${\bf P} \neq (0,0)$ will have more edge states than the number of edge electrons needed for charge neutrality, with the number of extra edge states scaling with $N$. As a consequence, a neutral TCI with  ${\bf P} \neq (0,0)$ has electrons that delocalize along the boundary in a metallic state and, being gapless, the notion of a corner filling anomaly is lost. Only if ${\bf P} = (0,0)$, will both the bulk and the edges be generically insulating (Appendix~F shows this characteristic in the simulation of the Hamiltonian in Eq.~\ref{eq:Simulation}) allowing for a well-defined corner filling anomaly, and consequently well-defined corner charges. Neutrality is then achieved only up to the corner filling anomaly (that does \emph{not} scale with $N$). Although each of the generators $h^{(4)}_{1b}$ and $h^{(4)}_{2c}$ has ${\bf P}=(\frac{e}{2},\frac{e}{2})$, the combined TCI in Eq.~\ref{eq:Simulation} has ${\bf P}=(0,0)$, and therefore its edges are also insulating, leading to the well-defined corner charge of Fig~\ref{fig:CornerCharge}(a).
 
 To generalize the properties illustrated in this example, consider a $C_n$-symmetric TCI with vanishing polarization forming a lattice in the shape of a regular polygon having $m$ corners, where $m$ is a multiple of $n$.
 The vanishing polarization will ensure that all the bulk and edge energy bands below the Fermi level are completely filled. When the filling anomaly is zero, the TCI is neutral, but if it is not, there will be a charge imbalance that localizes at corners. In this second case, the $C_n$ symmetry of the lattice enforces the existence of at least one set of $n$-fold degenerate states localized at corners. Since the degenerate corner states can be above or below the Fermi level, the total charge imbalance is not unique. We can say at most that the amount number of charge robustly protected by the bulk phase is $\eta$ modulo $n$ (Eq.~\ref{eq:FillingAnomaly}). This charge is distributed in equal parts in each of the symmetry-related sectors. Sectors subtended by an angle of $\frac{2\pi}{n}$ rad in our lattices will then have a total (electronic and ionic) charge of
\begin{align} Q_{sector} =\frac{\eta}{n} |e|.
\label{eq:Q}
\end{align} 
which is well-defined only modulo $|e|$. We more generically refer to sectors instead of corners because, depending on the chosen global geometry of the lattice, the charge on a single corner may not be quantized. For example, if a $C_2$- ($C_3$-) symmetric bulk Hamiltonian is put on a rectangular (hexagonal) lattice, only the sum of the charges in two adjacent corners will be fractionally quantized. Also, see Ref.~\onlinecite{vanderbilt2015} for a concrete example of an insulator that has zero filling anomaly and consequently zero total charge at each symmetry-related sector but nevertheless has small residual charges at each corner of opposite sign.

Just as in the case of polarization, the corner filling anomaly comes from the bulk of the crystal. This allows us to also develop a microscopic picture that relates the corner fractional charge in a $\frac{2\pi}{n}$ sector to the local distribution of Wannier centers around corner unit cells. In the extremely dimerized limits (as in the case of our generators, Figs.~\ref{fig:PrimitiveGenerators1} and~\ref{fig:PrimitiveGenerators2}), the Wannier orbitals are cut in equal parts by the unit cell's boundaries. The fractional number of electrons in a $\frac{2\pi}{n}$ sector (modulo $1$) can then be obtained by counting the portion of \emph{bulk} Wannier orbitals falling into the corner unit cells at that sector. Adding symmetry-preserving hopping terms to the Hamiltonian that take it away from the extremely dimerized limit can modify the distribution of Wannier centers in the lattice, with the most dramatic change happening at the corners, and the least change happening near the center of the lattice. This results in the spreading of the corner charge into the bulk with exponentially decreasing amplitude away from the corners (see Appendix~H). The integrated charge over the $\frac{2\pi}{n}$ sector, however, remains quantized. 

This microscopic picture explains the lack of quantization at individual corners, -- but the strict quantization over symmetry-related sectors -- by taking into account the shape of the Wannier orbital. This is discussed in detail in Appendix~G. Remarkably, this microscopic picture also stipulates the existence of particular cases of $C_2$-symmetric TCIs that, when put in lattices with 4 corners, exhibit strict quantization of fractional charge at each individual corner. This is the case of generator $h^{(6)}_{3c}$ when put in a parallelogram lattice, as shown in Appendix~G. Finally, the microscopic method allows us to assign fractional corner charge even in TCIs with non-vanishing polarization. These corner charges are not physically meaningful on their own, but their value is useful because they can result in well-defined corner charge in combination with other TCIs that make the total polarization vanish. We denote these ill-defined corner charges as \emph{nominal corner charge}; i.e., they are corner charges in systems that also have a bulk polarization. These corner charge values will be useful for the construction of the index theorems for corner charge in Section~\ref{sec:IndexConstruction}, but cannot be observed unless the polarization is ultimately removed.

\section{Construction of the Topological Indices\\ for Electronic Corner Charge}
\label{sec:IndexConstruction}

In Sections \ref{sec:AnomalyEdges} and \ref{sec:AnomalyCorners} we saw that the fractionalization of edge and corner charges proceeds from a filling anomaly that is intimately related to the positions of the Wannier centers in $C_n$-symmetric TCIs. Furthermore, we also saw that the fractional boundary electronic charges can be captured by inspection if we consider electronic configurations in the zero-correlation length limit because then the bulk Wannier centers are located at maximal Wyckoff positions of the lattice even with open boundary conditions. 

\begin{figure}[t!]
\centering
\includegraphics[width=\columnwidth]{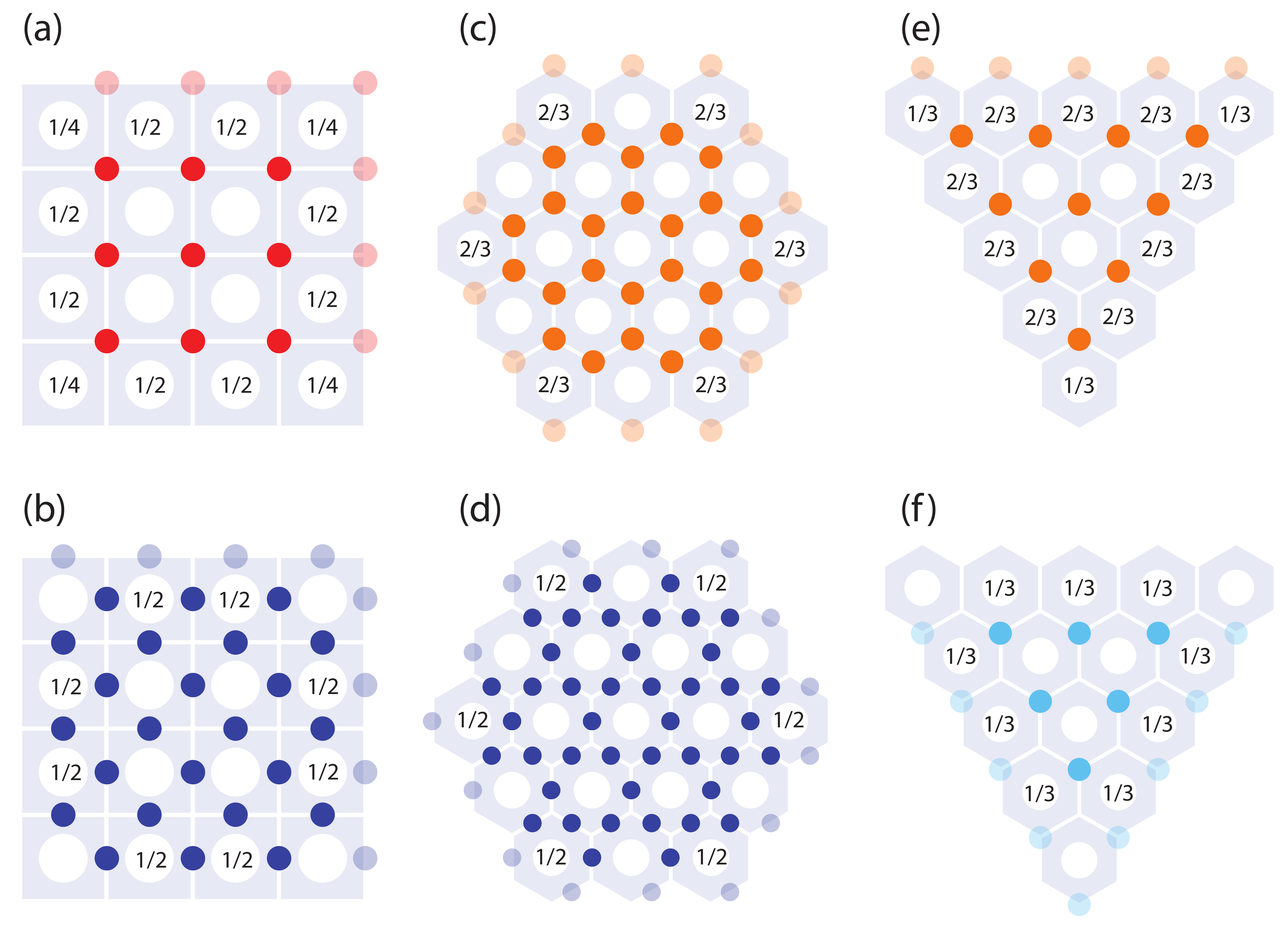}
\caption{Edge and corner fractional charges for TCIs with Wannier centers at maximal Wyckoff positions for (a,b) $C_4$-symmetric, (c,d) $C_6$-symmetric, and (e,f) $C_3$-symmetric lattices. (a) One electron at position $b$. (b) Two electrons at positions $c$ and $c'$. (c) Two electrons at positions $b$ and $b'$. (d) Three electrons at positions $c$, $c'$, and $c''$.  (e) One electron at position $b$. (f) One electron at position $c$. Solid colored circles represent bulk electrons; dimmed colored circles represent boundary electrons for a particular choice of $C_n$-symmetry breaking; white circles represent atomic ions. Bulk unit cells are always neutral. Electronic charges at edge and corner unit cells after the removal of the symmetry breaking electrons are indicated mod $1$ (in units of the electron charge $e$).}
\label{fig:WyckoffBoundaries1}
\end{figure}

In order to construct topological indices for the electronic corner charge (akin to those in Eq.~\ref{eq:WeakInvariantsFromRotationInvariants} for edge charge), we first consider all possible Wannier configurations that respect $C_n$-symmetry in the zero-correlation length limit. This is shown in Fig.~\ref{fig:WyckoffBoundaries1}. The electronic edge and corner charge can then be derived from Fig.~\ref{fig:WyckoffBoundaries1} pictorially by counting the fraction of Wannier orbitals falling in each unit cell. This information, along with the Wannier center description of the primitive generators defined in Section~\ref{sec:PrimitiveGenerators}, allows us to then extract both the electronic edge and corner charges of each generator, which we will need to construct the topological indices for the quanta of charge at corners. 

In what follows, we consider all minimal and inequivalent Wannier configurations given a $C_n$ crystalline symmetry (by minimal we mean that we will put only one Wannier orbital at each Wyckoff position).
 In $C_4$-symmetric TCIs, there are two possible Wannier configurations, one with one Wannier orbital at Wyckoff position $b$ [Fig.~\ref{fig:WyckoffBoundaries1}(a)] and a second one with two Wannier orbitals, one at $c$ and another one at $c'$ [Fig.~\ref{fig:WyckoffBoundaries1}(b)]. Both configurations have polarization ${\bf P}=(\frac{e}{2},\frac{e}{2})$, leading to the fractional charge on the edges. However, when we consider corners, a crucial distinction emerges; Wannier orbitals at Wyckoff positions $b$ have fractional corner charge of $\frac{e}{4}$, while those at positions $c,c'$ have no expected fractional corner charge. For $C_2$-symmetric TCIs, in addition to the two configurations allowed for $C_4$-symmetric TCIs, there is a third configuration having one Wannier orbital at Wyckoff positions $d$ (not pictured), which render $\frac{e}{2}$ edge charge along one pair of edges due to ${\bf P}=(0,\frac{e}{2})$, but no corner charge. For $C_6$-symmetric TCIs, there are two configurations, both having ${\bf P}=(0,0),$ and consequently leading to a vanishing edge charge. At corners, however, the charge is fractionalized. The first configuration has two Wannier orbitals, one at Wyckoff positions $b$ and another one at $b'$ [Fig.~\ref{fig:WyckoffBoundaries1}(c)]. This configuration leads to corner charge in multiples of $\frac{2e}{3}$. The second configuration has three Wannier orbitals, each of them at Wyckoff positions $c,c'$ and $c''$, respectively [Fig.~\ref{fig:WyckoffBoundaries1}(d)]. This second configuration leads to corner charge in multiples of $\frac{e}{2}$. The combination of these two systems can consequently give rise to corner charge in multiples of $\frac{e}{6}$. In $C_3$-symmetric TCIs, there are also two configurations: one with one Wannier orbital at Wyckoff positions $b$ [Fig.~\ref{fig:WyckoffBoundaries1}(e)] and a second one with Wannier orbital at $c$  [Fig.~\ref{fig:WyckoffBoundaries1}(f)]. They have polarizations of  ${\bf P}=(\frac{2e}{3},\frac{2e}{3})$ and  ${\bf P}=(\frac{e}{3},\frac{e}{3})$, respectively. Both configurations then give rise to the edge charge. At corners, however, the configuration with one Wannier orbital at Wyckoff position $b$ does not have fractional charges, while the one having the Wannier orbital at $c$ does.
  By these considerations, the \emph{nominal electronic corner charge} for the primitive generators are found to be those in Table~\ref{tab:PrimitiveGenerators}.

\begin{table}[t]
\begin{tabular}{c|cccccc}
\hline
\hline
symm. & generator & \multicolumn{3}{c}{Invariants} & \quad ${\bf P}$  & $Q_{corner}$\\
\hline
\hline
 &   & $[X^{(2)}_1]$ & $[M^{(4)}_1]$ & $[M^{(4)}_2]$ & & \\
\hline
$C_4$ &$h_{1b}^{(4)}$ & -1 & 1  &  0 &\quad $(\frac{e}{2},\frac{e}{2})$ & $\frac{e}{4}$ \\
&$h_{2b}^{(4)}$ & 2 & 0  &  0 &\quad (0,0) & $\frac{e}{2}$ \\
&$h_{2c}^{(4)}$ & 1 & 1 & -1 &\quad $(\frac{e}{2},\frac{e}{2})$ & 0 \\
\hline
\hline
 &  & $[X^{(2)}_1]$ & $[Y^{(2)}_1]$ & $[M^{(2)}_1]$ &  & \\
\hline
$C_2$ & $h_{1b}^{(4)}$ &-1 &-1 & 0 & \quad $(\frac{e}{2},\frac{e}{2})$ & $\frac{e}{2}$ \\
& $h_{2c}^{(4)}$ & 1 & 1 & 2 & \quad $(\frac{e}{2},\frac{e}{2})$ & $0$ \\
& $h_{1d}^{(2)}$ & 0 & 1 & 1 & \quad $(0,\frac{e}{2}) $ & $0$ \\
\hline
\hline
 &   & $[M^{(2)}_1]$ & $[K^{(3)}_1]$ & & & \\
\hline
$C_6$ & $h_{4b}^{(6)}$ & 0 & 2 & & \quad $(0,0)$ & $\frac{e}{3}$\\
& $h_{3c}^{(6)}$ & 2 & 0 & & \quad $(0,0)$ & $\frac{e}{2}$\\
\hline
\hline
 &  & $[K^{(3)}_1]$ & $[K^{(3)}_2]$ & & &\\
\hline
$C_3$ & $h_{2b}^{(3)}$ & 1 & -1 & & \quad $(\frac{e}{3},\frac{e}{3})$ & $\frac{2e}{3}$\\
& $h_{2c}^{(3)}$ & 1 & 0 & & \quad $(\frac{2e}{3},\frac{2e}{3})$ & 0 \\
\hline
\hline
\end{tabular}
\caption{
Topological invariants, polarization ${\bf P}$, and nominal electronic corner charge of the primitive generators that span the classifications of $C_n$-symmetric TCIs. The values of ${\bf P}=p_1 {\bf a_1} + p_2 {\bf a_2}$ are given in pairs $(p_1,p_2)$. The unit lattice vectors are ${\bf a_1=\hat{x}}$, ${\bf a_2=\hat{y}}$ for $C_4$ and $C_2$-symmetric lattices, and ${\bf a_1}={\bf \hat{x}}$, ${\bf a_2}=\frac{1}{2}{\bf \hat{x}}+\frac{\sqrt{3}}{2}{\bf \hat{y}}$ for $C_3$ and $C_6$-symmetric lattices.}
\label{tab:PrimitiveGenerators}
\end{table}
This information characterizes the corner properties of TCIs in class AI having additional $C_n$ symmetry, and hence allows us to build index theorems that determine the fractional electronic corner charge. This relies on the fact that for a Hamiltonian $h^{(n)}_3=h^{(n)}_1 \oplus h^{(n)}_2$, (i) its boundary electronic charge is $Q_3=Q_1+Q_2$ (mod $e$), and (ii) its invariants are $\chi^{(n)}_3=\chi^{(n)}_1+\chi^{(n)}_2$. The index for the electronic corner charge of a $C_n$-symmetric insulator is then given by a linear combination of the invariants that form the vector $\chi^{(n)}$. For example, $C_4$-symmetric TCIs have three invariants. The electronic corner charge is given by $Q^{(4)}_{corner}= \alpha_1 [X^{(2)}_1]+\alpha_2[M^{(4)}_1]+\alpha_3[M^{(4)}_2]$. To find the coefficients $\alpha_{i=1,2,3}$, we solve for $Q_i= \chi^{(4)}_{ij} \alpha_j$, where $Q_i$ is $i$th element in the vector of corner charges formed by the last column in Table~\ref{tab:PrimitiveGenerators}, and $\chi^{(4)}_{ij}$ is the $(i,j)$th element in the matrix formed by the three columns labeled $[X^{(2)}_1]$, $[M^{(4)}_1]$, and $[M^{(4)}_2]$ in Table~\ref{tab:PrimitiveGenerators}. This approach gives
\begin{align}
Q^{(4)}_{corner}&=\frac{e}{4}([X^{(2)}_1]+2[M^{(4)}_1]+3[M^{(4)}_2])\;\; \mbox{mod }e\nonumber\\
Q^{(2)}_{corner}&=\frac{e}{4}(-[X^{(2)}_1]-[Y^{(2)}_1]+[M^{(2)}_1])\;\; \mbox{mod }e\nonumber\\
Q^{(6)}_{corner}&=\frac{e}{4}[M^{(2)}_1]+\frac{e}{6}[K^{(3)}_1]\;\; \mbox{mod }e\nonumber\\
Q^{(3)}_{corner}&=\frac{e}{3}[K^{(3)}_2]\;\; \mbox{mod }e
\label{eq:CornerIndices}
\end{align}
where the superindex $n$ in $Q^{(n)}_{corner}$ labels the $C_n$ symmetry. In the $C_n$-symmetric classification, $Q^{(n)}_{corner}$ is a $\mathbb{Z}_n$ topological index.
We could refer to the indices in Eq.~\ref{eq:CornerIndices} as \emph{secondary topological indices} because they require the primary topological index --- the polarization ${\bf P}$ --- to vanish in order to give a protected, corner-localized quantized feature. 

As an example of the application of the indices in Eq.~\ref{eq:CornerIndices}, let us return to the eight-band model considered above in  Eq.~\ref{eq:Simulation}, which has electronic corner charge of $\frac{e}{4},$ and a total (electronic \emph{and} ionic) charge density shown in Fig.~\ref{fig:CornerCharge}(a). By itself,  the model $h^{(4)}_{1b}$ at $\frac{1}{4}$ filling that forms one block of the eight band system has edge states owing to its ${\bf P}=(\frac{e}{2},\frac{e}{2})$ polarization. Not all the edge states can be occupied at this filling while preserving the symmetry, however, and the edge is generically metallic (see Appendix~F for details). We can remove the polarization by the addition of $h^{(4)}_{2c}$, the second block of the eight band model, which at $\frac{1}{2}$ filling also has ${\bf P}=(\frac{e}{2},\frac{e}{2})$. Under any $C_4$ symmetry-preserving coupling terms $\gamma_c$ that keep the energy gap open, the primary index of the combined insulator (Eq.~\ref{eq:Simulation}) at $\frac{3}{8}$ filling is ${\bf P}=(0,0)$, but its secondary index is $Q^{(4)}_{corner}=\frac{e}{4}$ (first equation in Eq.~\ref{eq:CornerIndices}). To confirm that this charge is generically stable, we add general random hopping terms to the Hamiltonian up to nearest-neighbor unit cells that preserve only TRS and $C_4$ symmetry and numerically verify that the $\frac{e}{4}$ electronic charge remains strictly quantized (see Appendix~F). In contrast, if we add perturbations that break $C_4$ symmetry down to $C_2$ symmetry ($C_2$ symmetry keeps bulk polarization quantized to zero), the quantization of charge at each corner in the lattice is lost. However, the sum of electronic corner charge of two adjacent corners (i.e., in a region covering half the lattice) is $\frac{e}{2}$, in agreement with the value predicted by the secondary index in the second equation of Eq.~\ref{eq:CornerIndices}.

The indices in Eq.~\ref{eq:CornerIndices} can be used to generate other corner charges. The total (ionic and electronic) fractional charge of $\frac{|e|}{2}$ in Fig.~\ref{fig:CornerCharge}(b) was obtained with a Hamiltonian deformable to $h^{(4)}_{1b} \oplus h^{(4)}_{1b}$ at $\frac{1}{4}$ filling, while the corner charges of $\frac{|e|}{6}$ and $\frac{|e|}{3}$ in Fig.~\ref{fig:CornerCharge}(c,d) were obtained by Hamiltonians deformable to $h^{(6)}_{4b} \oplus h^{(6)}_{3c}$ at $\frac{7}{12}$ filling and to $h^{(3)}_{2b} \oplus h^{(3)}_{2c}$ at $\frac{2}{3}$ filling, respectively. In all cases, the polarization of the Hamiltonians is ${\bf P}=(0,0),$ and the electronic corner charge indices in Eq.~\ref{eq:CornerIndices} give $Q_{corner}=\frac{e}{2}$, $\frac{5e}{6}$, and $\frac{2e}{3}$, respectively. The total charge density, which takes into account the ionic contributions, results in the fractional charges shown in Fig~\ref{fig:CornerCharge}. 
Since the electronic charge fractionalization is a property of the bulk, a fast way to determine the filling anomaly is to remove all boundary Wannier centers in the lattice (e.g., removing all dimmed circles in Fig.~\ref{fig:FillingAnomaly2D}). The resulting charge imbalance mod $n$ then gives the filling anomaly. Table~\ref{tab:AnomaliesModels} shows the charge imbalance by removal of all boundary Wannier centers, filling anomalies, and total (electronic and ionic) corner charge values over $\frac{2\pi}{n}$ spatial sectors for $C_n$-symmetric TCIs used in the simulations in Fig.~\ref{fig:CornerCharge}.
\begin{table}[t]
\centering
\begin{tabular}{c|ccc}
\hline\hline
insulator  & charge imbalance & $\eta$ & $Q_{sector}$ \\
\hline
$h^{(4)}_{1b}\oplus h^{(4)}_{2c}$   &  $4N-1$ & 3 & $\frac{3|e|}{4}$\\
$h^{(4)}_{1b}\oplus h^{(4)}_{1b}$  &  $4N-2$ & 2 & $\frac{|e|}{2}$\\
$h^{(6)}_{4b}$ & $6N-4$ & 2 & $\frac{|e|}{3}$\\
$h^{(6)}_{3c}$ & $6N-3$ & 3 & $\frac{|e|}{2}$\\
$h^{(3)}_{2b}\oplus h^{(3)}_{2c}$  &  $6N-2$ & 1 & $\frac{|e|}{3}$\\
\hline\hline
\end{tabular}
\caption{Charge imbalance (upon removal of all boundary Wannier centers), filling anomaly $\eta$ (Eq.~\ref{eq:FillingAnomaly}) , and total sector charge $Q$ (Eq.~\ref{eq:Q}) for some models having vanishing bulk polarization. Calculations are assuming that $C_n$-symmetric Bloch Hamiltonians are put in $C_n$-symmetric lattices, respectively.}
\label{tab:AnomaliesModels}
\end{table}

\section{Fractional corner charge in TCIs without a Wannier representation}
\label{sec:FragilePhases}
The secondary index theorems in Eq.~\ref{eq:CornerIndices} were derived using a basis of primitive generators that admit Wannier representations. We chose that basis to make transparent the derivation of the indices. However, the indices themselves transcend the basis and indicate the fractionalization of electronic corner charge even in TCIs that are not Wannier-representable, for example, in fragile TCIs~\cite{vishwanath2018,cano2018a,bohuon2018,bradlyn2018}. Recently, corner states and corner fractional charges have been found in fragile TCIs~\cite{wang2018,ahn2018,wieder2018,wieder2018b}, and the existence of this fractionalization has been associated with quantized nested Berry phases~\cite{wang2018,ahn2018,wieder2018,wieder2018b} originally proposed in Ref.~\onlinecite{benalcazar2017quad} for the characterization of corner charges in quantized quadrupole insulators. Unlike atomic insulators, fragile TCIs do not admit the construction of Wannier centers. However, they have the property that upon the addition of atomic TCIs, the combined system becomes Wannier-representable. We can write this relation as $H_{AI_2} \sim H_{FT} \oplus H_{AI_1}$, where $AI_{i=1,2}$ are atomic TCIs and $FT$ is the fragile TCI. The electronic corner charges of these TCIs must then obey $Q_{AI_2} = Q_{FT} + Q_{AI_1}$, which implies that, since both $Q_{AI_1}$ and $Q_{AI_2}$ are quantized, $Q_{FT}$ will also be quantized. Moreover, due to the algebraic structure of our classification~\cite{teo2013,benalcazar2014}, it follows that the classes of these TCIs in their $C_n$ classification obey $\chi^{(n)}_{AI_2} = \chi^{(n)}_{FT} + \chi^{(n)}_{AI_1}$. The same algebraic structure stipulates that the secondary indices must obey $Q^{(n)}_{corner}(\chi^{(n)}_{AI_2})=Q^{(n)}_{corner}(\chi^{(n)}_{FT})+Q^{(n)}_{corner}(\chi^{(n)}_{AI_1})$. Since for the atomic TCIs we know that $Q^{(n)}_{corner}(\chi^{(n)}_{AI_i})=Q_{AI_i}$, for $i=1,2$, it follows that $Q^{(n)}_{corner}(\chi^{(n)}_{FT}) = Q_{FT}$. Thus, our indices in Eq.~\ref{eq:CornerIndices} correctly determine the quantization of electronic fractional charge in fragile phases. A concrete example of the corner charge in a fragile phase is shown in Appendix~I for one of the phases described in the recent preprint of Ref.~\onlinecite{liu2018shift}. There, we (i) calculate the indices from a decomposition into atomic TCIs, (ii) directly evaluate the secondary index from the topological invariants of the fragile phase, and (iii) compare these results with numerical simulations.

\section{Fractional charge at topological defects}
\label{sec:Defects}
\begin{figure}[t]
\centering
\includegraphics[width=\columnwidth]{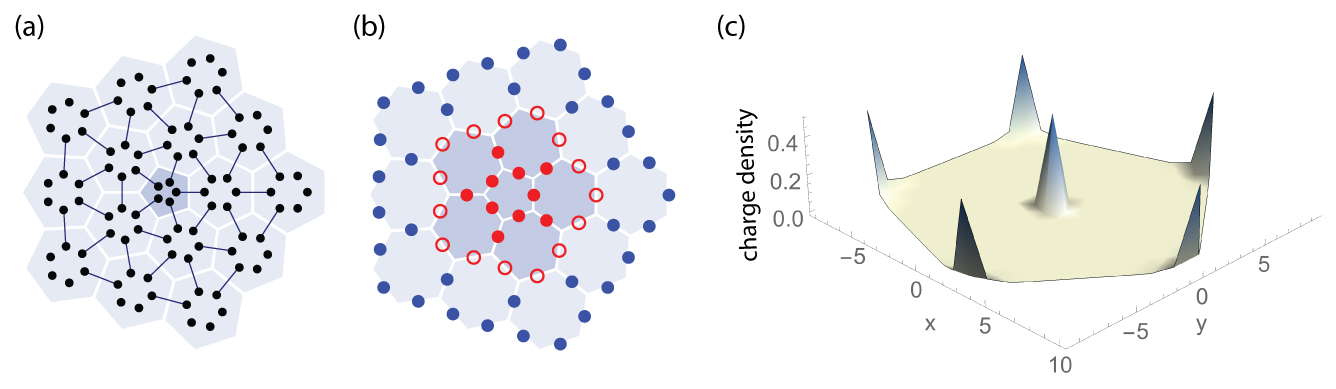}
\caption{Quantized fractionalization of charge at the core of disclinations. (a) Disclination in the lattice of primitive generator $h^{(6)}_{3c}$. (b) Wannier centers for the lattice in (a). There is an overall fractional electronic charge (each hollow circle contributes $\frac{e}{2}$ charge) within the region of darker unit cells which enclose the core of the disclination. (c) Charge density for the disclination in (a). All corners and the core of the disclination have charges of $\frac{|e|}{2}$. The simulation is done over 276 unit cells with added intra-unit cell hoppings between nearest neighbors of $\frac{1}{4}$ the amplitude of the inter-unit cell hoppings.}
\label{fig:Disclination}
\end{figure}

First-order TCIs manifest fractional charges at dislocations, following the topological index $Q_{dislocation} = {\bf P \cdot B}$, where ${\bf P}$ is the polarization, Eq.~\ref{eq:polarization}, and ${\bf B}$ is the Burgers vector that characterizes the dislocation~\cite{teo2010}. Higher-order TCIs have ${\bf P=0}$, and thus do not manifest fractional charges at dislocations. In this section we will see that instead they manifest fractional charges at the core of disclinations. Indeed, it is known that topological disclination defects that induce a curvature singularity in the lattice of $C_n$-symmetric topological superconductors can trap Majorana bound states~\cite{teo2013,benalcazar2014}. Here, we find that these defects also trap fractional charges in higher-order TCIs. In Fig.~\ref{fig:Disclination} we show a disclination with a Frank angle of $-\frac{2\pi}{6}$ rad in the primitive model $h^{(6)}_{3c}$. Inducing such a disclination converts the hexagon of Fig.~\ref{fig:PrimitiveGenerators2}(d) into the pentagon of Fig.~\ref{fig:Disclination}(a). The five corners in the pentagon give rise to an overall corner charge of $\frac{5e}{2}$. Thus, the core of the disclination must trap a fractional charge of $\frac{e}{2}$. Indeed, the Wannier center configuration shown in Fig.~\ref{fig:Disclination}(b) for the lattice in  Fig.~\ref{fig:Disclination}(a) reveals that, for any area comprised of unit cells containing the core of the disclination, a total fractional number of electrons are enclosed. Fig.~\ref{fig:Disclination}(c) shows a plot of the charge density for the lattice in Fig.~\ref{fig:Disclination}(a) but to which additional hopping terms inside the unit cell were added of weak enough amplitude so as to not cause a phase transition. This plot indeed presents the expected charge distribution. If the intra-unit cell couplings are larger than the inter-unit cell couplings, a bulk phase transition occurs, leading to vanishing corner charges and integer charge at the core of the disclination.

Generalizing this principle of charge conservation (mod $|e|$), our corner charge indices can be immediately used to generate indices for the fractional charge at the core of disclinations in a $C_n$-symmetric insulator:
\begin{align}
Q_{disclination}=-\frac{\Omega}{2\pi/n} Q_{corner}\;\;\mbox{mod }|e|.
\end{align}
We also note that inducing this disclination disrupts the chiral symmetry in the primitive model $h^{(6)}_{3c}$. Thus, although the pristine insulator has zero energy states localized at corners \cite{noh2018}, there are no such states at the core of the disclination. Despite this, the fractional charge trapped at the core of the defect is robustly quantized to $\frac{e}{2}$, suggesting that disclinations are bulk probes of TCIs with $Q^{(n)}_{corner} \neq 0$~\cite{teo2013,ruegg2013,ruegg2013b,benalcazar2014,teo2017}, just as dislocations are bulk probes of TCIs with ${\bf P} \neq {\bf 0}$~\cite{teo2017,ran2009,ran2010,teo2010,juricic2012,GopalakrishnanTeoHughes13,hughes2014,paulose2015}. 

\section{Discussion and Conclusion}
\label{sec:Conclusions}

In this paper, we have shown that electronic charge fractionalizes in multiples of $\frac{e}{n}$ at the corners of $C_n$ symmetric TCIs with vanishing polarization. We built topological indices for the quanta of corner charge in terms of the band representations at high symmetry points of the Brillouin zone. These constitute \emph{secondary topological indices}, Eq.~\ref{eq:CornerIndices}, that signal the presence of higher order topology in TCIs with vanishing polarization. When TCIs admit a Wannier representation, we find an clear relation between the existence of fractional corner charge and the positions of the Wannier centers in the bulk of the crystal. More generally, however, a Wannier representation is not guaranteed, but a filling anomaly can still persist, which in turn robustly protects the fractionalization of corner charge. Since the fractionalization of corner charge is ultimately related to the Wannier centers of the electrons within the crystal, we anticipate that the same principles derived in this study will lead to the characterization of corner charge in other classes of the ten-fold classification. For example, adding spin will double the corner charge quantization due to Kramers' degeneracy. However, deriving index theorems in these classes will not be as straightforward because symmetry representations at high symmetry points do not suffice to determine the Wannier centers in spinful systems.

In practice, we expect solid state TCIs with non-zero $Q^{(n)}_{corner}$ indices to prefer to be neutral. Despite the overall neutrality, we still expect the corner charges to be observable. The excess or deficit charge could be compensated for in several ways. In any realistic crystal, there will be impurities and, since the filling anomaly due to corners only indicates $O(1)$ uncompensated charges, impurities could absorb the charges needed to realize neutrality. This will affect the corner charge at most by an integer when the impurity is localized very near the corner and thus the fractional part of the charge will be preserved. Another scenario for localized charges is in TCIs with mid-gap topological modes associated with the fractional charge. In these systems, the symmetry could be mildly broken explicitly or spontaneously, allowing for a ground state filling of the mid-gap modes that is globally neutral. Then, the corner charges will also be shifted by an integer and the fractional portion of the charge is undisturbed. If instead there are no mid-gap topological or impurity states, we could imagine that the overall excess charge at the corners can be compensated by an occupation/de-occupation of eigenstates in the conduction or valence bulk bands. The resulting effect is a near-to-quantized corner charge, with a bulk interior containing the opposite charge, as shown in Appendix~J. The corrections to both the corner charge and the background bulk charge scale as $O(1/N^2)$ for a lattice with $N$ unit cells per side and thus closely approximate exact quantization in the thermodynamic limit.

While we expect there to be electronic material realizations of systems with fractional corner charges, we believe that the most straightforward realization of our models is in metamaterial systems. Since our classifications are for spinless systems that preserve time-reversal symmetry, the hopping terms in the Hamiltonian do not require any additional phase factors and can be engineered using only evanescently coupled modes. Thus, our generators can be easily implemented in a wide range of metamaterial platforms, as in the works in Refs.~\onlinecite{noh2018, imhof2018, serragarcia2018, peterson2018, xue2018,khanikaev2018}. In the experiments in Refs.~\onlinecite{noh2018, imhof2018, serragarcia2018, peterson2018}, the expected corner properties were observed spectroscopically through the appearance of corner states protected  to be at mid-gap by chiral symmetry, $\Pi h({\bf k})=-h({\bf k}) \Pi$, for some chiral operator $\Pi$. We illustrated here, however, that the true signature of fractionalization of corner charge is a bulk topological property of the subspace of occupied bands, and does not need to manifest in connection with corner-localized mid-gap states. In the absence of mid-gap states, more sophisticated experiments, for example, exploring the spatial distribution of all the states in an energy band, can reveal the fractional signatures at corners even in metamaterials.  Such experiments could easily explore the properties of disclinations as well by introducing such defects in resonator arrays or photonic crystals. 

\emph{Acknowledgements. --} We thank Aris Alexandradinata, B. Andrei Bernevig,  Eugeniu Plamadeala, Emil Prodan, and Benjamin Wieder for useful discussions. In particular, we thank Andrei Bernevig and Benjamin Wieder for detailed discussion and for sharing their manuscript, Ref.~\onlinecite{wieder2018b}, during the preparation of this article. W.A.B. thanks the support of the Eberly Postdoctoral Fellowship at the Pennsylvania State University. W.A.B. and T.L.H. thank the U.S. National Science Foundation under grant DMR-1351895 and the Sloan Foundation for support. TL thanks  US National Science Foundation (NSF) Emerging Frontiers in Research and Innovation (EFRI) grant EFMA-1627184.

\bibliography{references}
\end{document}


\title{Quantization of fractional corner charge\\
in $C_n$-symmetric higher-order topological crystalline insulators:\\ Supplementary information}
\author{Wladimir A. Benalcazar}
\affiliation{Department of Physics, The Pennsylvania State University, University Park, PA 16802, USA}
\affiliation{Department of Physics and Institute for Condensed Matter Theory, University of Illinois at Urbana-Champaign, IL 61801, USA}
\author{Tianhe Li}
\author{Taylor L. Hughes}
\affiliation{Department of Physics and Institute for Condensed Matter Theory, University of Illinois at Urbana-Champaign, IL 61801, USA}
\date{\today}
\maketitle

\tableofcontents

\begin{appendices}
\section{Construction of the classification}
\label{sec:ConstructionofClassification}
\begin{figure}[t]
\centering
\includegraphics[width=\columnwidth]{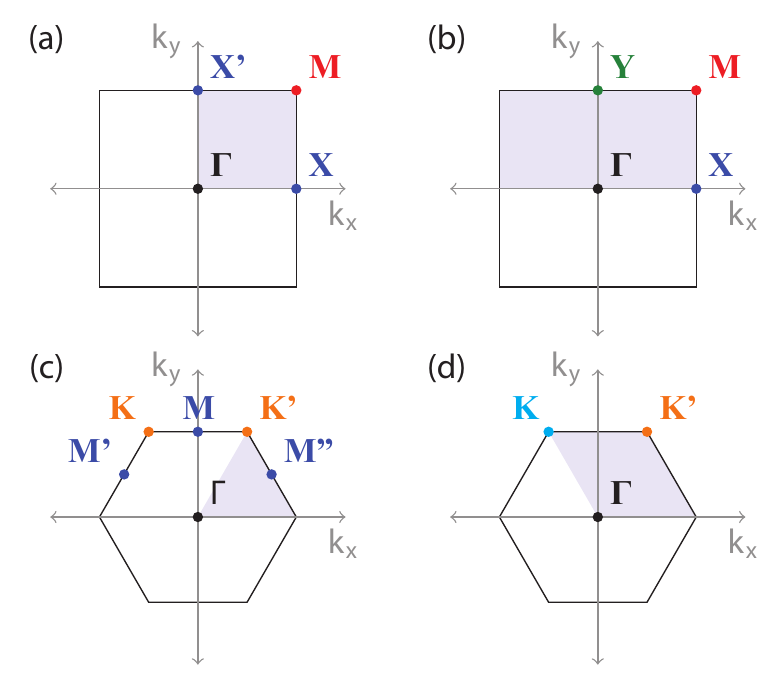}
\caption{Brillouin zone of crystals with $C_4$, $C_2$, $C_6$, and $C_3$ symmetries and their rotation invariant points. (a) ${\bf M}$ is a fourfold HSP, ${\bf X}$ and ${\bf X'}$ are twofold HSPs. (b) ${\bf X}$, ${\bf Y}$, and ${\bf M}$ are twofold HSPs. (c) ${\bf M}$, ${\bf M'}$, and ${\bf M''}$ are twofold HSPs; ${\bf K}$ and ${\bf K'}$ are threefold HSPs. (d) ${\bf K}$ and ${\bf K'}$ are threefold HSPs. For $C_n$-symmetric crystals, ${\bf \Gamma}$ is an n-fold HSP.}
\label{fig:BZInvariantPoints}
\end{figure}

In this Section, we classify TCIs in class AI of the 10-fold classification \cite{altland1997}. Insulators in this class have time-reversal symmetry (TRS) with a Bloch Hamiltonian satisfying
\begin{align}
\Theta h({\bf k}) \Theta^{-1} = h(-{\bf k}),
\end{align}
where $\Theta=K$ is the antiunitary time reversal operator, and $K$ is complex conjugation. The operator obeys $\Theta^2=1$. Preserving TRS leads to a vanishing Hall conductance, indicated by a vanishing Chern number, $Ch=0$. 
In addition, crystalline symmetries expand the classification of topological phases. In particular, we focus on crystals preserving rotation symmetry
\begin{align}
\hat{r}_n h({\bf k}) \hat{r}_n^\dagger = h (R_n {\bf k}),
\label{eq:rotation_symmetry}
\end{align}
where $\hat{r}_n$ is the $n$-fold rotation operator which obeys $\hat{r}_n^n=1$ and $R_n$ is the $n$-fold rotation matrix acting on the momentum vector ${\bf k}$. The rotation operator also obeys
\begin{align}
[\hat{r}_n, \Theta]=0.
\end{align}
At the high symmetry points (HSPs) of the BZ, ${\bf \Pi}^{(n)}$, i.e., at the points in the BZ that map back to themselves upon a rotation, $R_n {\bf \Pi}^{(n)} = {\bf \Pi}^{(n)}$ modulo a reciprocal lattice vector (Fig.~\ref{fig:BZInvariantPoints}), we have, from \eqref{eq:rotation_symmetry},
\begin{align}
[\hat{r}_n,h({\bf \Pi}^{(n)})]=0.
\end{align}

\begin{figure}[t]
\centering
\includegraphics[width=\columnwidth]{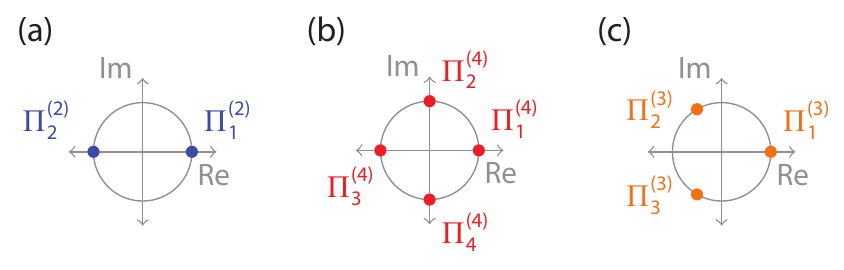}
\caption{Rotation eigenvalues, defined in Eq. \eqref{eq:RotationEigenvaluesapp}, at (a) a twofold HSP, (b) a fourfold HSP, and (c) a threefold HSP.}
\label{fig:RotationEigenvalues}
\end{figure}

Thus, the energy eigenstates at the HSPs can be chosen to be eigenstates of the rotation operator. Let us denote the eigenvalues of $\hat{r}_n$ at the HSP ${\bf \Pi}^{(n)}$ as
\begin{align}
\Pi^{(n)}_p = e^{2\pi i(p-1)/n}, \quad \mbox{for } p=1,2,\ldots n
\label{eq:RotationEigenvaluesapp}
\end{align}
as illustrated in Fig.~\ref{fig:RotationEigenvalues}.

The rotation eigenvalues at two HSPs of a given subspace of energy bands allows us to compare their representation. If different representations of a rotation symmetry exist between two HSPs of the BZ, the energy bands have non-trivial topology (we include non-trivial obstructed atomic limits in this definition of non-trivial topology). Accordingly, we use the rotation eigenvalues at two momenta, ${\bf \Pi}^{(n)}$ and ${\bf \Gamma}=(0,0)$, to define the integer topological invariants
\begin{align}
[\Pi^{(n)}_p] \equiv \# \Pi^{(n)}_p - \# \Gamma^{(n)}_p,
\label{eq:RotationInvariants}
\end{align} 
where $\# \Pi^{(n)}_p$ is the number of energy bands below the energy gap with eigenvalue $\Pi^{(n)}_p$. Not all these invariants are independent, however. First, rotation symmetry can force representations at certain HSPs to be the same. This is shown in Section~\ref{sec:ConstraintsRotationSymmetry}. We will see that in $C_4$-symmetric crystals, rotation symmetry forces the representations at ${\bf X}$ and ${\bf X'}$ to be equal, while in $C_6$-symmetric crystals, this symmetry forces equal representations at ${\bf X}$, ${\bf X'}$, and ${\bf X''}$, as well as at ${\bf K}$ and ${\bf K'}$. Furthermore, there are redundancies in the invariants due to (i) the fact that the number of bands in consideration is constant across the BZ, from which it follows that  $\sum_p \# \Pi^{(n)}_p = \sum_p \# \Gamma^{(n)}_p$, or $\sum_p [\Pi^{(n)}_p]=0$, and (ii) the existence of TRS in the TCIs, which implies that rotation eigenvalues are either real or they come in complex conjugate pairs, as shown in Section~\ref{sec:ConstraintsTRS}. 
To go over these constraints systematically, in what follows we first look at time-reversal symmetry, then rotation symmetry, and finally the interplay of the two of them. Using these constraints, we will then construct the complete set of invariants for each classification.

\subsection{Time-reversal symmetry}
\label{sec:ClassificationTRS}
A time-reversal symmetric TCIs obeys
\begin{align}
\Theta h({\bf k}) \Theta^{-1} = h(-{\bf k}).
\end{align}
Here, $\Theta=K$ is the time reversal operator, which consists only on complex conjugation $K$. The operator obeys $\Theta^2=1$. Acting on an energy eigenstate, we have
\begin{align}
h(-{\bf k}) \Theta \ket{u_{\bf k}^n}=\Theta h({\bf k})  \ket{u_{\bf k}^n}= \epsilon_n({\bf k}) \Theta \ket{u_{\bf k}^n}
\label{eq:TRS}
\end{align}
Thus, $\Theta \ket{u_{\bf k}^n}$ is an eigenstate of $h(-{\bf k})$ with energy $\epsilon_n({\bf k})$. This means that we can write the expansion
\begin{align}
\Theta \ket{u_{\bf k}^n} &=\sum_m \ket{u_{-{\bf k}}} V_{\bf k}^{mn}
\label{eq:TRSexpansion}
\end{align}
where
\begin{align}
V^{mn}_{\bf k}=\bra{u_{-{\bf k}}^m}\Theta \ket{u_{\bf k}^n}=\braket{u_{-{\bf k}}^m}{u_{\bf k}^{n*}}
\end{align}
is the (unitary) sewing matrix. Using Eq.~\eqref{eq:TRS}, let us operate as follows
\begin{align}
h(-{\bf k})\Theta \ket{u_{\bf k}^n}&=\epsilon_n({\bf k}) \Theta \ket{u_{\bf k}^n}\nonumber\\
&=\epsilon_n({\bf k})\sum_m \ket{u_{-{\bf k}}} V_{\bf k}^{mn}.
\label{eq:TRSaux1}
\end{align}
On the other hand, using the expansion in \eqref{eq:TRSexpansion}, we have that
\begin{align}
h(-{\bf k})\Theta \ket{u_{\bf k}^n}&=h(-{\bf k})\sum_m \ket{u_{-{\bf k}}^m} V_{\bf k}^{mn}\nonumber\\
&=\sum_m \epsilon_m(-{\bf k}) \ket{u_{-{\bf k}}^m} V_{\bf k}^{mn},
\label{eq:TRSaux2}
\end{align}
Comparing \eqref{eq:TRSaux1} and \eqref{eq:TRSaux2}, it follows that
\begin{align}
\sum_m \ket{u_{-{\bf k}}^m} V_{\bf k}^{mn} (\epsilon_n({\bf k})-\epsilon_m(-{\bf k}))=0
\end{align}
for every $n$. Furthermore, since the eigenstates form an orthonormal basis, the expression above implies that
\begin{align}
V_{\bf k}^{mn} (\epsilon_n({\bf k})-\epsilon_m(-{\bf k}))=0
\end{align}
for every $m$ and $n$. This means that the sewing matrix $V_{\bf k}^{mn}$ only connects states at ${\bf k}$ and $-{\bf k}$ having the same energy. At the time-reversal invariant points (TRIP) of the Brillouin zone $\bf k^\star=\Gamma, X, Y$ and ${\bf M}$, the sewing matrix $V^{mn}_{\bf k^\star}$ is block-diagonal in energy-degenerate states, and has values 
\begin{align}
V^{mn}_{\bf k^\star} = \braket{u^m_{\bf k^\star}}{u^{n*}_{\bf k^\star}}.
\end{align}

\subsection{Rotation symmetry}
We proceed in a similar way as for time reversal symmetry. Rotation symmetry is expressed as
\begin{align}
\hat{r} h({\bf k}) \hat{r}^\dagger = h(R{\bf k})
\end{align}
Here, $\hat{r}$ is the n-fold rotation operator, which obeys $\hat{r}^n=1$. Acting on an energy eigenstate, we have
\begin{align}
h(R{\bf k}) \hat{r} \ket{u_{\bf k}^n}=\hat{r} h({\bf k})  \ket{u_{\bf k}^n}= \epsilon_n({\bf k}) \hat{r}\ket{u_{\bf k}^n}.
\end{align}
Thus, $\hat{r} \ket{u_{\bf k}^n}$ is an eigenstate of $h(R{\bf k})$ with energy $\epsilon_n({\bf k})$. We can then write the expansion
\begin{align}
\hat{r} \ket{u_{\bf k}^n} &=\sum_m \ket{u_{-{\bf k}}} B_{\bf k}^{mn},
\end{align}
where
\begin{align}
B^{mn}_{\bf k}&=\bra{u_{R{\bf k}}^m} \hat{r} \ket{u_{\bf k}^n}
\end{align}
is the rotation sewing matrix. An analysis analogous as that in section \ref{sec:ClassificationTRS} leads to the following expression
\begin{align}
B_{\bf k}^{mn} (\epsilon_n({\bf k})-\epsilon_m(R{\bf k}))=0
\end{align}
for every $m$ and $n$. This means that the sewing matrix $B_{\bf k}^{mn}$ only connects states at ${\bf k}$ and $R{\bf k}$ having the same energy. 

\subsection{Invariant points under rotation}
\label{sec:ConstraintsRotationSymmetry}
Now we consider points ${\bf k}={\bf \Pi}$ such that
\begin{align}
R {\bf \Pi} = {\bf \Pi}
\end{align}
in the BZ. In $C_4$-symmetric TCIs there are two 2-fold HSPs: ${\bf X}$ and ${\bf X'}$, and two 4-fold HSPs: ${\bf M}$ and ${\bf \Gamma}$.  In $C_2$-symmetric TCIs there are four 2-fold HSPs: ${\bf X}$, ${\bf Y}$, ${\bf M}$ and ${\bf \Gamma}$. In $C_6$ symmetric TCIs there are three 2-fold HSPs: ${\bf M}$, ${\bf M'}$, and ${\bf M''}$, two 3-fold HSPs: ${\bf K}$ and ${\bf K'}$, and one 6-fold HSP: ${\bf \Gamma}$. Finally, in $C_3$ symmetric TCIs there are only three 3-fold HSPs: ${\bf K}$, ${\bf K'}$, and ${\bf \Gamma}$. These points are shown in Fig~\ref{fig:BZInvariantPoints} for all the crystalline symmetries. At these points we have $\hat{r} h({\bf \Pi}) \hat{r}^\dagger = h({\bf \Pi})$, or
\begin{align}
[\hat{r},h({\bf \Pi})]=0.
\end{align}
Thus, it is possible to choose a basis in which the energy eigenstates are also eigenstates of the rotation operator,
\begin{align}
\hat{r} \ket{u_{\bf \Pi}^n}=r_{\bf \Pi}^n \ket{u_{\bf \Pi}^n},
\end{align}
These eigenvalues take the form specified in Eq.~\eqref{eq:RotationEigenvaluesapp} and allow for the construction of the invariants in Eq.~\eqref{eq:RotationInvariants}.
 
Now, we show that the rotation eigenvalues of HSPs that are related by symmetry are equal. Consider the rotation by an angle $\phi$ in a crystal with $C_{2\pi/\phi}$ symmetry. This rotation symmetry relates HSPs that are invariant under rotations by a larger angle $\theta=n \phi$, for $n$ integer. Call these HSPs ${\bf \Pi}_\theta$. Here, we are interested in knowing how the rotation eigenvalues of ${\bf \Pi}_\theta$ and $R_\phi {\bf \Pi}_\theta$ are related. In particular, this applies to two cases: (1) In $C_6$ symmetric crystals, $\phi=2\pi/6$. For $\theta_1=2\pi/3=2\phi$ we have ${\bf K}=R_\phi {\bf K'}$, while for $\theta_2=\pi=3\phi$ we have ${\bf M'}=R_\phi {\bf M}=R_\phi^2 {\bf M''}$; (2) in $C_4$-symmetric crystals, $\phi=\pi/2$, for $\theta=\pi=2\phi$ we have ${\bf X'}=R_\phi {\bf X}$.
Let us start by asking what we get from operating $\hat{r}_\theta \ket{u_{{R_\phi}{\bf \Pi_\theta}}^n}$. Since ${R_\phi}{\bf \Pi_\theta}$ is invariant under $\hat{r}_\theta$, we have
\begin{align}
\hat{r}_\theta \ket{u_{{R_\phi}{\bf \Pi_\theta}}^n}=r_{R_\theta {\bf \Pi_\theta}}^n \ket{u_{{R_\phi}{\bf \Pi_\theta}}^n}
\label{eq:app_rotation_eigenvalues}
\end{align}
in an obvious notation.
Now, since ${R_\phi}{\bf \Pi_\theta}$ and ${\bf \Pi_\theta}$ are related by $C_{2\pi/\phi}$ symmetry, we can expand
\begin{align}
\hat{r}_\phi \ket{u_{{\bf \Pi}_\theta}^n}=\sum_m \ket{u_{R_\phi {\bf \Pi}_\theta}^m} B_{{\bf \Pi}_\theta}^{mn},
\label{eq:app_expansion}
\end{align}
where $B_{{\bf \Pi}_\theta}^{mn}=\bra{u_{R_\phi {\bf \Pi}_\theta}^m}\hat{r}_\phi \ket{u_{{\bf \Pi}_\theta}^n}$ is the sewing matrix for $\hat{r}_\phi$. Following an analysis similar to that of the previous two sections, we arrive at the expression
\begin{align}
(r_{R_\phi {\bf \Pi}_\theta}^n-r_{{\bf \Pi}_\theta}^m) \left(B_{{\bf \Pi}_\theta}^{\dagger}\right)^{mn}=0
\end{align}
for all $m$ and $n$. Now, the sewing matrix will have non-zero elements for equal energies at the two different points in the BZ $R_\phi {\bf \Pi}_\theta$ and ${\bf \Pi}_\theta$. Thus, for $\epsilon_m(R_\phi {\bf \Pi}_\theta)=\epsilon_n({\bf \Pi}_\theta)$, we need $r^m_{{\bf \Pi}_\theta}=r^n_{R_\phi {\bf \Pi}_\theta}$, i.e., the rotation spectra at $R_\phi {\bf \Pi}_\theta$ and ${\bf \Pi}_\theta$ are equal.
In particular we have the relations
\begin{align}
\{r^n_{\bf X}\}& \stackrel{C_4}{=} \{r^n_{\bf X'}\}\nonumber\\
\{r^n_{\bf K}\}&\stackrel{C_6}{=}\{r^n_{\bf K'}\}\nonumber\\
\{r^n_{\bf M}\}& \stackrel{C_6}{=} \{r^n_{\bf M'}\} \stackrel{C_6}{=} \{r^n_{\bf M''}\}.
\end{align}
This implies that the rotation invariants defined in Eq.~\eqref{eq:RotationInvariants} obey
\begin{align}
[X_p]& \stackrel{C_4}{=} [X'_p]\nonumber\\
[K_p]&\stackrel{C_6}{=}[K'_p]\nonumber\\
[M_p]& \stackrel{C_6}{=} [M'_p] \stackrel{C_6}{=} [M''_p]
\label{eq:ConstraintRotation}
\end{align}
for all allowed values of $p$.

\subsection{Constraints on the rotation eigenvalues due to time-reversal symmetry}
\label{sec:ConstraintsTRS}
Finally, we look at the interplay between TRS and rotation symmetry. The two operators commute
\begin{align}
[\Theta, \hat{r}]=0.
\end{align}
Thus, on one hand we have
\begin{align}
\Theta \left( \hat{r} \ket{u_{{\bf k}}^l} \right) &= \Theta \left( \sum_n \ket{u_{R {\bf k}}^n} B_{\bf k}^{nl} \right)\nonumber\\
&= \sum_{m,n}\ket{u_{-R {\bf k}}^m} V_{R {\bf k}}^{mn} B_{\bf k}^{nl*}. 
\end{align}
On the other we have
\begin{align}
\hat{r} \left( \Theta \ket{u_{{\bf k}}^l} \right) &= \hat{r} \left( \sum_m \ket{u_{- {\bf k}}^n} V_{\bf k}^{nl} \right)\nonumber\\
&= \sum_{m,n} \ket{u_{-R {\bf k}}^m} B_{- {\bf k}}^{mn} V_{\bf k}^{nl}. 
\end{align}
In the last expression we have used the fact that $R (- {\bf k})= - R {\bf k}$. From these two expresions we conclude that
\begin{align}
\sum_{n} \left( V_{R {\bf k}}^{mn} B_{\bf k}^{nl*}  - B_{- {\bf k}}^{mn} V_{\bf k}^{nl} \right) = 0
\label{eq:VB-BV}
\end{align}
for all $m$, $l$. As noted earlier, of particular interest are the HSPs. At these points, $B_{{\bf \Pi}}^{mn}=r^n_{{\bf \Pi}}\delta_{mn}$ in the gauge in which $\{ \ket{u^n_{{\bf \Pi}}}\}$ are rotation eigenstates. Then, at these points, Eq.~\ref{eq:VB-BV} results in
\begin{align}
V_{{\bf \Pi}}^{ml} \left( r^{l*}_{{\bf \Pi}} - r^m_{-{\bf \Pi}} \right) = 0
\end{align}
for all $l$, $m$. Thus, if $V_{{\bf \Pi}}^{ml} \neq 0$, $r^{l*}_{{\bf \Pi}} = r^m_{-{\bf \Pi}}$. This is possible only if $\epsilon_m(-{\bf \Pi})=\epsilon_l({\bf \Pi})$. Thus, we have that, under time-reversal symmetry,
\begin{align}
\{r^n_{\bf \Pi}\}& \stackrel{TRS}{=} \{r^{n*}_{-{\bf \Pi}}\}.
\end{align}
More specifically, for equal energies at ${\bf k}={\bf \Pi}$ and ${\bf k}=-{\bf \Pi}$, their rotation eigenvalues are complex conjugates of each other. If, on the other hand, $\epsilon_m(-{\bf \Pi}) \neq \epsilon_l({\bf \Pi})$, we have that $V_{{\bf \Pi}}^{ml}=0$, which means that there is no restriction on the rotation eigenvalues. In particular, at time-reversal invariant points (TRIP) which are also HSPs, ${\bf \Pi}=-{\bf \Pi}$, we have that $r^{l*}_{{\bf \Pi}}=r^m_{{\bf \Pi}}$ for equal energies $\epsilon_m({\bf \Pi})= \epsilon_l({\bf \Pi})$. This imposes the following constraints on the rotation eigenvalues: (1) for a non-degenerate state labeled by $n$, $r^{n*}_{{\bf \Pi}}=r^{n}_{{\bf \Pi}}$, i.e., its rotation eigenvalue is real: $r^{n}_{{\bf \Pi}}=\pm 1$ and (2) for two degenerate states $n=1,2$ one could have $r^1_{{\bf \Pi}}=\lambda$ and $r^2_{{\bf \Pi}}=\lambda^*$, so that $r^{1*}_{{\bf \Pi}}=\lambda^*=r^{2}_{{\bf \Pi}}$ and $r^{2*}_{{\bf \Pi}}=\lambda=r^{1}_{{\bf \Pi}}$, that is, in energy-degenerate states, the rotation eigenvalues can be complex, but have to come in complex conjugate pairs. As said before, these constraints follow for HSPs that are also TRIP. This is the case for all the HSPs except  ${\bf K}$ and ${\bf K'}$, which map into each other under time-reversal. 

This implies that the rotation invariants defined in Eq.~\eqref{eq:RotationInvariants} obey
\begin{align}
[M_2^{(4)}]& \stackrel{C_4}{=} [M_4^{(4)}]\nonumber\\
[K^{(3)}_1]&\stackrel{C_6}{=}[K^{(3)}_2]\nonumber\\
[K^{(3)}_2]& \stackrel{C_3}{=} [K'_3]\nonumber\\
[K_3]& \stackrel{C_3}{=} [K'_2].
\label{eq:ConstraintTRS}
\end{align}

\subsection{Complete set of topological invariants}

Due to rotation and time-reversal symmetries, the rotation invariants in Eq.~\eqref{eq:RotationInvariants} must obey the relations in Eqs.~\ref{eq:ConstraintRotation} and \ref{eq:ConstraintTRS}. Additionally, due to the fact that the number of occupied bands is constant across the BZ, we have the constraint
\begin{align}
\sum_p [\Pi^{(n)}_p]=0.
\end{align}
Applying these three sets of constraints, the resulting topological classes of crystals with rotation symmetry $C_n$ are given by the indices $\chi^{(n)}$, as follows,
\begin{align}
\chi^{(4)}&= \left( [X^{(2)}_1],[M^{(4)}_1],[M^{(4)}_2]\right)\nonumber\\
\chi^{(2)}&= \left( [X^{(2)}_1],[Y^{(2)}_1],[M^{(2)}_1]\right)\nonumber\\
\chi^{(6)}&=\left( [M^{(2)}_1],[K^{(3)}_1] \right)\nonumber\\
\chi^{(3)}&=\left( [K^{(3)}_1], [K^{(3)}_2] \right).
\label{eq:ClassificationIndicesapp}
\end{align}

\section{Polarization and rotation symmetry}
\label{sec:PolarizationAndRotationSymmetry}
In this section, we review the quantization of polarization due to $C_n$ symmetry. We will then derive expressions of polarization in terms of the rotation invariants defined in Eq.~\eqref{eq:ClassificationIndicesapp}. A general discussion of the relation between point group symmetry and polarization can be found in Refs.~\onlinecite{fang2012b,benalcazar2017quadPRB}. We follow closely the discussion in Ref.~\onlinecite{fang2012b,benalcazar2017quadPRB}.
\subsection{Quantization of polarization}
\begin{figure}[t!]
\centering
\includegraphics[width=\columnwidth]{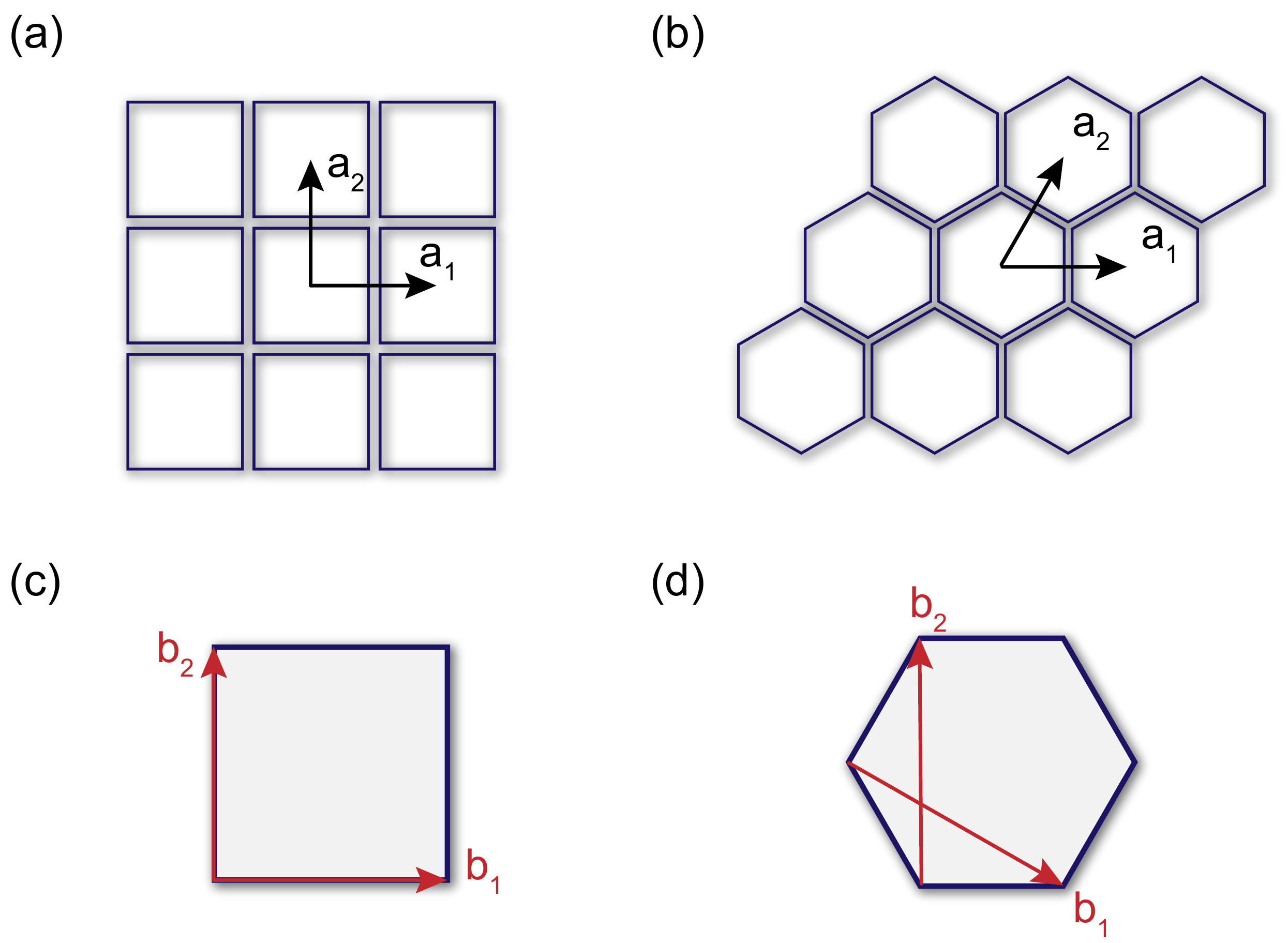}
\caption{Schematically showing our choice of lattice vectors $\bd{a}_1,\bd{a}_2$ for (a) $C_4, C_2$ TCIs and (b) $C_3,C_6$ TCIs. (c) The Brillouin zone (BZ) and reciprocal lattice vectors for $C_4,C_2$ symmetric crystals, $\bd{b}_1=2\pi(1,0),\,\,\bd{b}_2=2\pi(0,1)$. (d) BZ and reciprocal lattice vectors for $C_3,C_6$ symmetric crystals, $\bd{b}_1=2\pi(1,\frac{-1}{\sqrt{3}}),\,\,\bd{b}_2=2\pi(0,\frac{2}{\sqrt{3}})$. }
\label{lattic}
\end{figure}

We denote the lattice vectors in real space as $\bd{a}_1,\bd{a}_2$ and the corresponding reciprocal lattice vectors in the $\kk$ space as $\bd{b}_1,\bd{b}_2$. The reciprocal lattice vectors satisfy
\begin{align}
\bd{a}_i\cdot\bd{b}_j=2\pi\delta_{ij}.
\label{Eq:bvector}
\end{align}
Without loss of generality, we choose our lattice vectors and reciprocal lattice vectors for each symmetry to be those shown in Fig.~\ref{lattic}.
The conventional modern definition of polarization per unit cell in 2D crystals is~\cite{king-smith1993}
\begin{align}
\bd{P}&=-\frac{e}{S}\int_{\text{BZ}}\tr\bd{\mathcal{A}(\kk)}d^2\kk,
\label{Eq:defineP}
\end{align}
where $S$ is the area of the BZ, and $\mathcal{A}$ is the Berry connection, which has components $\bd{\mathcal{A}}^{\alpha\beta}(\kk)=-i\bra{u^\alpha(\kk)}\nabla_{\kk}\ket{ u^\beta(\kk)}$ defined at each $\kk$ point in the BZ. We parameterize the BZ as $\kk=s_1\bd{b}_1+s_2\bd{b}_2$ so that the integral in Eq.~\eqref{Eq:defineP} is
\begin{align}
\bd{P}
=-e\int_0^1ds_1\int_0^1ds_2~ \tr \bd{\mathcal{A}}(s_1\bd{b}_1+s_2\bd{b}_2),
\end{align}
where the determinant of the Jacobian matrix that transforms the variables of integration from $dk_x dk_y$ to $ds_1ds_2$ cancels the area of the BZ.
We define the quantity
\begin{align}
\mu_i\equiv
-\frac{e}{2\pi}\int_0^1 ds_1\int_0^1 ds_2\tr[\bd{\mathcal{A}}(s_2\bd{b}_2+s_1\bd{b}_1 )]\cdot\bd{b}_i 
\label{eq:definemu}
\end{align}
so that the projection of polarization along the reciprocal lattice vector $\bd{b}_i$ is
\begin{align}
\bd{P}\cdot \bd{b}_i=2\pi\mu_i\nonumber.
\end{align}
In real space, we can express the polarization in terms of lattice vectors, $\bd{P}=(p_1\bd{a}_1+p_2\bd{a}_2)$ modulo integer linear combinations of lattice vectors. Following Eq.~\eqref{Eq:bvector}, the projection of $\bd{P}$ along the reciprocal lattice vector is
\begin{align}
\bd{P}\cdot \bd{b}_i=(p_1\bd{a}_1\cdot\bd{b}_i+p_2\bd{a}_2\cdot\bd{b}_i)=2\pi p_i,
\end{align}
therefore,
\begin{align}
p_1=\mu_1\mod e,\quad p_2=\mu_2 \mod e.
\end{align}
Now we analyze the role of rotation symmetries. Under a rotation operation $\hat{r}_n$, the lattice vectors transform as $\bd{a}'_i=T_n^{ij}\bd{a}_j$ (in the following, we will assume the summation over repeated indices). The polarization becomes
\begin{align}
\bd{P}=p_i\bd{a}_i\rightarrow p_iT_n^{ij}\bd{a}_j.
\label{eq:transofp1}
\end{align}
If the model is $C_n$-symmetric, the change in polarization after a $C_n$ rotation can only be multiples of lattice vectors
\begin{align}
\bd{P}=p_i\bd{a}_i\rightarrow(p_i+n_i)\bd{a}_i,
\label{eq:transofp2}
\end{align}
where $n_i\in\mathbb{Z}e,\,i=1,2 $. Comparing Eq.~\eqref{eq:transofp1} and Eq.~\eqref{eq:transofp2}, we find the constraints on the polarization due to rotation symmetry:
\begin{align}
 p_jT_n^{ji}=(p_i+n_i).
\label{eq:transofp}
\end{align}
Without loss of generality, we choose the lattice vectors for $C_2$ and $C_4$-symmetric TCIs to be $\bd{a}_1=(1,0),\bd{a}_2=(0,1)$ [see Fig.~\ref{lattic}(a)], and for $C_3$ and $C_6$ symmetric TCIs to be $\bd{a}_1=(1,0),\bd{a}_2=(\frac{1}{2},\frac{\sqrt{3}}{2})$ [see Fig.~\ref{lattic}(b)].
We summarize the transformation matrix $\bd{T}_n$ for our choice of lattice vectors $\bd{a}_1,\bd{a}_2$ under the $C_n$ rotations in Table.~\ref{tab:transmat}. 
Plugging the matrices $T_n^{ji}$ in Eq.~\eqref{eq:transofp}, we can solve the polarization components $p_1,p_2$,
\begin{align}
&p_1=-\frac{n_1}{2},\,\,p_2=-\frac{n_2}{2}&  \text{ for $C_2$ symm.}\nonumber\\
&p_1=\frac{n_2-n_1}{2},\,\,p_2=-\frac{n_1+n_2}{2}& \text{ for $C_4$ symm.}\nonumber\\
&p_1=\frac{n_2-n_1}{3},\,\,p_2=-\frac{2n_2+n_1}{3} & \text{ for $C_3$ symm.}\nonumber\\
&p_1=-n_2-n_1,\,\,p_2=n_2.& \text{ for $C_6$ symm.}
\label{eq:realspacepolar}
\end{align}
\begin{table}[t]
\begin{tabular}{c c c c c}
\hline\hline
symm. & $C_2$ & $ C_4$ & $C_3$ & $C_6$\\
\hline
$\bd{T}_n$ & 
$\left(\begin{array}{cc}-1& 0\\ 0 & -1\end{array}\right)$ &
$\left(\begin{array}{cc} 0& 1\\ -1 & 0\end{array}\right)$ &
$\left(\begin{array}{cc}-1& 1\\ -1 & 0\end{array}\right)$ &
$\left(\begin{array}{cc} 0& 1\\ -1 & 1\end{array}\right)$ \\
\hline
$\bd{T'}_n$ & 
$\left(\begin{array}{cc}-1& 0\\ 0 & -1\end{array}\right)$ &
$\left(\begin{array}{cc} 0& 1\\ -1 & 0\end{array}\right)$ &
$\left(\begin{array}{cc}0& 1\\ -1 & -1\end{array}\right)$ &
$\left(\begin{array}{cc} 0& 1\\ 1 & 0\end{array}\right)$ \\
\hline\hline
\end{tabular}
\caption{The transformation matrix $\bd{T}_n$ for lattice vectors $\bd{a}_1,\bd{a}_2$ and the transformation matrix $\bd{T}'_n$ for reciprocal lattice vectors $\bd{b}_1,\bd{b}_2$ under $C_n$ rotations.}
\label{tab:transmat}
\end{table}
Since $p_1,p_2$ are defined modulo $e$, the constraints from above the equations imply that, with $C_2$ or $C_4$ symmetries, the polarization components $p_1,p_2$ are quantized to be $0$ or $\frac{e}{2},$ while with $C_3$ symmetry, $p_1,p_2$ are quantized to be $0,\frac{e}{3},\frac{2e}{3}$. With $C_6$ symmetry, the polarization components are always $0$ (mod $e$). Furthermore, with $C_4, C_3$ symmetry, the difference of the two polarization components $p_1-p_2$ is a multiple of the integer charge $n_2$. Therefore, the two polarization components are the same, 
\begin{align}
p_1=p_2  \mod  e,\quad\text{ for } C_4, C_3 \text{ symm.}
\label{eq:p1equalp2}
\end{align}

\begin{figure}[t]
\centering
\includegraphics[width=\columnwidth]{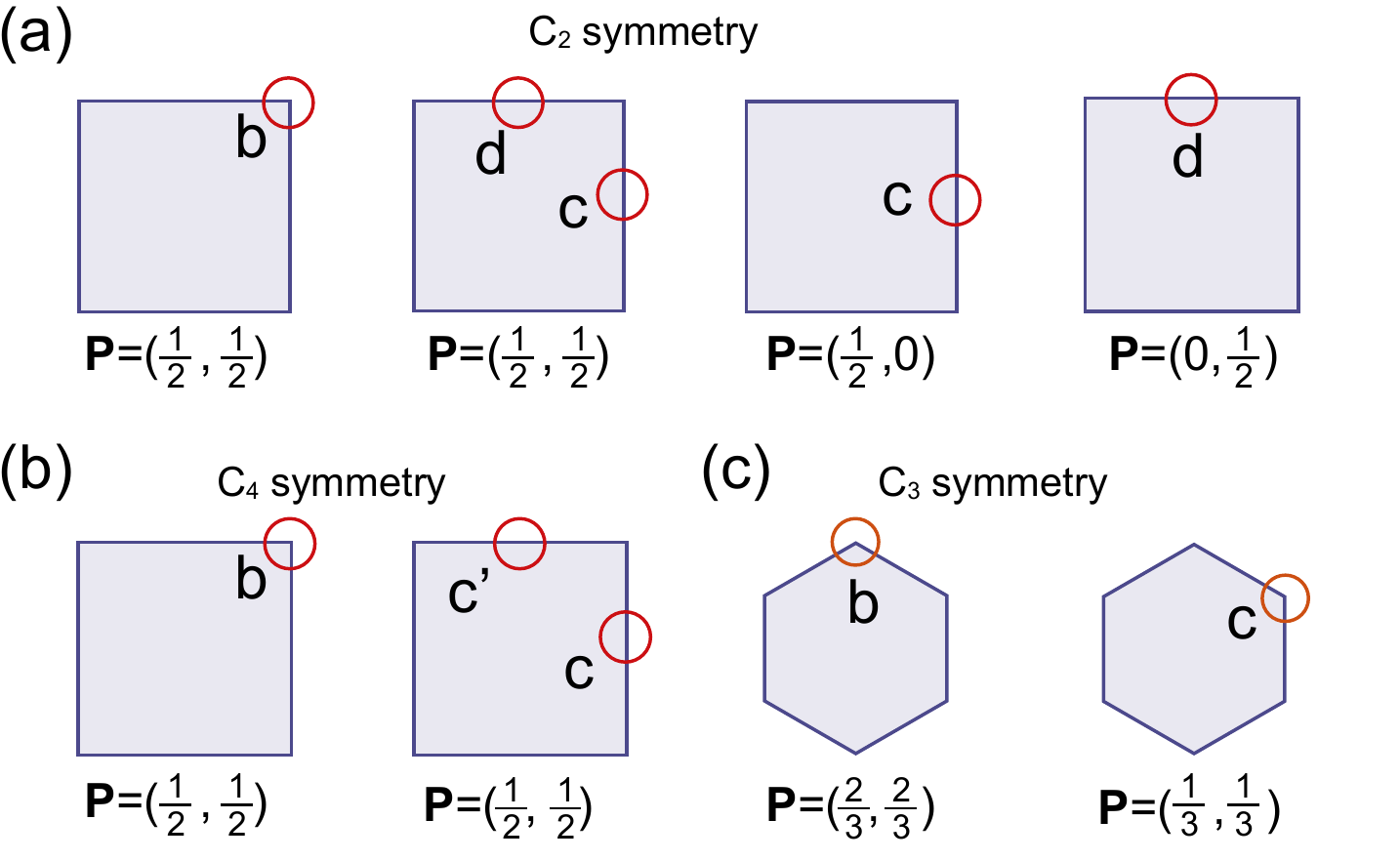}
\caption{The configurations of the center of negative charges per unit cell with nontrivial polarization. The red circles represent the center of negative charges. The blue squares represent unit cells with $C_2, C_4$ symmetry and blue hexagons represent unit cells with $C_3$ symmetry. The English letters $a,b,c,d$ inside the unit cell indicate the maximal Wyckoff positions.}
\label{Wyckoff}
\end{figure}

The quantization of the polarization means that with nontrivial polarization, the center of negative charges coincides with maximal Wyckoff positions in each unit cell, as shown in Fig.~\ref{Wyckoff}.
In $C_4$-symmetric TCIs, the only allowed non-trivial polarization is $(\frac{e}{2},\frac{e}{2})$, which corresponds to either one center of negative charge located at the maximal Wyckoff position $b,$ or two centers located at the maximal Wyckoff positions $c$ and $c'$ respectively.
In $C_2$-symmetric TCIs, due to the absence of the constraint $p_1=p_2$, the possible nontrivial polarizations can be $(\frac{e}{2},\frac{e}{2}),(\frac{e}{2},0),(0,\frac{e}{2})$. The first possibility is $C_4$-symmetric and corresponds to the same Wyckoff positions as we discussed above. The latter two possibilities break $C_4$ symmetry and correspond to the center of negative charges located at the maximal Wyckoff positions $c$ or $d$ respectively. 
Finally, in $C_3$ symmetric TCIs, the only possible non-trivial polarizations are $(\frac{e}{3},\frac{e}{3})$ and $(\frac{2e}{3},\frac{2e}{3})$, with the center of negative charges located at the maximal Wyckoff position $b$ or $c$ respectively.
\subsection{Polarization as a topological index}
\label{subsec:PolarZnIndex}
\begin{figure}[t]
\centering
\includegraphics[width=\columnwidth]{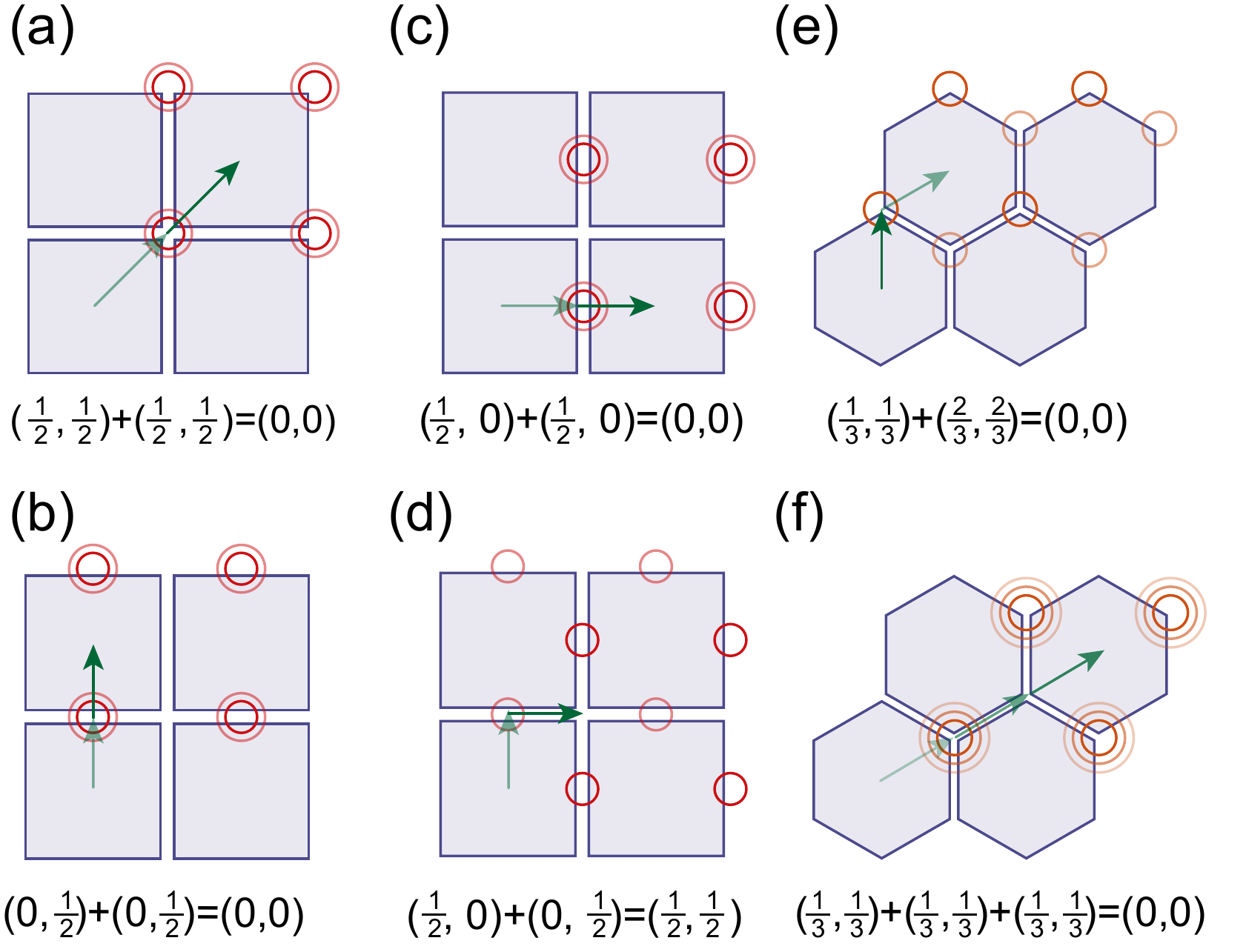}
\caption{Stacking of models with non-trivial polarization. One red circle represents the center of negative charges and different opacity indicates different models. The equations below each plot denote the addition of polarization (in the units of electronic charge $e$).}
\label{polar_quant}
\end{figure}

In this section, we will illustrate that the polarization behaves as a topological index. Stacking two models with polarization $\bd{P}_1,\bd{P}_2$ results in an TCI with polarization $\bd{P}=\bd{P}_1+\bd{P}_2$.
In Fig.~\ref{polar_quant} (a), we stack two $C_4$-symmetric TCIs with polarization $(\frac{e}{2},\frac{e}{2})$ (represented by the light and dimmed red circles) together. The green arrows with different opacity indicate the polarizations coming from different models. It is clear to see that the overall polarization has $p_1=p_2=e$ and hence  is trivial. 
Thus, the polarization $\bd{P}$ forms a $\mathbb{Z}_2$ index for $C_4$-symmetric TCIs. In Fig.~\ref{polar_quant} (b)[(c)], we stack two $C_2$-symmetric TCIs with polarization $(\frac{e}{2},0)$ [$(0,\frac{e}{2})$] together, each of which has only one non-trivial polarization component. By stacking two such models, the overall polarization is also trivial.  In Fig.~\ref{polar_quant} (d), we show the stacking of two $C_2$-symmetric TCIs with polarizations $(\frac{e}{2},0),(0,\frac{e}{2})$. The resulting model has non-trivial polarization component in both directions. From plots (b),(c),(d), we see that each of the two polarization components $p_1,p_2$ forms a $\mathbb{Z}_2$ index independently. Therefore, the polarization $\bd{P}$ forms a $\mathbb{Z}_2\times\mathbb{Z}_2$ index for $C_2$-symmetric TCIs. In Fig.~\ref{polar_quant} (e), we show stacking of a $C_3$ symmetric TCIs with polarization $(\frac{e}{3},\frac{e}{3})$ and a $C_3$ symmetric TCIs with polarization $(\frac{2e}{3},\frac{2e}{3})$. The overall polarization becomes trivial. In Fig.~\ref{polar_quant} (f), we stack three $C_3$ symmetric TCIs with polarization $(\frac{e}{3},\frac{e}{3})$ together and the overall polarization also becomes trivial. Therefore, the polarization $\bd{P}$ for $C_3$ symmetric TCIs forms a $\mathbb{Z}_3$ index.
\subsection{Relation between polarization and rotational topological invariants}
\begin{figure}
\centering
\includegraphics[width=\columnwidth]{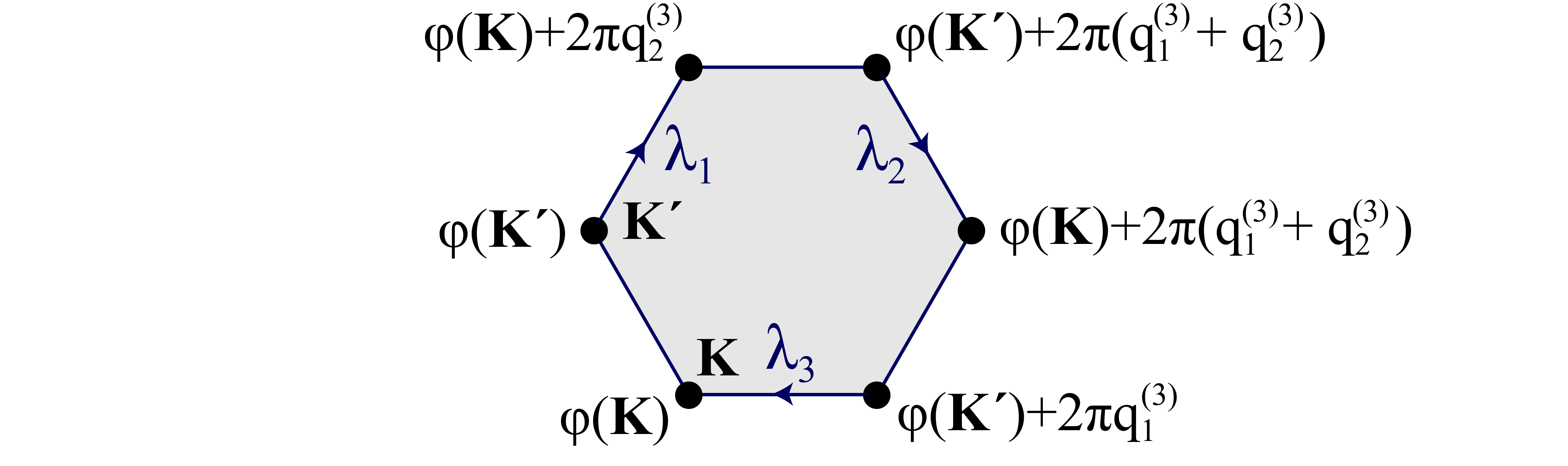}
\caption{Brillouin zone of a $C_3$ symmetric TCIs and the phases of the determinant of the sewing matrix $B_{\hat{r}_3,{\bf k}}$ at HSPs. Arrows indicate the integral paths in Eq.\eqref{eq:integral_sewing}.}
\label{sewing_phase}
\end{figure}

We now show that the quantized polarizations can be written in terms of topological rotation invariants defined at the HSPs $\bd{\Pi}^{(n)}$. 
Consider a Bloch Hamiltonian $h(\kk)$ satisfying the $n-$fold rotation symmetry
\begin{align}
\hat{r}_n h(\kk) \hat{r}_n^{-1}=h(R_n\kk),
\end{align}
where $\hat{r}_n$ and $R_n$ are the $n-$fold rotation operations that act on a Hamiltonian and a vector in ${\bf k}$ space, respectively. This rotation symmetry allows us to relate a rotated eigenstate of $h(\kk)$, $\hat{r}_n\ket{u^n_{\kk}}$ to an eigenstate at $R_n\kk$ with the same energy,
\begin{align}
\hat{r}_nh(\kk)\ket{u_{\kk}^n}=h(R_n\kk)\hat{r}_n\ket{u_{\kk}^n}=\epsilon_{n,\kk}\hat{r}_n\ket{u_{\kk}^n}.
\end{align}
Therefore, we can expand the state $\hat{r}_n\ket{u_{\kk}^n}$ in terms of the eigenstates $\ket{u^m_{R_n\kk}}$ at $R_n\kk$,
\begin{align}
\hat{r}_n\ket{u_{\kk}^n}&=\ket{u_{R_n\kk}^m}\bra{u_{R_n\kk}^m}\hat{r}_n\ket{u_{\kk}^n}
\nonumber\\
&\equiv\ket{u_{R_n\kk}^m}B_{\hat{r}_n,\kk}^{mn},
\label{eq:sewingmatrix}
\end{align}
where the summation over the repeated indices $m$ is over the occupied subspace (we will assume the same summation rule for repeated band indices unless otherwise specified).
In Eq.~\eqref{eq:sewingmatrix}, we defined the sewing matrix,
\begin{align}
B_{\hat{r}_n,\kk}^{mn}\equiv\bra{u_{R_n\kk}^m}\hat{r}_n\ket{u_{\kk}^n},
\end{align} 
which connects the eigenstates at $\kk$ with those at $R_n\kk$. 
Notice that the sewing matrix is periodic in $\kk$ space
\begin{align}
B_{\hat{r}_n,\kk}=B_{\hat{r}_n,\kk+\bd{G}},
\label{eq:sewingmatrixperiodicity}
\end{align}
where $\bd{G}$ is a reciprocal lattice vector. Recalling Eq.~\eqref{eq:definemu}, by the chain rule we have
\begin{align}
&i\bd{\mathcal{A}}(s_2\bd{b}_2+s_1\bd{b}_1)\cdot\bd{b}_i\nonumber\\
=&\bra{u(\kk(s_1,s_2))}\nabla_{\kk}\ket{u(\kk(s_1,s_2))}\cdot\bd{b}_i\nonumber\\
=&\bra{u(s_1\bd{b}_1+s_2\bd{b}_2)}\frac{\pa}{\pa s_i}\ket{u(s_1\bd{b}_1+s_2\bd{b}_2)}.
\end{align}
Therefore, the polarization component $p_i$ can be computed as,
\begin{align}
p_i=\mu_i=\frac{ie}{2\pi}\int_0^1ds_1\int_0^1ds_2\tr[\bra{u_{\bd{S}}}\frac{\pa}{\pa s_i}\ket{u_{\bd{S}}}].
\end{align}
For simplicity in notation, we used $\bd{S}=(s_1,s_2)$ to indicate a point $\kk= s_1\bd{b}_1+s_2\bd{b}_2$ in the BZ. Under a $C_n$ rotation, the reciprocal lattice vectors transform as $\bd{b}'_i=[T'_n]^{ij}\bd{b}_j$, where $[T_n']^{ij}$ is a matrix element in the transformation matrix of the reciprocal lattice basis. Note that, in general, the transformation matrix $\bd{T}_n'$ for reciprocal lattice vectors is different from the transformation matrix $\bd{T}_n$ for the lattice vectors and depends on the choice of lattice vectors. We summarize $\bd{T}_n'$ under each $C_n$ rotation for our choice of lattice vectors in Table~\ref{tab:transmat}. Therefore, a $\kk$ point transforms as $R_n\kk=s_iR_n\bd{b}_i=s_i [T_n']^{ij}\bd{b}_j$ under rotation of $C_n$. The new coefficient after the rotation is $s_j'=s_i[T_n']^{ij}$.
Using Eq.~\eqref{eq:sewingmatrix}, and inserting an identity operator $\hat{r}_n^\dagger \hat{r}_n$ inside the bracket, one can rewrite the expression of $p_i$,
\begin{align}
p_i=&\frac{ie}{2\pi }\int_0^1 ds_2\int_0^1 ds_1 \bra{u^m_{\bd{S'}}}B^{\dagger \alpha m}_{\hat{r}_n,\bd{S}}\frac{\pa}{\pa s_i}B^{n\alpha}_{\hat{r}_n,\bd{S}}\ket{u^n_{\bd{S'}}}
\nonumber\\
=&\frac{ie}{2\pi }\int_0^1 ds_2\int_0^1 ds_1\tr\left[\bra{u_{\bd{S'}}}\frac{\pa}{\pa s_i}\ket{u_{\bd{S'}}}\right.\nonumber\\
&+\left.B^\dagger_{\hat{r}_n,\bd{S}}\frac{\pa}{\pa s_i}B_{\hat{r}_n,\bd{S}}\right].
\label{eq:rewritep}
\end{align}
Using the chain rule, the first term can be further calculated,
\begin{align}
&\bra{u_{\bd{S'}}}\pa_{s_i}\ket{u_{{\bd{S'}}}}
=\bra{u_{\bd{S'}}}\pa_{s'_j}\ket{u_{{\bd{S'}}}}\frac{\pa s'_j}{\pa s_i}\nonumber\\
=&\bra{u(\kk)}\pa_{\kk}\ket{u(\kk)}\cdot \bd{b}_j[T_n']^{ij}
=i\bd{\mathcal{A}}(\kk)\cdot\bd{b}_j[T_n']^{ij}.
\end{align}
Compared with Eq.~\eqref{eq:definemu}, after the integral, the contribution from the first term in Eq.~\eqref{eq:rewritep} can be summarized as $p_j[T_n']^{ij}$. Using the identity
\begin{align}
\tr\left[B^\dagger_{\hat{r}_n,\bd{S}}\frac{\pa}{\pa s_i}B_{\hat{r}_n,\bd{S}}\right]=\frac{\pa}{\pa{s_i}}\ln\det \left[B_{\hat{r}_n,\bd{S}}\right],
\end{align}
the integral of the second term over $s_i$ in Eq.~\eqref{eq:rewritep} is the phase difference of $\det [B_{\hat{r}_n,\bd{S}}]$ on the BZ boundaries, which is forced to be quantized in units of $2\pi i$ due to the periodicity of the sewing matrix [see Eq.~\eqref{eq:sewingmatrixperiodicity}]. Since the Chern number is vanishing in time-reversal symmetric TCIs, we can choose a smooth gauge for eigenstates across the entire BZ. Therefore, the phase of $\det [B_{\hat{r}_n,\bd{S}}]$ is continuous across the BZ. For a quantized value, continuity means the quantity must remain constant, otherwise, the discreteness breaks the continuity. As a result, the phase difference of $\det [B_{\hat{r}_n,\bd{S}}]$ between the BZ boundaries $s_i=0,s_i=1$ is a constant along $s_j$. Therefore,
\begin{align}
\int_0^1 ds_1\int_0^1 ds_2\frac{\pa}{\pa{s_i}}\ln\det B_{\hat{r}_n,\bd{S}}=2i\pi q_i^{(n)},\quad q^{(n)}_i\in\mathbb{Z}
\label{eq:integerq}
\end{align}
where $2i\pi q^{(n)}_i$ is the phase difference of $\det [B_{\hat{r}_n,\bd{S}}]$ between $\kk=s_j\bd{b}_j$ and $\kk=s_j\bd{b}_j+\bd{b}_i$. Combining the two terms together we have
\begin{align}
p_i=p_j[T_n']^{ij}-eq^{(n)}_i.
\label{eq:transp}
\end{align}
Generally, one can solve Eq.~\eqref{eq:transp} for $p_1,p_2$ if one knows the phase difference of $\det[B_{\hat{r}_n,\kk}]$ at the BZ boundary:
\begin{align}
\bm{p}=e(\bd{T}'_n-\bm{\mathbb{I}})^{-1}\bd{q}^{(n)},
\label{eq:qtomu}
\end{align}
where the boldface $\bd{T'},\bm{p},\bd{q}$ represent the transformation matrix, and the column vectors $(p_1,p_2)^T$ and $(q_1,q_2)^T$, respectively. Now we can solve for the polarization components for each symmetry. 
For $C_2$ symmetry, the transformation matrix of the lattice vectors under a counterclockwise rotation of $180^{\circ}$ is,
\begin{align}
\bm{T}'_2=\left(
\begin{array}{cc}
-1 & 0\\
0 & -1\\
\end{array}
\right).
\end{align}
According to Eq.~\eqref{eq:qtomu}, the polarization components are,
\begin{align}
p_1=-\frac{eq^{(2)}_1}{2},\quad p_2=-\frac{eq^{(2)}_2}{2}.
\label{eq:C2qtomu}
\end{align}
We can divide the integral in Eq.~\eqref{eq:integerq} into two parts:\,
\begin{align}
2i\pi q_i^{(2)}=&\int_0^{1/2}ds_i\frac{\pa}{\pa s_i}\ln \det\left[B_{\hat{r}_2,(s_i,\frac{1}{2})}\right]\nonumber\\
&+\int_{1/2}^1ds_i\frac{\pa}{\pa s_i}\ln \det\left[B_{\hat{r}_2,(s_i,\frac{1}{2})}\right].
\label{eq:integralofq}
\end{align}
As argued before, $q_i^{(2)}$ does not depend on $s_j$. We take $s_j=1/2$  in Eq~\eqref{eq:integerq}.
The sewing matrix for $\hat{r}_2$ rotation satisfies
\begin{align}
B_{\hat{r}_2,\bd{S}}=&\bra{u_{-\bd{S}}}\hat{r}_2\ket{u_{\bd{S}}}=\bra{u_{-\bd{S}}}\hat{r}_2^\dagger\ket{u_{\bd{S}}}\nonumber\\
B_{\hat{r}_2,-\bd{S}}=&\bra{u_{\bd{S}}}\hat{r}_2\ket{u_{-\bd{S}}}=B_{\hat{r}_2,\bd{\bd{S}}}^\dagger.
\end{align}
Using this relation, the second term in Eq~\eqref{eq:integralofq} can be transformed into
\begin{align}
&\int_{1/2}^1ds_i\frac{\pa}{\pa s_i}\ln \det\left[B_{\hat{r}_2,(s_i,\frac{1}{2})}\right]\nonumber\\
=&\int_{1/2}^1ds_i\frac{\pa}{\pa s_i}\ln \det\left[B^\dagger_{\hat{r}_2,(-s_i,-\frac{1}{2})}\right]\nonumber\\
=&\int_{-1}^{-1/2}ds_i\frac{\pa}{\pa s_i}\ln \det\left[B^\dagger_{\hat{r}_2,(s_i,-\frac{1}{2})}\right]\nonumber\\
=&\int_{0}^{1/2}ds_i\frac{\pa}{\pa s_i}\ln \det\left[B_{\hat{r}_2,(s_i,\frac{1}{2})}\right].
\end{align}
From the second line to the last line, we used the periodicity of the sewing matrix: $B_{\hat{r}_2,(s_1,s_2)}=B_{\hat{r}_2,(s_1+1,s_2+1)}$. Now the integral becomes
\begin{align}
2i\pi q_1^{(2)}=&2\int_{0}^{1/2}ds_1\frac{\pa}{\pa s_1}\ln \det\left[B_{\hat{r}_2,(s_1,\frac{1}{2})}
\right]\nonumber\\
=&2\left(\ln \det\left[B_{\hat{r}_2,\bd{M}}]-\ln \det[B_{\hat{r}_2,\bd{Y}}\right]\right)\nonumber\\
2i\pi q_2^{(2)}=&2\int_{0}^{1/2}ds_2\frac{\pa}{\pa s_2}\ln \det[B_{\hat{r}_2,(\frac{1}{2},s_2)}]\nonumber\\
=&2(\ln \det\left[B_{\hat{r}_2,\bd{M}}\right]-\ln \det\left[B_{\hat{r}_2,\bd{X}}\right]).
\label{eq:qforC2}
\end{align}
The phase of $\det[B_{\hat{r}_n,\bd{\Pi}^{(n)}}]$ at a HSP $\bd{\Pi}^{(n)}$satisfies 
\begin{align}
\ln \det\left[B_{\hat{r}_n,\bd{\Pi}^{(n)}}\right]=i\sum_{p=1}^{n}\left(\#\Pi_p^{(n)}\right)\frac{2(p-1)\pi}{n}.
\label{eq:RIPSewing}
\end{align}
Using Eq.~\eqref{eq:RIPSewing}, the expressions for $q_1^{(2)}$ and $q_2^{(2)}$ in Eq.~\eqref{eq:qforC2} reduce to
\begin{align}
q_1^{(2)}=&\# M^{(2)}_2-\#Y_2=[M^{(2)}_1]+[Y^{(2)}_1] \textrm{ mod } 1\nonumber\\
q_2^{(2)}=&\# M^{(2)}_2-\#X_2=[M^{(2)}_1]+[X^{(2)}_1] \mod  1,
\end{align}
where we used the identity $[\Pi^{(2)}_1]+[\Pi^{(2)}_2]=0$. Using Eq.~\eqref{eq:C2qtomu}, we can finally express the polarization components $p_1,p_2$ of $C_2$-symmetric TCIs in terms of the rotation invariants
\begin{align}
\Aboxed{
p_1&=\frac{e}{2}\left([M_1^{(2)}]+[Y^{(2)}_1]\right) \mod e}\nonumber\\
\Aboxed{p_2&=\frac{e}{2}\left([M_1^{(2)}]+[X^{(2)}_1]\right)\mod e.}
\label{eq:PolarC2}
\end{align}
For TCIs with $C_4$ symmetry, the polarization components satisfy the same expression since $C_2$ symmetry is naturally obeyed in such TCIs. However, we can further write the expression in terms of $C_4$ rotation invariants by using the relation, 
$\#\Pi_1^{(2)}=\#\Pi_1^{(4)}+\#\Pi_3^{(4)}$,
\begin{align}
p_1=p_2=&\frac{e}{2}\left([M_1^{(4)}]+[M_3^{(4)}]+[X^{(2)}_1]\right)\nonumber\\
=&e\left(-[M_2^{(4)}]+\frac{1}{2}[X^{(2)}_1]\right)\nonumber\\
\Aboxed{p_1=p_2=&\frac{e}{2}[X^{(2)}_1] \mod e.}
\label{eq:PolarC4}
\end{align}
For TCIs with $C_3$ symmetry, the transformation matrix is:
\begin{align}
T_3'=\left(
\begin{array}{cc}
0 & 1\\
-1 & -1\\
\end{array}
\right).
\end{align}
Solving for $p_1,p_2$ we have
\begin{align}
p_1&=-\frac{e(2q^{(3)}_1+q^{(3)}_2)}{3}\nonumber\\
p_2&=\frac{e(q^{(3)}_1-q^{(3)}_2)}{3}.
\label{ptophase}
\end{align}
Let us denote the phase of $\det B_{\hat{r}_3,\kk}$ as
\begin{align}
\varphi(\kk)=-i\ln\det B_{\hat{r}_3,\kk},
\end{align}
then upon a translation by a reciprocal lattice vector, we have
\begin{align}
\varphi(\kk+\bd{b}_1)&=-i\ln\det B_{\hat{r}_3,\kk}+2\pi q^{(3)}_1\nonumber\\
\varphi(\kk+\bd{b}_2)&=-i\ln\det B_{\hat{r}_3,\kk}+2\pi q_2^{(3)}.
\end{align}
A particular choice of these phases at each HSP is shown in Fig.~\ref{sewing_phase}.  The sewing matrix for $C_3$ symmetry satisfies
\begin{align}
B_{\hat{r}_3,\kk}=&\bra{u_{R_3\kk}}(r^\dagger_3)^2\ket{u_{\kk}}\nonumber\\
=&\left<u_{R_3\kk}\left|r^\dagger_3\right|u_{R_3^2\kk}\right>\left<u_{R_3^2\kk}\left|r^\dagger_3\right|u_{\kk}\right>\nonumber\\
=&B^{\dagger}_{\hat{r}_3,R_3\kk}B^{\dagger }_{\hat{r}_3,R_3^{-1}\kk},
\end{align}
where $\ket{u_{R_3^2\kk}}\bra{u_{R_3^2\kk}}=\mathbb{I}=\hat{P}_{unocc}(R_3^2\kk)+\hat{P}_{occ}(R_3^2\kk),$ and we used the fact that, for a gapped system, $\hat{P}_{unocc}(R_3^2\kk)\hat{r}_3^\dagger\ket{u_{R_3\kk}}=0$.
Using this property, we can divide the integral of $d\ln\det B_{\hat{r}_3,\kk}$ along path $\lambda_1$ in Fig.~\ref{sewing_phase} into two parts,
\begin{align}
&\varphi(\bd{K})+2\pi q^{(3)}_2-\varphi(\bd{K}')\nonumber\\
=&\int_{\lambda_1}d\ln\det B_{\hat{r}_3,\kk}
=\int_{\lambda_1}d\ln\left(\det B^\dagger_{\hat{r}_3,R_3\kk}\det B^\dagger_{\hat{r}_3,R^{-1}_3\kk}\right)\nonumber\\
=&\left(\int_{\lambda_2}d\ln\det B^\dagger_{\hat{r}_3,\kk}+\int_{\lambda_3}d\ln\det B^\dagger_{\hat{r}_3,\kk}\right)\nonumber\\
=&-[\varphi(\bd{K})-\varphi(\bd{K}')+\varphi(\bd{K})-\varphi(\bd{K}')-2\pi q^{(3)}_1].
\label{eq:integral_sewing}
\end{align}
Further simplifying Eq.~\eqref{eq:integral_sewing}, we have
\begin{align}
2\pi(q^{(3)}_2-q^{(3)}_1)=3[\varphi(\bd{K}')-\varphi(\bd{K})].
\end{align}
Plugging into Eq.~\eqref{eq:RIPSewing}, we have
\begin{align}
\varphi(\bd{K'})&=\frac{2\pi}{3}(\#K'_2-\#K'_3)\nonumber\\
\varphi(\bd{K})&=\frac{2\pi}{3}(\#K_2-\#K_3).
\end{align}
Using the relation between eigenvalues at $\bd{K}$ and $\bd{K'}$, $\#K_2=\#K'_3,\,\,\#K_3=\#K'_2,\,\, \#\Gamma_2=\#\Gamma_3$, we find that
\begin{align}
q^{(3)}_2-q^{(3)}_1=&2(\#K_3-\#K_2)\nonumber\\
=&2[(\#K_3-\#\Gamma_3)-(\#K_2-\#\Gamma_2)]\nonumber\\
=&2([K_3]-[K^{(3)}_2]).
\end{align}
Using the constraint from the conservation of number of bands across the BZ, $[K^{(3)}_1]+[K^{(3)}_2]+[K^{(3)}_3]=0$, we rewrite the above equation in terms of the rotation invariants defined in Eq~\eqref{eq:ClassificationIndicesapp},
\begin{align}
q^{(3)}_2-q^{(3)}_1=-2([K^{(3)}_1]+2[K^{(3)}_2]).
\label{eq:PolarC3}
\end{align}
Therefore, using Eq~\eqref{ptophase}, the polarization component $p_2$ is
\begin{align}
p_2=\frac{2e}{3}\left([K^{(3)}_1]+2[K^{(3)}_2]\right) \mod e.
\end{align}
Since the two components are the same up to multiple of $e$ as shown in Eq.~\eqref{eq:p1equalp2}, the polarization for $C_3$ symmetric TCIs can be expressed as
\begin{align}
\Aboxed{p_1=p_2=\frac{2e}{3}\left([K^{(3)}_1]+2[K^{(3)}_2]\right) \mod e.}
\end{align}
\\
When an TCI has $C_6$ symmetry, we have the additional constraints $\#K_{i}=\#K'_{i},~i=1,2,3$. With the conservation of the number of bands $[K^{(3)}_1]+[K^{(3)}_2]+[K^{(3)}_3]=0$ we have,
\begin{align}
\Aboxed{p_1=p_2=e\left([K^{(3)}_1]+2[K^{(3)}_2]\right)=0 \mod e,}
\label{eq:PolarC6}
\end{align}
i.e., the polarization is always vanishing in $C_6$-symmetric TCIs. This result is also consistent with our earlier analysis (see Eq.~\eqref{eq:realspacepolar}, where we have shown that the polarization components are always trivial integer values in $C_6$ symmetric TCIs).

\section{Unit cells and maximal Wyckoff positions}
\label{sec:UnitCells}
\begin{figure}[h!]
\centering
\includegraphics[width=\columnwidth]{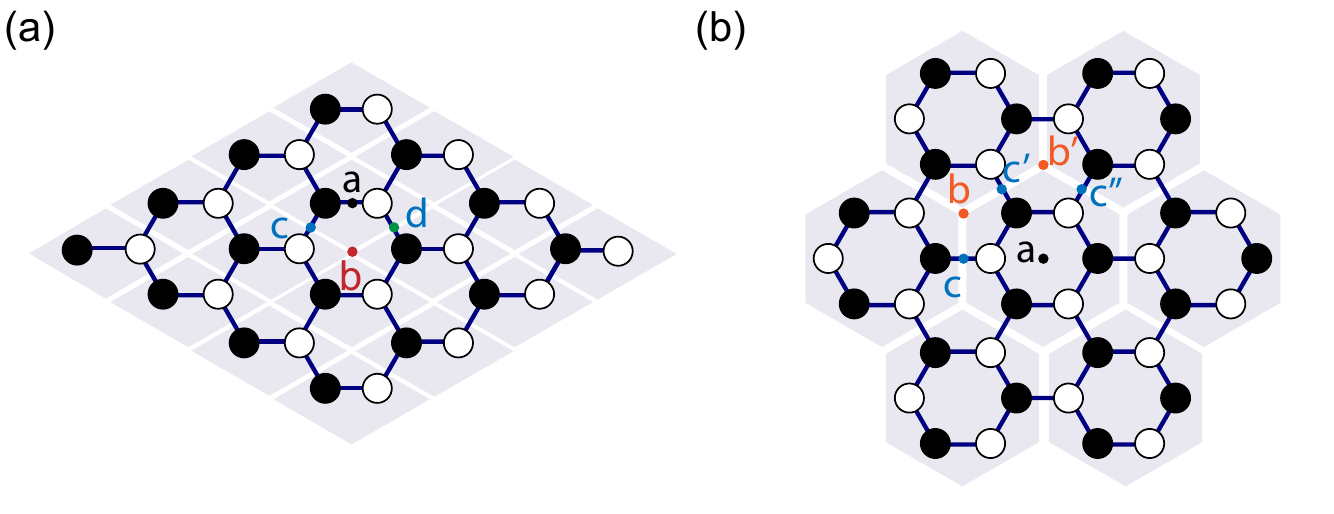}
\caption{Honeycomb lattice with two finite geometries. Black and white circles represent atoms, blue links represent electronic hopping terms. Parallelograms in (a) and hexagons in (b) are choices of unit cells compatible with the boundary conditions of the lattice. In both cases, maximal Wyckoff positions are indicated for the unit cell at the center of the lattice by colored dots.}
\label{lattice_choice}
\end{figure}

A lattice with no boundaries has multiple choices of unit cells. When a boundary is open, however, unit cells are restricted to be compatible with the boundaries. Only under this compatibility a bulk-boundary correspondence can be established. 
In $C_n$-symmetric lattices, given a choice of unit cell, there are special high-symmetry points within the unit cell, called  \emph{maximal Wyckoff positions}, that are invariant under rotations (about the center of the unit cell) up to lattice translations. Let us take the $C_6$ symmetric honeycomb lattice as an example (Fig.~\ref{lattice_choice}). In Fig.~\ref{lattice_choice}(a), we put the honeycomb lattice on a finite parallelogram with the zigzag-type edge. The unit cells are parallelograms containing the horizontally closest two atoms. Since the unit cells are only invariant under a $C_2$ rotation, we have four maximal Wyckoff positions, $a$ at the center of the unit cell, $b$ at its corner, and $c, d$ at the middle points of its edges. Suppose that now we have the lattice boundaries shown in Fig.~\ref{lattice_choice} (b). A choice of unit cells compatible with these boundaries consist of hexagons, each containing six atoms. At these $C_6$-symmetric unit cells, we have three maximal Wyckoff positions: the $C_6$-symmetric point $a$ at the center of the unit cell, the $C_3$-invariant points $b, b'$ at the corners of unit cell, and the $C_2$-invariant points $c,c',c''$ in the middle of the edges of the unit cell. Notice that in both cases the Wyckoff position $a$ is at the center of the unit cell. Furthermore, for any choice of unit cell, the atoms always fully fall inside a unit cell. Therefore, the atoms either locate at the maximal Wyckoff position $a$, or at non-maximal Wyckoff positions related by symmetry, e.g. in Fig.~\ref{lattice_choice}(a) the two atoms within each unit cell are related by a $C_2$ rotation, while in \ref{lattice_choice}(b) the six atoms are related by $C_6$ rotation. In any case, the center of positive ionic charge is always at the maximal Wyckoff position~$a$.

\section{Tight-binding models for the primitive generators}
\label{sec:TBModels}
In this section we provide detailed discussion of the tight-binding models that give rise to the primitive generators that span the classifications of TCIs with $C_n$ symmetry. We use the capital $H$ to represent the general tight-binding Hamiltonians of the models and the lower case $h$ to represent the generators in the Main Text that are chosen from a certain topological phases of the general Hamiltonians $H$.

\subsection{Twofold and fourfold symmetry}
\begin{figure}[!t]
	\centering
    \includegraphics[width=\columnwidth]{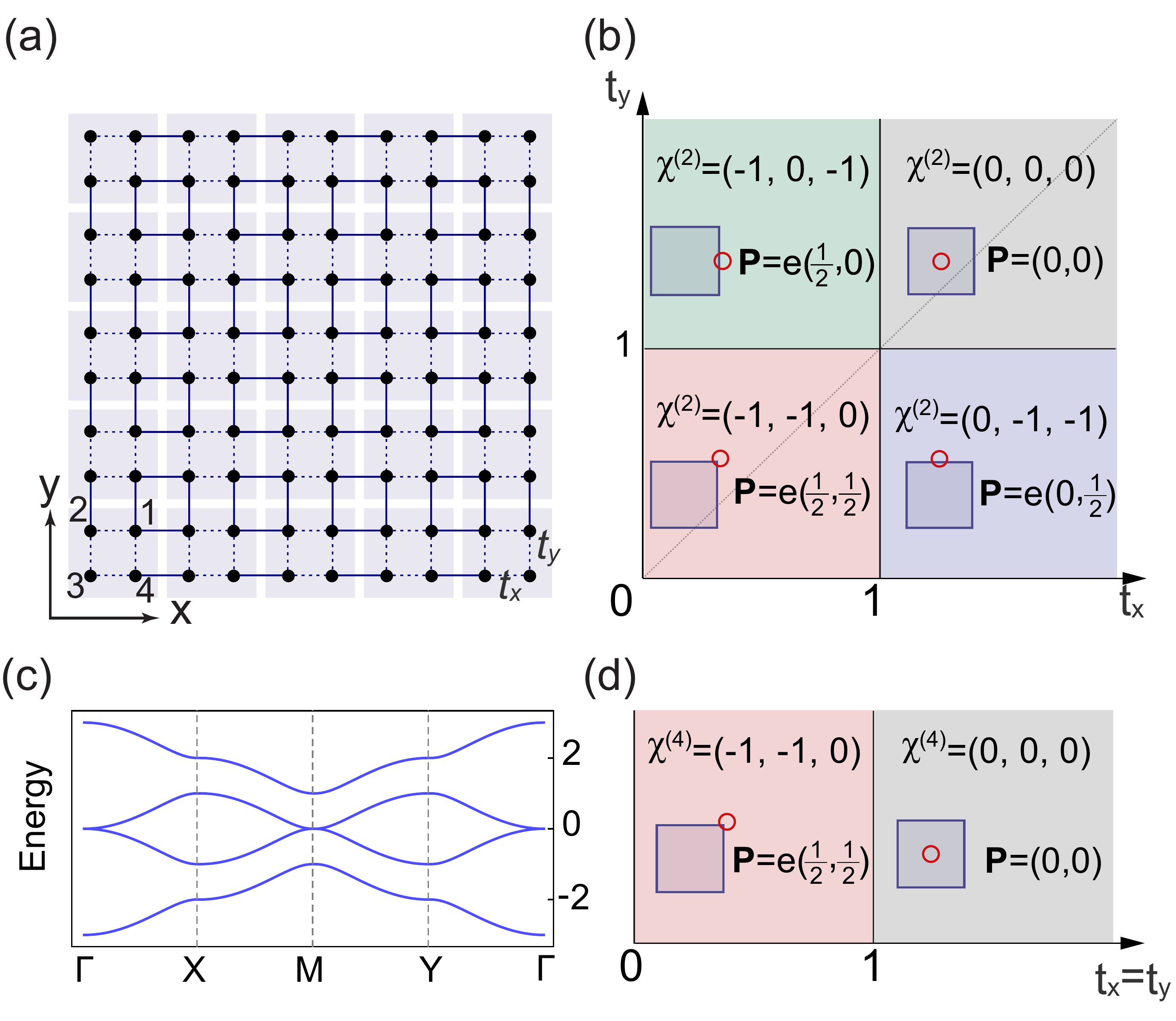}
	\caption{
	 (a) Lattice configurations of $H_1^{(2)}(\kk,t_x,t_y)$. The light blue squares represent unit cells and black dots are sites. Dashed lines and solid lines represent hoppings within and between unit cells, respectively. (b) The phase diagram for the lowest band of $H_1^{(2)}(\kk,t_x,t_y)$ as a function of the intra-cell hopping amplitude $t_x,t_y$. In each phase, we indicate the rotation invariants, the polarization and the Wannier centers (red circles) in one bulk unit cell (blue square). (c) The bulk band structure along high symmetry lines in the BZ when $t_x=t_y=0.5$. (d) The phase diagram of $H_1^{(4)}(\kk,t_x=t_y)$. }
	\label{C41}
\end{figure}

The classification of time-reversal invariant TCIs with $C_4,C_2$ symmetries is given by the topological indices
\begin{align}
\chi^{(4)}&=([X^{(2)}_1],[M_1^{(4)}],[M_2^{(4)}])\nonumber\\
\chi^{(2)}&=([X^{(2)}_1],[Y^{(2)}_1],[M^{(2)}_1])\nonumber
\end{align}
The lattice configuration from which the generator $h_{1b}^{(4)}$ can be obtained is shown in Fig.~\ref{C41}(a). It has four sites per unit cell, each of which has hoppings between neighboring sites along horizontal and vertical directions. We set the inter-cell hopping amplitude to $1$ (from now on, we will set the inter-cell hopping amplitude to $1$ unless otherwise noted) and the intra-cell hopping amplitude along the $x(y)$ direction to $t_x(t_y)$. Using the basis for each site as labeled in Fig.~\ref{C41}~(a), the Bloch Hamiltonian for this lattice is
\begin{align}
H_1^{(2)}(\kk,t_x,t_y)&=
\left(
\begin{array}{cccc}
0  & e^{ik_x} &  0  & e^{ik_y}   \\
e^{-ik_x} &  0  & e^{ik_y} &  0  \\
0  &  e^{-ik_y}&  0  & e^{-ik_x} \\
e^{-ik_y} &  0  & e^{ik_x} &  0
\end{array}
\right)\nonumber\\
&+\left(
\begin{array}{cccc}
0 & t_x & 0 & t_y\\
t_x & 0 & t_y & 0  \\
0 & t_y & 0 & t_x \\
t_y & 0 & t_x & 0  \\
\end{array}
\right).
\end{align}
For generic values of $t_x, t_y$, this model is $C_2$-symmetric. However, when $t_x=t_y$, the model is $C_4$-symmetric. We call the $C_4$-symmetric Hamiltonian
\begin{align}
H_1^{(4)}(\kk,t)= H_1^{(2)}(\kk,t_x=t_y=t).
\label{equation:h1b}
\end{align}
It obeys $\hat{r}_4 H_1^{(4)}(\kk,t)\hat{r}^\dagger_4=H_1^{(4)}(R_4\kk,t)$, where
\begin{align}
\label{rot_c4}
\hat{r}_4=\left(\begin{array}{cccc}
0 & 0 & 0 & 1\\
1 & 0 & 0 & 0\\
0 & 1 & 0 & 0\\
0 & 0 & 1 & 0\\
\end{array}\right),
\end{align}
and $R_4$ is the rotation operation rotates the crystal momenta by $\frac{\pi}{2}$, i.e., $R_4(k_x,k_y)=(k_y,-k_x)$. The rotation operator obeys $\hat{r}_4^4=1$ and has eigenvalues $\{r_4\}=\{+1,+i,-1,-i\}$. 

The bulk spectrum of $H_1^{(2)}(\kk,t_x,t_y)$ is gapped at $\frac{1}{4}-$filling and $\frac{3}{4}-$filling as long as $t_x,t_y\neq1$. We show the bulk band spectrum of $H_1^{(4)}(\kk,t=0.5)$ in Fig.~\ref{C41}~(c). The second and third bands are degenerate at the $\bd{\Gamma}$ and $\bd{M}$ points. The degeneracy is protected by TRS and $C_4$ symmetry. When $t_x\neq t_y$, we lose the protection of the degeneracy at the $\bd{\Gamma}$ and $\bd{M}$ points, and the degeneracy moves to other points in the BZ. When either one of $t_x, t_y$ approaches $1$, the gaps at $1/4-$filling and $3/4-$filling close at the $\bd{M}$ point and a phase transition occurs. 
In Fig.~\ref{C41}(b), we show the phase diagram for $H_1^{(4)}(\kk,t_x,t_y)$. Additionally, Fig.~\ref{C41}(b) shows the $C_2$ rotation invariants, the Wannier centers and the polarizations in each phase when the first band is occupied. When $t_x,t_x>1$, the model has trivial rotation invariants, $\chi^{(2)}=(0,0,0)$ and the polarization is zero. When $t_x,t_y<1$, $\chi^{(2)}=(-1,-1,0)$ and the polarization is ${\bf P}=\frac{e}{2} \bd{a}_1+\frac{e}{2}\bd{a}_2$ and the Wannier center is located at the maximal Wyckoff position $b$. 
When either one of the intra-cell hopping amplitudes is larger than $1$, the polarization component along that direction becomes trivial. For example, if $t_y>1,t_x<1$, the polarization is $\bd{P}=\frac{e}{2}\bd{a}_1$. In this case, the Wannier center is located at the Wyckoff position $c$. Similarly, the polarization for when $t_x>1,t_y<1$ is $\bd{P}=\frac{e}{2}\bd{a}_2$ and the Wannier center is located at Wyckoff position $d$. 
Along the diagonal line (the gray dotted line) in Fig.~\ref{C41}~(b), $t_x=t_y$ and the model is $C_4$-symmetric. We show the phase diagram for $H^4(\kk,t)$ with the lowest band filled in Fig.~\ref{C41} (d). The phase with $t_x=t_y<1$ belongs to class $\chi^{(4)}=(-1,-1,0)$. It has the polarization $\bd{P}=\frac{e}{2}(\bd{a}_1+\bd{a}_2)$ and one Wannier center located at Wyckoff position $b$. 
We choose the $\chi^{(4)}=(-1,-1,0)$ phase (at $\frac{1}{4}-$filling) to serve as the generator $h_{1b}^{(4)}$ [Fig.~2(c) in the Main Text].

\begin{figure}[!t]
	\centering
    \includegraphics[width=\columnwidth]{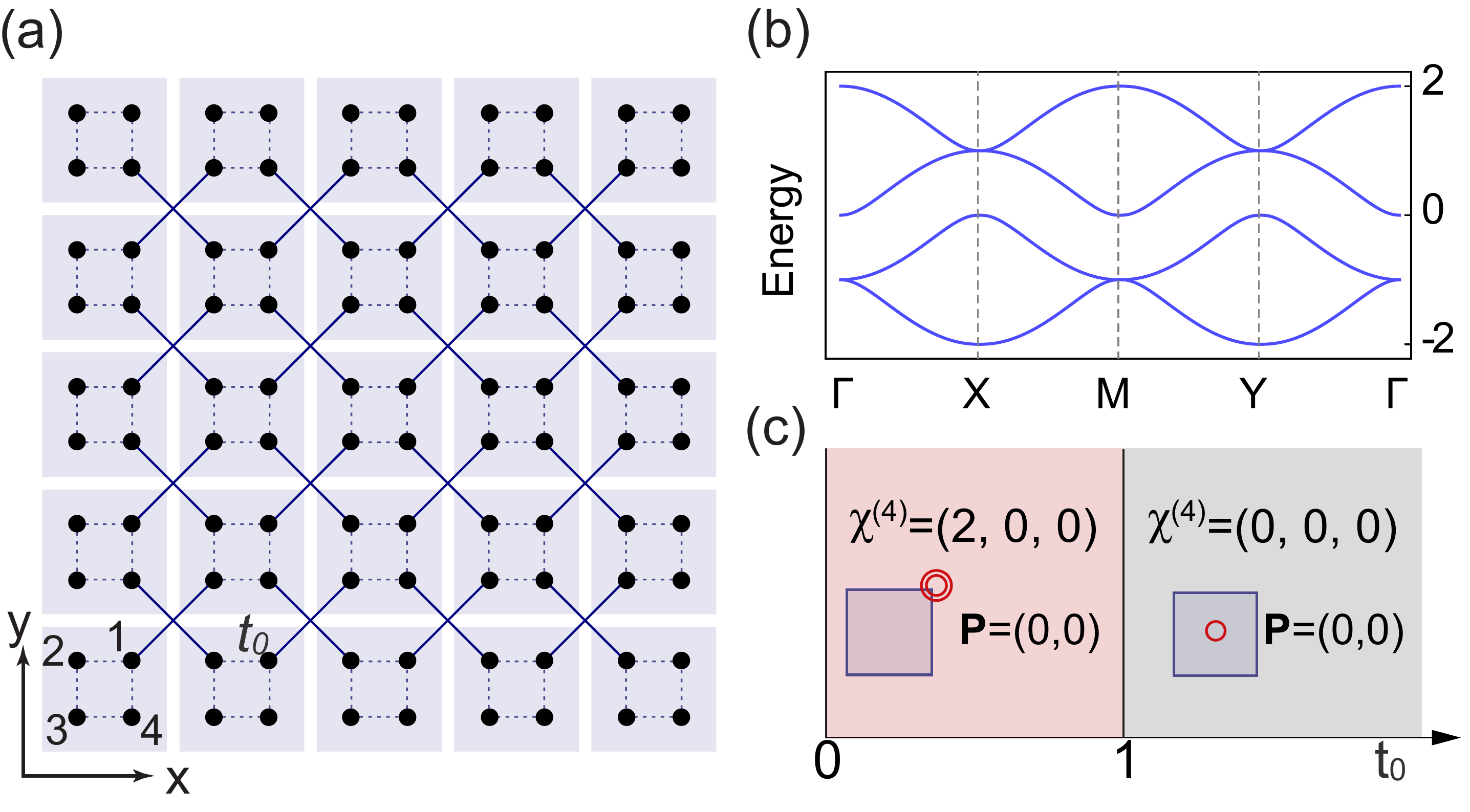}
	\caption{
	 (a) Lattice configurations of $H_2^{(4)}(\kk,t_0)$. (b) The bulk band structure along high symmetry lines in the BZ when $t_0=0.5$. (c) The phase diagram for $H_2^{(4)}(\kk,t_0)$ with the lowest two bands filled as a function of intra-cell hoppings $t_0$.}
	\label{C42}
\end{figure}

The lattice configuration from which the generator $h_{2b}^{(4)}$ can be obtained is shown in Fig.~\ref{C42} (a). It has four sites per unit cell, next nearest neighboring hoppings between sites in different unit cells and nearest neighboring hoppings between sites within one unit cell. The Bloch Hamiltonian for this model is
\begin{align}
H_2^{(4)}(\kk,t_0)&=
\setlength{\arraycolsep}{0.1pt}
\begin{pmatrix}
0  &  t_0   &  e^{i(k_x+k_y)} &  t_0   \\
t_0  &  0   &  t_0  & e^{i(k_y-k_x)} \\
e^{-i(k_x+k_y)} &  t_0  & 0  & t_0 \\
t_0  & e^{i(k_x-k_y)} &  t_0 & 0
\end{pmatrix}.
\end{align}
This model is $C_4$-symmetric and has the fourfold rotation operator $\hat{r}_4$ defined in Eq.~\eqref{rot_c4}. When $t_0<1$, the bulk bands are gapped at half-filling. The spectrum for $H_2^{(4)}(\kk,t_0=0.5)$ is shown in Fig.~\ref{C42} (b). 
The degeneracy between the lower two bands at the $\bd{\Gamma}$ and $\bd{M}$ point is protected by $C_4$ symmetry and TRS. When $t_0=1$ the gap closes at the $\bd{M}$ point. When $t_0>1$, two gaps appear at $\frac{1}{4}-$filling and $\frac{3}{4}-$filling. However, the rotation invariants for each subspace are trivial in that phase. In Fig.~\ref{C42}(c), we show the phase diagram of $H_2^{(4)}(\kk,t_0)$. For the phase with $t_0<1$, we choose to fill the lowest two bands and it belongs to class $\chi^{(4)}=(2,0,0)$. For the phase with $t_0>1$, with the lowest band filled, all rotation invariants are zero. In the $\chi^{(4)}=(2,0,0)$ phase, the polarization is trivial and two Wannier centers are located at Wyckoff position $b$. We choose the $\chi^{(4)}=(2,0,0)$ phase (at half-filling) to serve as the generator $h_{2b}^{(4)}$ [Fig.~2(d) in the Main Text] of the $C_4$-symmetric classification.

\begin{figure}[!t]
	\centering
    \includegraphics[width=\columnwidth]{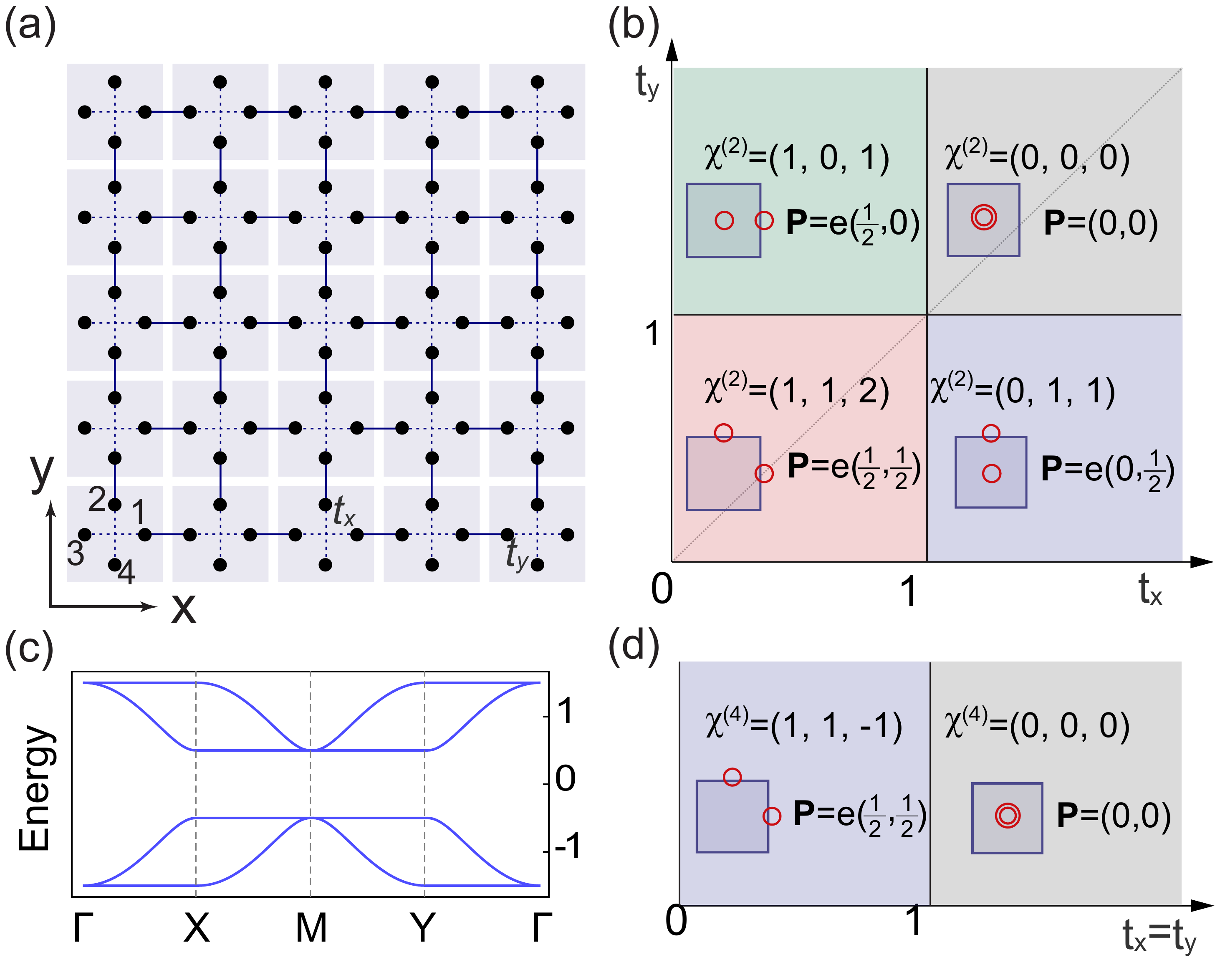}
	\caption{
	 (a) Lattice configurations of $H_3^{(2)}(\kk,t_x,t_y)$. (c) The bulk band structure along high symmetry lines in the BZ when $t_x=t_y=0.5$. (b) The phase diagram for the lower two bands of $H_3^{(2)}(\kk,t_x,t_y)$ as a function of the intra-cell hoppings $t_x,t_y$. (d) The phase diagram for the lowest two bands along the gray dashed line $t_x=t_y$ in (b). Along that line, the model restores $C_4$ symmetry.}
	\label{C43}
\end{figure}

A  general model from which generator $h_{2c}^{(4)}$ can be obtained is constructed by stacking two perpendicular SSH chains in horizontal and vertical directions as shown in Fig.~\ref{C43}(a). It has four sites per unit cell, hoppings between neighboring unit cells in both vertical and horizontal directions and next nearest neighboring hoppings inside each unit cell. The corresponding Bloch Hamiltonian is
\begin{align}
H_3^{(2)}(\bd{ k},t_x,t_y)&=\left(\begin{array}{cccc}
0 & 0 & e^{i k_x} & 0 \\
0 & 0 & 0 & e^{i k_y} \\
e^{-i k_x}& 0 & 0 & 0 \\
0 & e^{-i k_y}& 0 & 0
\end{array}\right)\nonumber\\
&+
\left(\begin{array}{cccc}
0 & 0 & t_x & 0 \\
0 & 0 & 0 & t_y \\
t_x & 0 & 0 & 0 \\
0 & t_y & 0 & 0
\end{array}\right).
\label{Eq:C4_3}
\end{align}

With generic values of $t_x, t_y$ this model is $C_2$-symmetric. When $t_x=t_y$, this model is $C_4$-symmetric. We call the $C_4$-symmetric Hamiltonian $H_3^{(4)}(\kk,t)=H_3^{(2)}(\kk,t_x=t_y=t)$, which has fourfold rotation operator defined in Eq.~\eqref{rot_c4}.
The bulk spectrum is gapped at half filling as long as $t_x,t_y\neq 1$. We show the bulk spectrum for $H_3^{(4)}(\kk,t=0.5)$ in Fig.~\ref{C43}(c). The degeneracy between the lower two bands at $\bd{M},$ and the degeneracy between the upper two bands at $\bd{\Gamma},$ are protected by $C_4$ symmetry and TRS. Once $t_x\neq t_y$, the degeneracy moves to other points in BZ. When $t_x=1\,(t_y=1)$ the gap closes along the BZ boundary $\bd{XM}\,\,(\bd{YM})$ and a phase transition occurs. In Fig.~\ref{C43}(b), we show the phase diagram of $H_3^{(4)}(\kk)$ with the lower two bands filled. When $t_x,t_y>1$, the gap reopens at half-filling. Occupying the lower two bands, the rotation invariants are trivial, $\chi_3^{(4)}=(0,0,0)$.  The phase with $t_x, t_y<1$ has nontrivial polarization components in both directions, $\bd{P}=\frac{e}{2}(\bd{a}_1+\bd{a}_2)$, and two Wannier centers located at the Wyckoff positions $c$ and $d$. As we increase the intra-cell hopping amplitude in one direction to be larger than $1$, the polarization component in that direction becomes trivial. For example, a topological phase having $t_y>1,t_x<1$ has only one nontrivial polarization component, $p_1=\frac{e}{2}$, and one Wannier center located at the maximal Wyckoff position $c$, the other located at the unit cell center $a$. Along the diagonal gray dotted line in Fig.~\ref{C43}(b), $t_x=t_y$ and the model is $C_4$-symmetric. The $C_4$-symmetric phase with $t_x=t_y<1$ belongs to class $\chi^{(4)}=(1,1,-1)$. It has the polarization $\bd{P}=\frac{e}{2}(\bd{a}_1+\bd{a}_2)$ and two Wannier centers located at Wyckoff positions $c$ and $d$, respectively. We take the $\chi^{(4)}=(1,1,-1)$ phase at half-filling to serve as the generator $h_{2c}^{(4)}$ [Fig.~2(e) in the Main Text].

\begin{figure}[!t]
	\centering
    \includegraphics[width=\columnwidth]{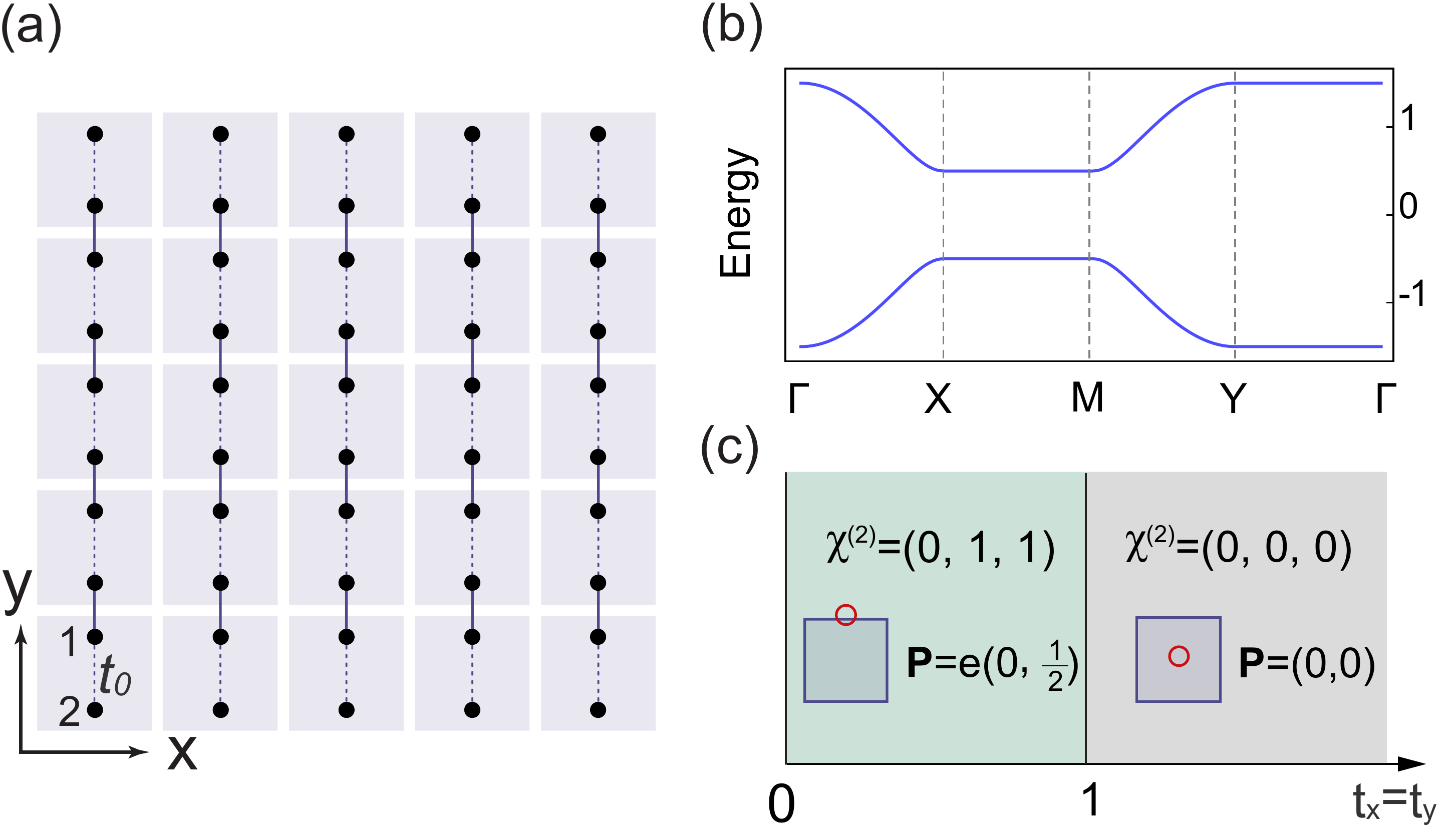}
	\caption{
	 (a) Lattice configurations of $H_4^{(2)}(\kk,t_0)$. (b) The bulk band structure along high symmetry lines in the BZ when $t_0=0.5$. (c) The phase diagram for $H_4^{(2)}(\kk,t_0)$ as tuning the intra-cell hopping strength $t_0$. In each phase, we indicate the rotation invariants, polarization and the Wyckoff position of the Wannier centers for the lower band.}
	\label{C21}
\end{figure}

In Fig.~\ref{C21}(a), we show a lattice configuration from which the generator $h_{1d}^{(2)}$ can be obtained. Each unit cell contains two sites and only vertical hoppings between nearest neighboring sites exist. The Bloch Hamiltonian is 
\begin{align}
H_{4}^{(2)}(\kk,t_0)=&
\left(
\begin{array}{cc}
0 &t_0+e^{-ik_y}\\
t_0+e^{ik_y} & 0
\end{array}\right).
\end{align}
It is $C_2$-symmetric, so that the Bloch Hamiltonian obeys $r_2H_{4}^{(2)}(\kk)\hat{r}_2^\dagger=H_{4}^{(2)}(-\kk)$, where the twofold rotation operator is $\hat{r}_2=\sigma_x$. 
In Fig.~\ref{C21}(b), we show the bulk energy spectrum for $H_2^{(2)}(\kk,t_0=0.5)$.
The bands are gapped as long as $t_0\neq 1$. We show the phase diagram of $H_2^{(2)}(\kk)$ with the lowest band filled in Fig.~\ref{C21}(c). In the phase with $t_0<1$, the model has polarization $\bd{P}=\frac{e}{2}\bd{a}_2$ and one Wannier center located at Wyckoff position $d$. We take the phase with $\bd{P}=\frac{e}{2}\bd{a}_2$ to serve as the generator $h_{1b}^{(2)}$ [Fig.~2(f) in the Main Text].
\subsection{Threefold and sixfold symmetry}
\begin{figure}[!t]
	\centering
    \includegraphics[width=\columnwidth]{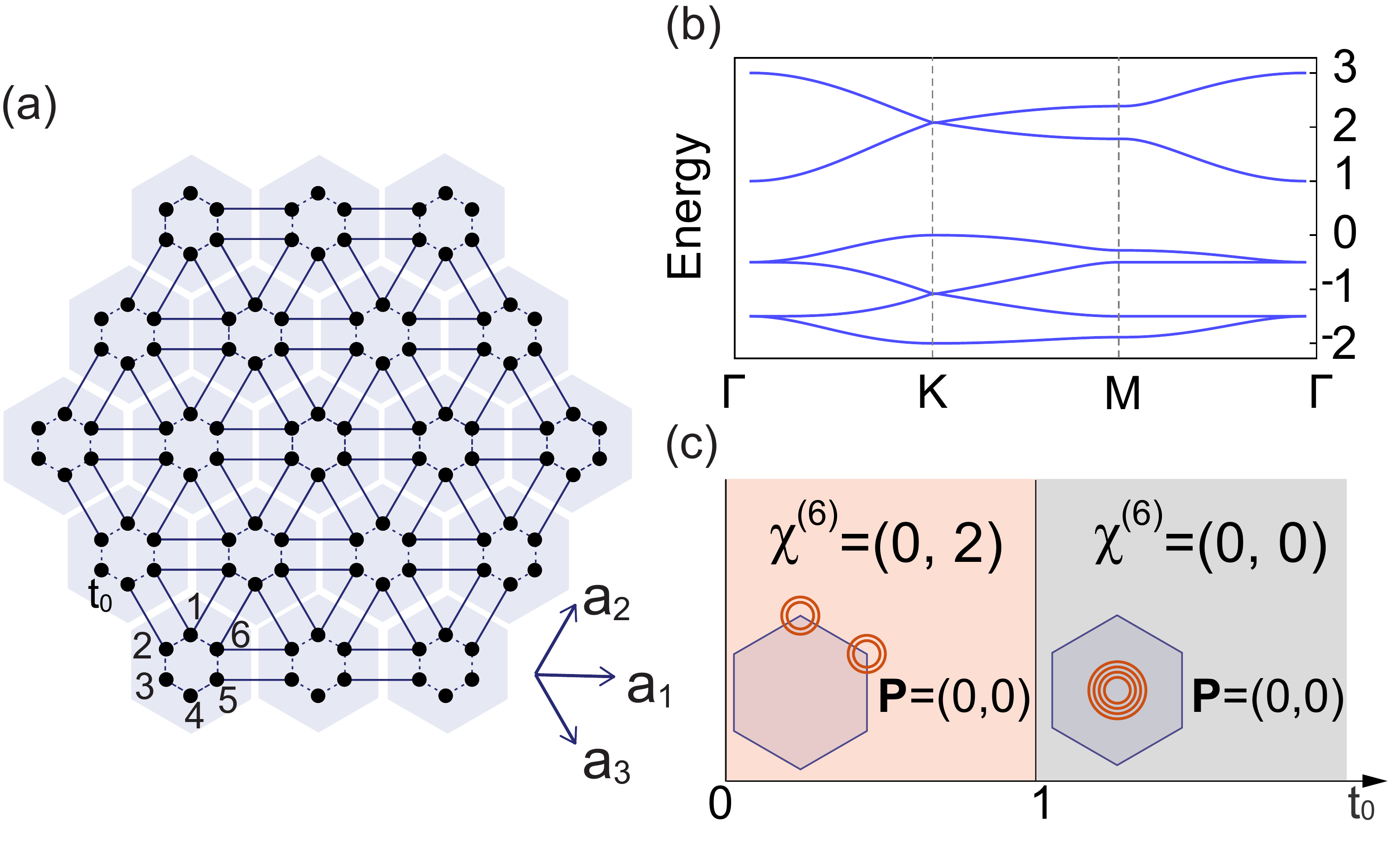}
	\caption{(a) Lattice configurations of $H_1^{(6)}(\kk,t_0)$. The light blue hexagons represent unit cells and black dots are sites. Dashed lines and solid lines represent hoppings within and between unit cells, respectively. (b) The bulk band structure along high symmetry lines in the BZ when $t_0=0.5$. (c) The phase diagram for  $H_2^{(6)}(\kk,t_0)$ as a function of the intra-cell hopping strength $t_0$. In each phase, we indicate the rotation invariants, polarization and the Wyckoff position of the Wannier centers.}
	\label{C62}
\end{figure}

The classification of time reversal symmetric TCIs with $C_6$ and $C_3$ symmetry is given by the topological indices
\begin{align}
\chi^{(6)}&=([M^{(2)}_1],[K^{(3)}_1])\nonumber\\
\chi^{(3)}&=([K^{(3)}_1],[K^{(3)}_2]).\nonumber
\end{align}
respectively. For all models that we discuss in this section, we choose the lattice vectors to be
\begin{align}
\bd{a}_1=(1,0),\,\, \bd{a}_2=\left(\frac{1}{2},\frac{\sqrt{3}}{2}\right).
\end{align}
The lattice configuration of a $C_6$ symmetric model from which the generator $h_{4b}^{(6)}$ is obtained is shown in Fig.~\ref{C62}(a). Each unit cell contains six sites and hoppings exist between nearest neighboring sites. The tight-binding Hamiltonian is
\begin{align}
\label{eq:H_61}
H_1^{(6)}(\kk)=&
\setlength{\arraycolsep}{0.1pt}
\begin{pmatrix}
0 & t_0 & e^{i\bd{k}\cdot \bd{a}_2} & 0 & e^{-i\bd{k}\cdot \bd{a}_3} & t_0\\
t_0 & 0 & t_0 & e^{-i\bd{k}\cdot \bd{a}_3} & 0 & e^{-i\bd{k}\cdot \bd{a}_1} \\
e^{-i\bd{k}\cdot \bd{a}_2} & t_0 & 0 & t_0 & e^{-i\bd{k}\cdot \bd{a}_1} & 0\\
0 & e^{i\bd{k}\cdot \bd{a}_3} &t_0 & 0 & t_0 & e^{-i\bd{k}\cdot \bd{a}_2}\\
e^{i\bd{k}\cdot \bd{a}_3} & 0 & e^{i\bd{k}\cdot \bd{a}_1} & t_0 & 0 & t_0\\
t_0 &e^{i\bd{k}\cdot \bd{a}_1} & 0 & e^{i\bd{k}\cdot \bd{a}_2} & t_0 & 0
\end{pmatrix},
\end{align} 
which has the $C_6$ rotation operator, 
\begin{align}
\label{rot_c6}
\hat{r}_6=\left(\begin{array}{cccccc}
0&0&0& 0&0&1\\
1&0& 0&0&0&0\\
0& 1&0&0&0&0\\
0& 0&1&0&0&0\\
0& 0&0&1&0&0\\
0& 0&0&0&1&0\\
\end{array}\right),
\end{align}
which obeys $\hat{r}_6^6=1$ and has eigenvalues $r_6=e^{2i\pi (p-1)/6}$ for $p=1,2\ldots,6$. In Fig.~\ref{C62}(b), we show the spectrum of $H_1^{(6)}(\kk,t_0=0.5)$. When $t_0<1$, the bulk spectrum is gapped at $\frac{2}{3}-$filling. 
When $t_0=1$ the bulk gap closes at the $\bd{\Gamma}$ point and a phase transition occurs. In Fig.~\ref{C62}(c), we show the phase diagram of $H_1^{(6)}(\kk,t_0)$. The $\chi^{(6)}=(0,2)$ phase, which exists for $t_0<1$, has four Wannier centers, two located at Wyckoff position $b$ and two at position $b'$. We take the $\chi^{(6)}=(0,2)$ phase (at $\frac{2}{3}-$filling) as the generator $h_{4b}^{(6)}$ [Fig.~3(c) in the Main Text] for the $C_6$ symmetric classification.

\begin{figure}[!t]
	\centering
    \includegraphics[width=\columnwidth]{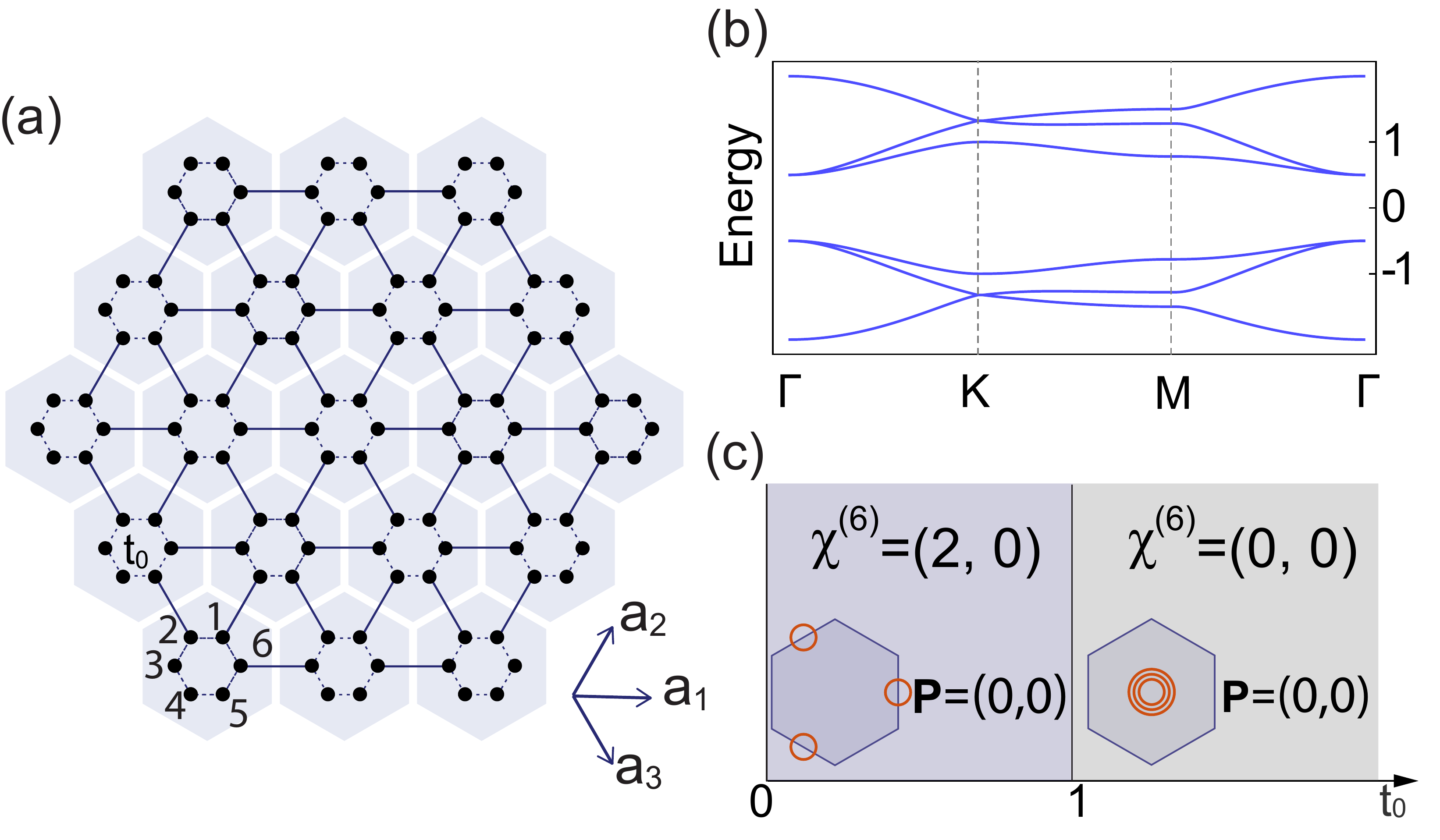}
	\caption{
	(a) Lattice configurations for $H_2^{(6)}(\kk,t_0)$. (b) The bulk band spectrum along high symmetry lines in the BZ when $t_0=0.5$. (c) The phase diagram for $H_2^{(6)}(\kk,t_0)$ as a function of the intra-cell hopping $t_0$. In each phase, we indicate the rotation invariants, polarization and the Wyckoff position of the Wannier centers.}
	\label{C61}
\end{figure}

The primitive generator $h_{3c}^{(6)}$ can be obtained from the the lattice configuration shown in Fig.~\ref{C61}(a). It is a hexagonal lattice that has six sites per unit cell and hoppings between nearest neighboring sites. Using the basis of sites as labeled in Fig.~\ref{C61}(a),
the Bloch Hamiltonian is
\begin{align}
\label{H_c6_2}
H_2^{(6)}(\kk)&=
\setlength{\arraycolsep}{0.1pt}
\begin{pmatrix}
0 & t_0 & 0 & e^{i\bd{k}\cdot \bd{a}_2} & 0 & t_0\\
t_0 & 0 & t_0 & 0 & e^{-i\bd{k}\cdot \bd{a}_3} & 0\\
0 & t_0 & 0 & t_0 & 0 & e^{-i\bd{k}\cdot \bd{a}_1}\\
e^{-i\bd{k}\cdot \bd{a}_2} & 0 & t_0 & 0 & t_0 &  0\\
0 & e^{i\bd{k}\cdot \bd{a}_3} & 0 & t_0 & 0 & t_0\\
t_0 & 0 & e^{i\bd{k}\cdot \bd{a}_1} & 0 & t_0 & 0
\end{pmatrix}.
\end{align}
We show the spectrum of $H_2^{(6)}(\kk,t_0=0.5)$ in Fig.~\ref{C61}(b). When $t_0<1$, the model is gapped at half filling, and belongs to class $\chi^{(6)}=(2,0)$. In this phase, it has three Wannier centers located at Wyckoff positions $c$, $c'$, and $c''$ in each unit cell.  We take this phase to serve as the generator $h_{3c}^{(6)}$ for the $C_6$ symmetric classification [Fig.~3(d) in the Main Text]. This phase has a degeneracy between the first and the second bands at the $\bd{K}$ and $\bd{K'},$ points as well as a degeneracy between the second and the third bands at $\bd{\Gamma}$ point. These degeneracies are protected by $C_6$ symmetry and TRS. At $t_0=1$, the gap closes at the $\bd{\Gamma}$ point. When $t_0>1$, the model is in the trivial phase, $\chi^{(6)}=(0,0)$.

\begin{figure}[!t]
	\centering
    \includegraphics[width=\columnwidth]{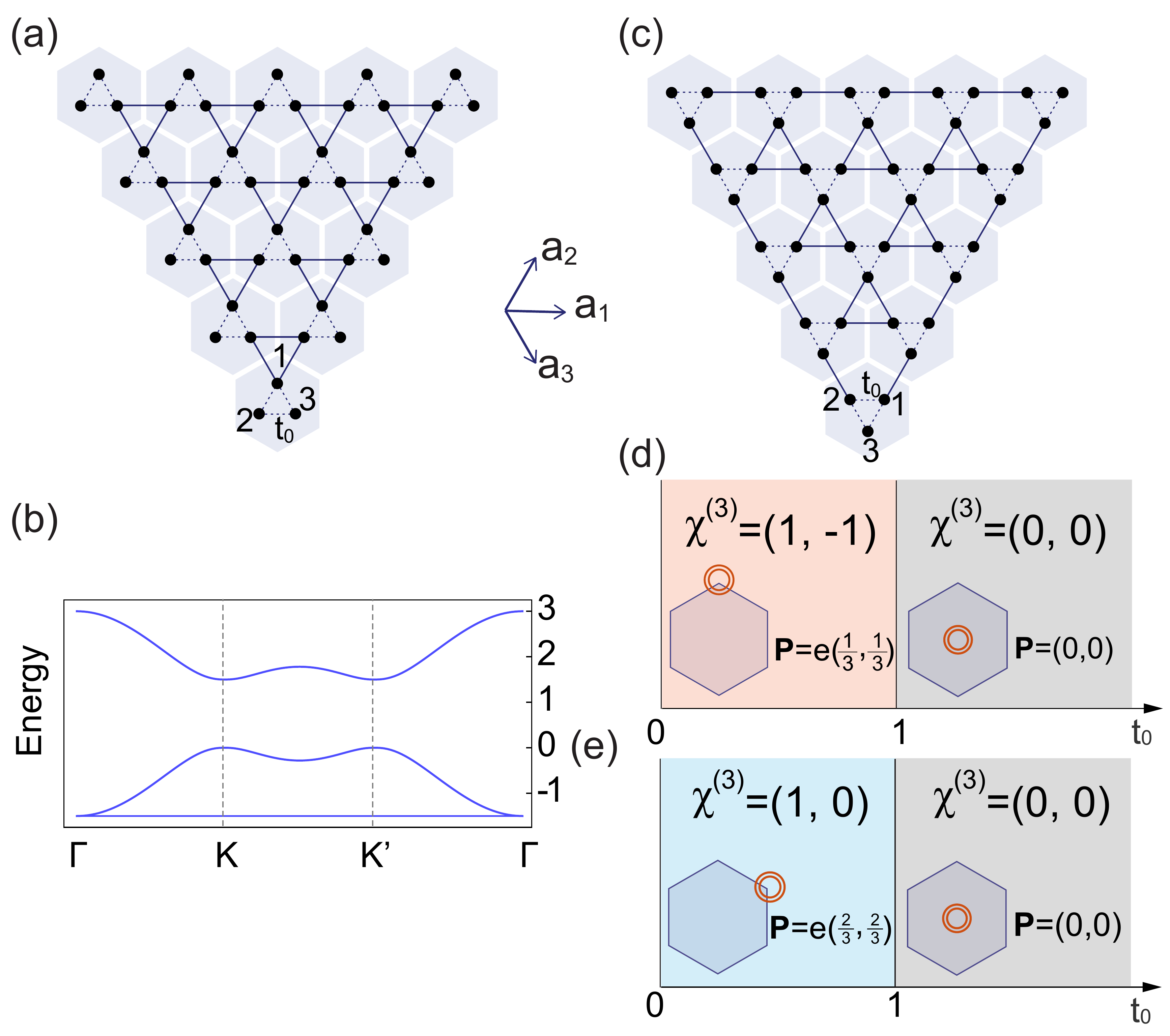}
	\caption{
	 (a) Lattice configurations of $H_1^{(3)}(\kk,t_0)$. (b) The bulk band structure along high symmetry lines in the BZ when $t_0=0.5$. (c) Lattice configurations of $H_2^{(3)}(\kk,t_0)$. (d) The phase diagram for $H_1^{(3)}(\kk,t_0)$ as function of the intra-cell hopping strength $t_0$. In each phase, we indicate the rotation invariants, polarization and the Wyckoff position of the Wannier centers for the lower two bands. (e) The phase diagram for $H_2^{(3)}(\kk,t_0)$.}
	\label{C31}
\end{figure}

The general models from which we can obtain the $C_3$ symmetric generators are built from the kagom\`e lattice. The configurations for two models are shown in Fig.~\ref{C31}(a),(b). Both of them have three sites per unit cell and hoppings between nearest neighboring sites. The Bloch Hamiltonian for the configuration in Fig.~\ref{C31}(a) is
\begin{align}
\label{H_c3_2=1}
H^{(3)}_1(\bd{k})=&
\left(\begin{array}{ccc}
0 &t_0+ e^{i \bd{k}\cdot \mathbf{a}_2}& t_0e^{-i \bd{k}\cdot \mathbf{a}_3} \\
t_0+e^{-i \bd{k}\cdot \mathbf{a}_2} & 0 & t_0+e^{-i \bd{k}\cdot \mathbf{a}_1}\\
t_0+e^{i \bd{k}\cdot \mathbf{a}_3} & t_0+e^{i \bd{k}\cdot\mathbf{a}_1}& 0 
\end{array}\right),
\end{align}
where $\bd{a}_3=\bd{a}_1-\bd{a}_2$.
With the labeling of sites in Fig.\ref{C31}(a), (b), the counter clockwise $C_3$ rotation operator is 
\begin{align}
\hat{r}_3=\left(\begin{array}{ccc}
 0 & 0 & 1\\
 1 & 0 & 0\\
 0 & 1 & 0
\end{array}\right).
\label{eq:rot_C3}
\end{align}
It satisfies $\hat{r}_3^3=1$ and has eigenvalues $r_3=1,e^{i\frac{2\pi}{3}}, e^{-i\frac{2\pi}{3}}$. In Fig.~\ref{C31} (b), we show the bulk energy spectrum for $H_1^{(3)}(\kk,t_0=0.5)$.
It is gapped  at $\frac{2}{3}$-filling as long as $t_0\neq1$. The lowest two bands are degenerate at the $\bd{\Gamma}$ point and the degeneracy is protected by TRS since the eigenvalues of $\hat{r}_3$ come in a complex conjugate pair, and hence the two bands are forced to have the same energy by TRS.
At $t_0=1$ the bulk gap closes at the $\bf{K}$ point and a phase transition occurs. In Fig.~\ref{C31}(d), we show the phase diagram of $H_1^{(3)}(\kk,t_0=0.5)$ with lowest two bands filled. When $t_0<1$, the model has nontrivial rotation invariants $\chi^{(3)}=(1,-1)$. In this phase, the polarization is $\bd{P}=\frac{e}{3}(\bd{a}_1+\bd{a}_2)$ and two Wannier centers located at the Wyckoff position $b$. We take the $\chi^{(3)}=(1,-1)$ phase (at $\frac{2}{3}$-filling) as the generator $h_{2b}^{(3)}$ [Fig.~3(e) in the Main Text] for the $C_3$ symmetric classification.

The Bloch Hamiltonian for lattice configuration in Fig.~\ref{C31}(c) is 
\begin{align}
\label{H_c3_2}
H_2^{(3)}(\bd{k})&=
\left(\begin{array}{ccc}
0 & t_0+ e^{i \bd{k}\cdot \mathbf{a}_1} &t_0+ e^{i \bd{k}\cdot \mathbf{a}_2} \\
t_0+e^{-i \bd{k}\cdot \mathbf{a}_1} & 0 & t_0+e^{-i \bd{k}\cdot \mathbf{a}_3}\\
t_0+e^{-i \bd{k}\cdot \mathbf{a}_2} & t_0+e^{i \bd{k}\cdot\mathbf{a}_3} & 0 
\end{array}\right).
\end{align}
The rotation operator is the same as defined in Eq.~\eqref{eq:rot_C3}. The band structure and phase transitions for this model are the same as $H_1^{(3)}(\kk)$. However, its topological phase is characterized by different topological indices. In Fig.~\ref{C31} (e), we show the phase diagram of $H_2^{(3)}$ with the lower two bands filled. When $t_0<1$, this model has nontrivial rotation invariants, $\chi^{(3)}=(1,0)$. The polarization in this phase is $\frac{2e}{3}(\bd{a}_1+\bd{a}_2)$, which corresponds to two Wannier centers located at Wyckoff position $c$. We take the $\chi^{(3)}=(1,0)$ phase (at $\frac{2}{3}-$filling) as the generator $h_{2c}^{(3)}$ [Fig.~3(f) in the Main Text] for the $C_3$ symmetric classification.

\section{Band representations induced from maximal Wyckoff positions}
\label{sec:BandRep}
Following the procedure in Ref.~\cite{cano2018}, we induce the band representations for Wannier orbitals located at all maximal Wyckoff positions for each $C_n$ symmetry. By comparing these representations with those of our primitive generators, we verify that all of our generators are Wannier-representable.
\subsection{Fourfold and twofold symmetry}
As shown in Fig.~2 of the Main Text, there are three maximal Wyckoff positions in the $C_4$ symmetric lattices, a (with multiplicity $1$), b (with multiplicity $1$) and c (with multiplicity $2$). When breaking the $C_4$ symmetry down to $C_2$ symmetry, there are four maximal Wyckoff positions per unit cell,  a (with multiplicity $1$), b (with multiplicity $1$), c (with multiplicity $1$) and d (with multiplicity $1$). We denote the orbital at a Wyckoff position $x$ by $x_l$, where $l$ represents the angular momentum for that orbital. For a position $x$ with its stabilizer group $C_n$, the angular momentum is $l=0,\ldots,n-1$. We summarize the results in Table.~\ref{Tab:Wy_C4} and~\ref{Tab:Wy_C2}. From these tables we find that all generators can be decomposed into Wannier orbitals located at the maximal Wyckoff positions with positive coefficients,
\begin{align}
&\left.\begin{array}{ccc}
h_{1b}^{(4)}&\sim & b_2\\
h_{2b}^{(4)}&\sim & b_1+b_3\\
h_{2c}^{(4)}&\sim & c_1+c_1'
\end{array}\right\} C_4\,\,\textrm{Symmetry}
\\
&\left.\begin{array}{ccc}
h_{1b}^{(4)}&\sim & b_0\\
h_{2c}^{(4)}&\sim & c_1 + d_1 \\
h_{1d}^{(4)}&\sim & d_1
\end{array}\right\} C_2\,\,\textrm{Symmetry}.
\end{align}
Therefore, generators for $C_2$ and $C_4$ symmetric classification are all Wannier representable.
\begin{table}[!h]
\centering
\begin{tabular}{c|cccc}
\hline\hline
Wyckoff positions& $\bd{\Gamma}$ & $\bd{M}$ & $\bd{X}$ & $\bd{Y}$\\
\hline
$a_l$ & $e^{\frac{i\pi l}{2}}$ & $e^{\frac{i\pi l}{2}}$
      & $(-1)^l$ & $(-1)^l$ \\
$b_l$ & $e^{\frac{i\pi l}{2}}$ & $-e^{\frac{i\pi l}{2}}$
	  & $(-1)^{(l+1)}$ & $(-1)^{(l+1)}$ \\
\multirow{2}{*}{$c_l$} 
     & $e^{\frac{i\pi }{2}(l+2)}$ & $e^{\frac{i\pi }{2}(l+1)}$ &$1$ & $1$\\
     & $ e^{\frac{i\pi l}{2}} $ & $e^{\frac{i\pi }{2}(l+3)}$ & $-1$ &$-1$\\
\hline\hline
\end{tabular}
\caption{Eigenvalues for $C_4$ ($C_2$) rotation operator at $\bd{\Gamma}$ and $\bd{M}$ ($\bd{X}$ and $\bd{Y}$) points for Wannier orbitals located at the maximal Wyckoff positions in $C_4$ symmetric lattices with angular momentum $l$. For position $a$ and $b$, $l$ runs from $0$ to $ 3$ and for position $c$, $l=0,1$.}
\label{Tab:Wy_C4}
\end{table}
\begin{table}[h!]
\centering
\begin{tabular}{c|cccc}
\hline\hline
Wyckoff positions & $\bd{\Gamma}$ & $\bd{M}$ & $\bd{X}$ & $\bd{Y}$\\
\hline
$a_l$ & $(-1)^l$ & $(-1)^l$
      & $(-1)^l$ & $(-1)^l$ \\
$b_l$ & $(-1)^l$ & $(-1)^l$
      & $(-1)^{l+1}$ & $(-1)^{l+1}$ \\
$c_l$ & $(-1)^l$ & $(-1)^{l+1}$
      & $(-1)^{l+1}$ & $(-1)^l$ \\
$d_l$ & $(-1)^l$ & $(-1)^{l+1}$
      & $(-1)^l$ & $(-1)^{l+1}$ \\
\hline\hline
\end{tabular}
\caption{Eigenvalues for $C_2$ rotation operator at $\bd{\Gamma, M, X}$ and $\bd{Y}$ points for Wannier orbitals located at the maximal Wyckoff positions in $C_2$ symmetric lattices with angular momentum $l$, $l=0,1$.}
\label{Tab:Wy_C2}
\end{table}

\subsection{Sixfold and Threefold symmetry}
As shown in Fig.~3 of the Main Text, there are three maximal Wyckoff positions in the $C_6$ symmetric lattices, $a$ (with multiplicity $1$), $b$ (with multiplicity $2$) and $c$ (with multiplicity $3$). When breaking the $C_6$ symmetry down to $C_3$ symmetry, there are three maximal Wyckoff positions, $a$ (with multiplicity $1$), $b$ (with multiplicity $2$) and $c$ (with multiplicity $3$). We summarize the eigenvalues of rotation operators corresponding the induced band representation at HSPs in Table.~\ref{Tab:Wy_C6} and~\ref{Tab:Wy_C3}.
\begin{table}[h!]
\centering
\begin{tabular}{c|cccc}
\hline\hline
Wyckoff positions & $\bd{\Gamma}$ & $\bd{K}$ & $\bd{M}$\\\hline
$a_l$   & $e^{\frac{i \pi l}{3}}$ & $e^{\frac{i 2\pi l}{3}}$ 
                & $(-1)^l$\\
\multirow{2}{*}{$b_l$} 
&$e^{\frac{i l \pi}{3}}$ & $e^{\frac{-i2\pi(l-1)}{3}}$ & $1$\\
&$-e^{\frac{i l \pi}{3}}$ & $e^{\frac{-i2\pi(l+1)}{3}}$ & $-1$\\
\multirow{2}{*}{$c_l$}  
& $e^{i\frac{l\pi}{3}}$ & $1$ & $(-1)^{l}$\\
& $-e^{i\frac{(l\pm1)\pi}{3}}$ & $e^{\pm\frac{i2\pi}{3}}$ & $(-1)^{l\pm1}$\\
\hline\hline
\end{tabular}
\caption{Eigenvalues for $C_6$ rotation operator at $\bd{\Gamma}$ points, $C_3$ rotation operator at $\bd{K}$ points and $C_2$ rotation operator at $\bd{M}$ points for Wannier orbitals located at the maximal Wyckoff positions with angular momentum $l$ in the $C_6$ symmetric lattices. For position $a$, $l=0,1,\ldots,5$, for position $b$, $l=0,1,2$ and for position $c$, $l=0,1$.}
\label{Tab:Wy_C6}
\end{table}

\begin{table}[h!]
\centering
\begin{tabular}{c|cccc}
\hline\hline
Wyckoff positions & $\bd{\Gamma}$ & $\bd{K}$ & $\bd{K'}$\\\hline
$a_l$   & $e^{\frac{i 2\pi l}{3}}$ & $e^{\frac{i 2\pi l}{3}}$ 
                & $e^{\frac{i 2\pi l}{3}}$\\
$b_l$ & $e^{\frac{i2l\pi}{3}}$ & $e^{\frac{i2(l+1)\pi}{3}}$ & $e^{\frac{i2(l-1)\pi}{3}}$\\
$c_l$ & $e^{\frac{i2l\pi}{3}}$ & $e^{\frac{i2(l-1)\pi}{3}}$ & $e^{\frac{i2(l+1)\pi}{3}}$\\
\hline\hline
\end{tabular}
\caption{Eigenvalues for $C_3$ rotation operator at $\bd{\Gamma,K}$ and $\bd{K'}$ points for Wannier orbitals located at the maximal Wyckoff positions with angular momentum $l$ in $C_3$ symmetric lattices. For each position, $l=0,1,2$.}
\label{Tab:Wy_C3}
\end{table}
Comparing the band representation of each Wyckoff position with the band representation of generators, we find the generators for $C_6$ and $C_3$ symmetric classifications can be decomposed as,
\begin{align}
&\left.\begin{array}{ccc}
h_{4b}^{(6)}&\sim & b_1+b_2+b_1'+b_2'\\
h_{3c}^{(3)}&\sim & c_1+c'_1+c''_1
\end{array}\right\} C_6\,\,\textrm{Symmetry}
\\
&\left.\begin{array}{ccc}
h_{2b}^{(3)}&\sim & b_1+b_2\\
h_{2c}^{(3)}&\sim & c_1 + c_2
\end{array}\right\} C_3\,\,\textrm{Symmetry}.
\end{align}
Since the coefficients for each Wannier orbital are positive, all generators in $C_3$ and $C_6$ symmetric classifications are Wannier representable.

\section{Numeric simulation of a TCI with fractional corner charge }
\label{sec:NumericSimulation}
\begin{figure}[!t]
	\centering
    \includegraphics[width=\columnwidth]{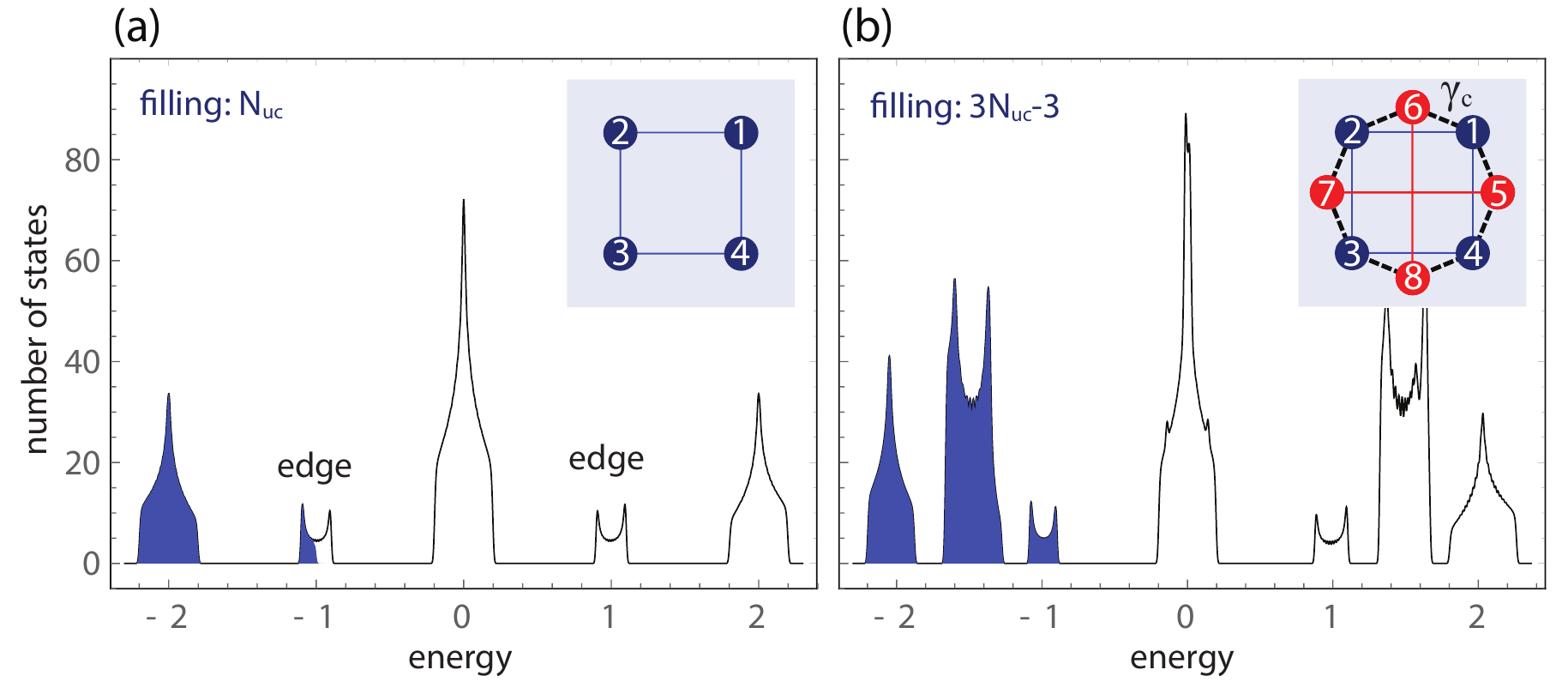}
	\caption{(a) Density of states for the lattice of $H^{(4)}_1({\bf k}, t=0.1)$ [Eq.~\eqref{equation:h1b}] with full open boundaries and with $N_{uc}$ unit cells. Intra-cell hopping terms are added as shown in the unit cell in the inset. The dark blue region denotes the filling at $N_{uc}$. (b) Similar density of states, but for Hamiltonian $H^{(4)}(\kk)$ [Eq.~\eqref{eq:simulation}]. The inset shows a unit cell and its intra-cell hopping terms. Numbers 1-4 (5-8) label the sites corresponding to $H^{(4)}_1({\bf k})$ [$H^{(4)}_3({\bf k})$]. Dark blue regions in the main plot denote the filling at $3N^2-3$. For the sake of clarity, we isolate edge energy bands from bulk energy bands by setting the inter-cell hoppings to 1 for $H^{(4)}_1(\kk)$ and to 1.5 for  $H^{(4)}_3(\kk)$. All intra-cell hoppings are set to 0.1.}
	\label{fig:simulation}
\end{figure}

In this section, we describe the simulation of the Hamiltonian that has charge density as indicated in Fig.~1(a) of the Main Text. As a starting point, we simulate $H^{(4)}_1({\bf k}, t=0.1)$ [Eq.~\eqref{equation:h1b}], which is in the same topological phase as the primitive generator $h^{(4)}_{1b}$ [indeed, $h^{(4)}_{1b}=H^{(4)}_1({\bf k}, t=0)$]. At $\frac{1}{4}$-filling, there is one electron per unit cell and its Wannier center is located at the maximal Wyckoff position $b$. Consequently, ${\bf P}=(\frac{e}{2},\frac{e}{2})$. Additionally, this Hamiltonian has a nominal corner charge of $\frac{e}{4}$, as seen in Table 1 of the Main Text. However, this TCI has gapless edges at this filling. This can be seen in the density of states plot of Fig.~\ref{fig:simulation}(a). Due to the existence of these metallic edges, the corner charge is ill-defined at this filling. To have well defined corner charges, the total polarization has to vanish. For that purpose, we stack the TCI $H^{(2)}_{3}({\bf k}, t_x=0.1,t_y=0.1)$ [Eq.~\eqref{Eq:C4_3}] at $\frac{1}{2}$-filling, which is deformable to the primitive generator $h^{(4)}_{2c}$ [indeed, $h^{(4)}_{2c}=H^{(2)}_3({\bf k}, t_x=0,t_y=0)$], and therefore has ${\bf P}=(\frac{e}{2},\frac{e}{2})$ but zero nominal corner charge. The overall Hamiltonian is
\begin{align}
H^{(4)}({\bf k})=\begin{pmatrix}
H^{(4)}_{1}({\bf k}, t=0.1) & \gamma_c\\
\gamma_c^\dagger & H^{(2)}_{3}({\bf k}, t_x = t_y = 0.1)
\end{pmatrix}
\label{eq:simulation}
\end{align}
The coupling terms $\gamma_c$ in principle can be any terms within the unit cell that respect $C_4$ symmetry and do not close the gap. We choose the  $\gamma_c$ hoppings to be those illustrated by the black dashed lines in the inset of Fig.~\ref{fig:simulation}(b) and we set their amplitude to $0.1$. Under these conditions, The Hamiltonian in Eq.~\ref{eq:simulation} is in the same class as the Hamiltonian in Eq.~(8) of the Main Text. Fig.~\ref{fig:simulation}(b) shows the density of states for the Hamiltonian in Eq.~\eqref{eq:simulation} when the lowest three bands are filled. Notice that this time the bulk and edge bands are fully filled, so that we have insulating bulk \emph{and} edges. The charge density at this filling is shown in Fig.~1(a) of the Main Text. As indicated in the Main Text, the other fractional charges in Fig.~1 are obtained by similar procedures.

\subsection{Breakdown of charge quantization at corners}
We added perturbations to the Hamiltonian in Eq.~\eqref{eq:simulation} to numerically verify that the corner charge remains quantized when the relevant symmetries are preserved and that quantization is lost when they are not. The perturbation Hamiltonian has the form
\begin{align}
h({\bf k}) = E_1 \cos k_x + E_2 \cos k_y + O_1 \sin k_x + O_2 \sin k_y
\label{eq:PertHamiltonian}
\end{align}
where $E_{1,2}$ and $O_{1,2}$ are 8 $\times$ 8 random matrices subject to the constraints imposed by symmetries. Due to time reversal symmetry,
\begin{align}
h^*(k_x,k_y) = h(-k_x,-k_y), \nonumber
\end{align}
we require
\begin{align}
E^*_{1,2} = E_{1,2} \quad O^*_{1,2} = -O_{1,2}.
\end{align} 
Additionally, to preserve $C_2$ symmetry,
\begin{align}
\hat{r}_2 h(k_x,k_y) \hat{r}^\dagger_2 = h(-k_x,-k_y), \nonumber
\end{align}
we require
\begin{align}
[\hat{r}_2, E_{1,2}]=0 \quad \{\hat{r}_2,O_{1,2}\}=0.
\label{eq:PertC2conditions}
\end{align}
Finally, to impose $C_4$ symmetry,
\begin{align}
\hat{r}_4 h(k_x,k_y) \hat{r}^\dagger_4 = h(k_y,-k_x), \nonumber
\end{align}
we require Eq.~\eqref{eq:PertC2conditions} in addition to 
\begin{align}
E_2 = \hat{r}_4 E_1 \hat{r}^\dagger_4 \quad O_2 = \hat{r}_4 O_1 \hat{r}^\dagger_4.
\end{align}
\begin{figure}[!t]
	\centering
    \includegraphics[width=\columnwidth]{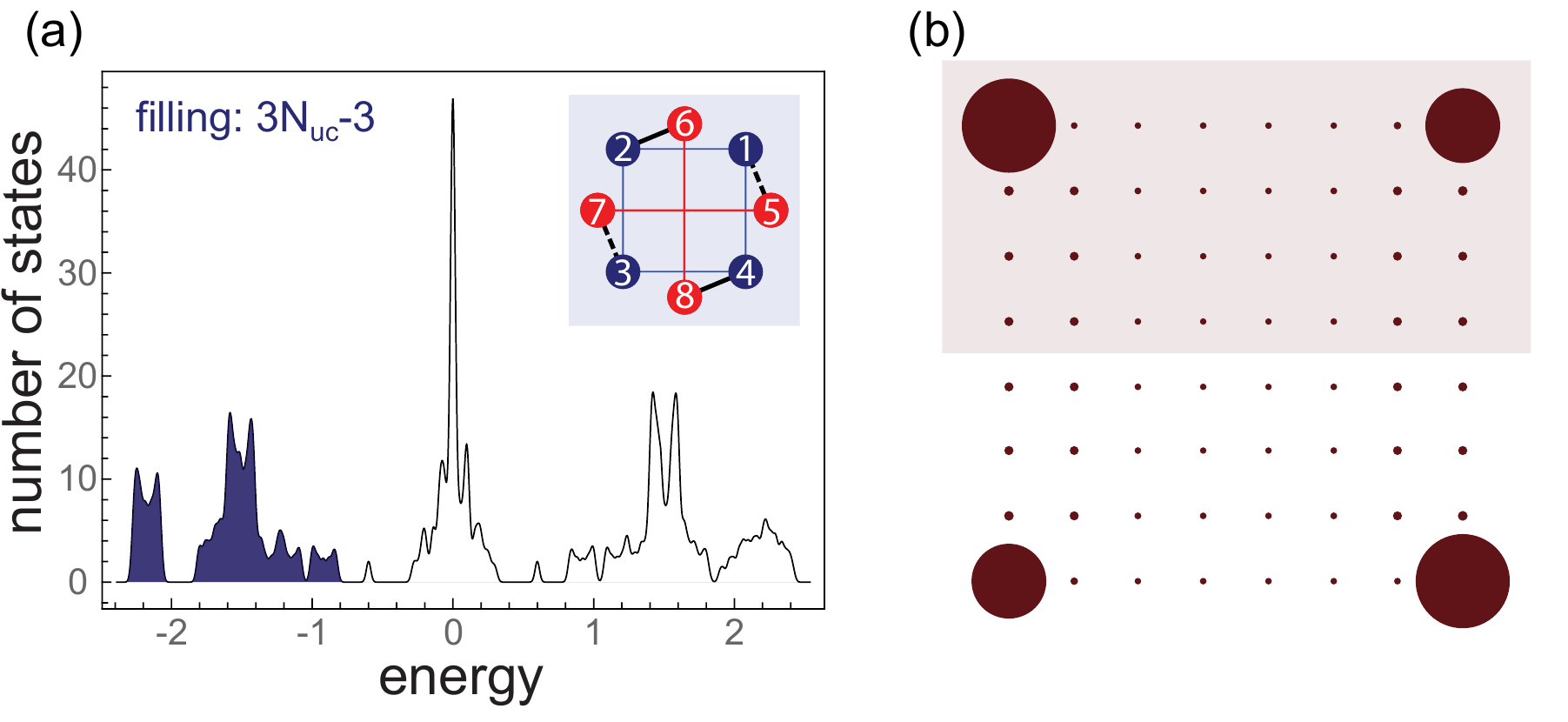}
	\caption{(a) Density of states for $H^{(4)}_1({\bf k}, t=0.1)\oplus H^{(4)}_3({\bf k}, t=0.1)$ with a $C_4$ symmetry breaking coupling terms as shown in the inset. In the inset, 1-4 (5-8) label $H^{(4)}_1({\bf k})$ [$H^{(4)}_3({\bf k})$] degrees of freedom. The solid and dashed black lines represent the coupling between  $H^{(4)}_1({\bf k})$ and $H^{(4)}_3({\bf k})$, whose amplitudes are set to be $0.1$ and $0.4$ to break the $C_4$ symmetry. Blue regions denote the filling at $3N_{uc}-3$. (b) The charge density for the same system at the filling of $3N_{uc}-3$. Each dot represents the charge density per one unit cell and the size of a dot is proportional to the absolute value of the net charge (total electronic charge subtracted by the background bulk electrons of $3e$ per unit cell).}
	\label{fig:corner_c2}
\end{figure}
When the perturbation in Eq.~$\eqref{eq:PertHamiltonian}$ preserves TRS and $C_4$ symmetry, the electronic corner charge remains quantized. However, when the perturbation breaks $C_4$ symmetry down to only $C_2$ symmetry, the quantization at each corner is lost. Fig.~\ref{fig:corner_c2} shows an example of non-quantized corner charge when the coupling terms $\gamma_c$ break $C_4$ symmetry down to only $C_2$ symmetry, as shown in the inset of Fig.~\ref{fig:corner_c2}(a). The charge density for this configuration is show in Fig.~\ref{fig:corner_c2}(b) for a filling of $3N_{uc}-3$ states (here $N_{uc}$ is the number of unit cells in the lattice). The corner charge, arranged in a $C_2$-symmetric pattern are \emph{not} quantized at each corner. However, the charge over a half of the lattice [shaded area in Fig.~\ref{fig:corner_c2}(b)], which contains two corners, is quantized to be $\frac{3e}{2}$.

\begin{figure}[!t]
	\centering
   \includegraphics[width=\columnwidth]{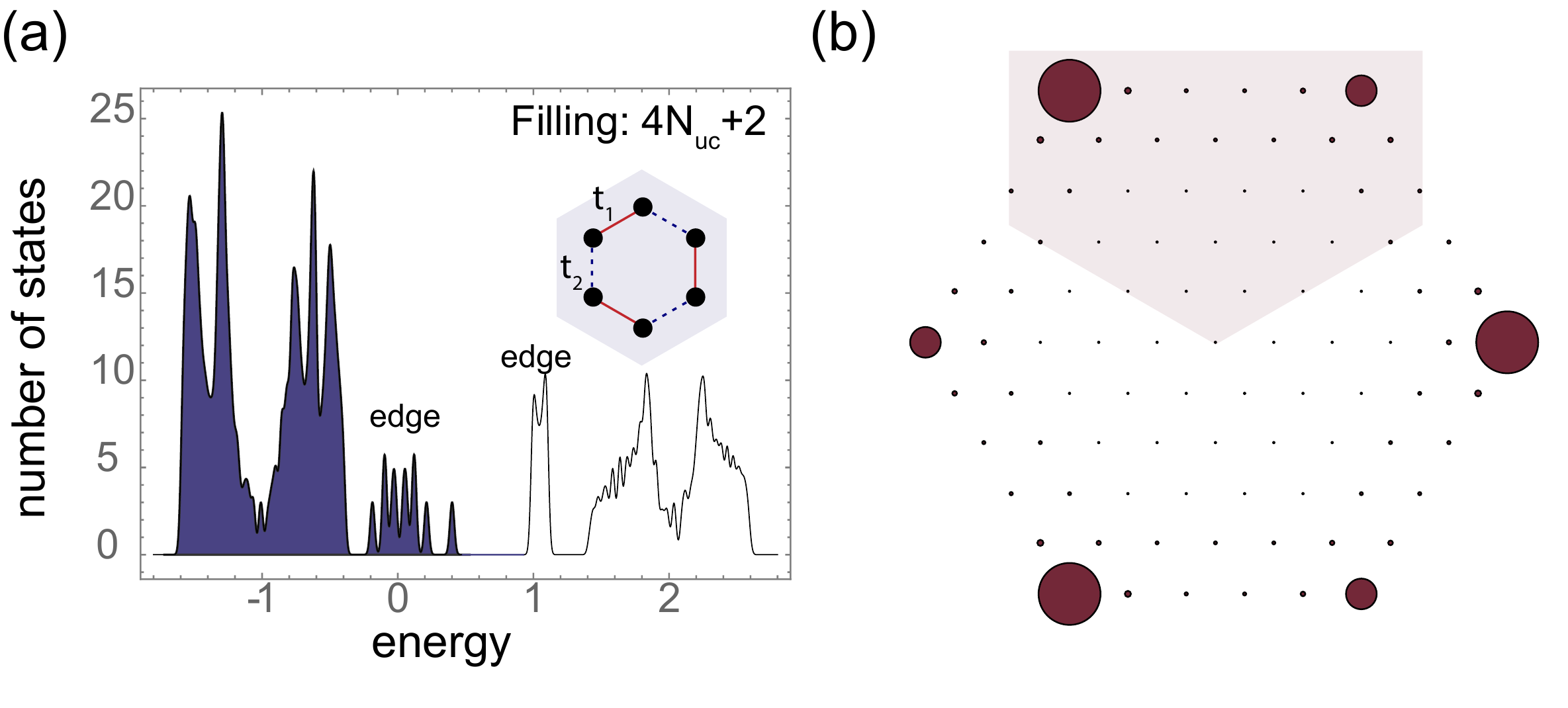}
	\caption{(a) Density of states for the generator $H^{(6)}_1({\bf k})$ with the modified intra-cell hopping terms as shown in the inset. $t_1$ and $t_2$ are set to be $0.1$ and $0.3$ respectively, which breaks the $C_6$ symmetry down to $C_3$ symmetry. (b) Charge density for the same system at the filling of $4N_{uc}+2$. Each dot represents the charge density per one unit cell and the size of the dot is proportional to the absolute value of the net charge (total electronic charge subtracted by the background bulk electrons of $4e$ per unit cell).}
	\label{simulations:corner_c3}
\end{figure}

Another case in which corner charge is not quantized occurs if a hexagonal lattice (a lattice with six equal sectors subtended by $\frac{2\pi}{6}$ rad) hosts a $C_3$-symmetric Hamiltonian. An example of this is shown in Fig.~\ref{simulations:corner_c3} for $H_1^{(6)}(\bf{k},t)$[Eq.~\eqref{eq:H_61}] with the intra-cell hopping terms being modified as illustrated in the inset of Fig.~\ref{simulations:corner_c3}(a): when $t_1\neq t_2$, the intra-cell hopping terms break $C_6$ symmetry down to $C_3$ symmetry. Fig.~\ref{simulations:corner_c3}(a) shows the density of states for a filling of $4N_{uc}+2$ states. The corresponding charge density is shown in Fig.~\ref{simulations:corner_c3}(b). The total extra charge of $2e$ localizes at the corners of the hexagonal lattice forming a $C_3$-symmetric pattern. However, the sum of charge over a $\frac{2\pi}{3}$ sector [shaded area in Fig.~\ref{simulations:corner_c3}(b)] is quantized to $\frac{2e}{3}$.

\section{Microscopic theory\\ of the corner charge}
\label{sec:QuantCornerCharge}
In the Main Text, we saw that the existence of a filling anomaly can be extracted from the positions of the electrons within the unit cell in obstructed atomic limits. That lead to the conclusion that in a lattice with global $C_n$ symmetry that hosts a $C_m$ symmetric Hamiltonian, for $n=4$, $m=4,2$ or $n=6$, $m=6,3,2$, the charge over sectors subtended by an angle of $\frac{2\pi}{m}$ are quantized in fractions of $e$ according to the secondary indices in Eq.~(10) of the Main Text. From the notion of filling anomalies and charge densities, we derived a pictorial representation of the bulk-boundary correspondence, by which the charge at any unit cell is given by the (possibly fractional) number of electrons that fall into it. Here, we elaborate on this idea to include the shape of the Wannier orbitals into this prescription. With this addition, it is possible to have a microscopic understanding of the situations in which $C_2$ and $C_3$ symmetries do not quantize charges at individual corners when embedded in lattices with global $C_4$ or $C_6$ symmetries (i.e. having 4 and 6 corners), respectively. 

\begin{figure}[!t]
	\centering
    \includegraphics[width=\columnwidth]{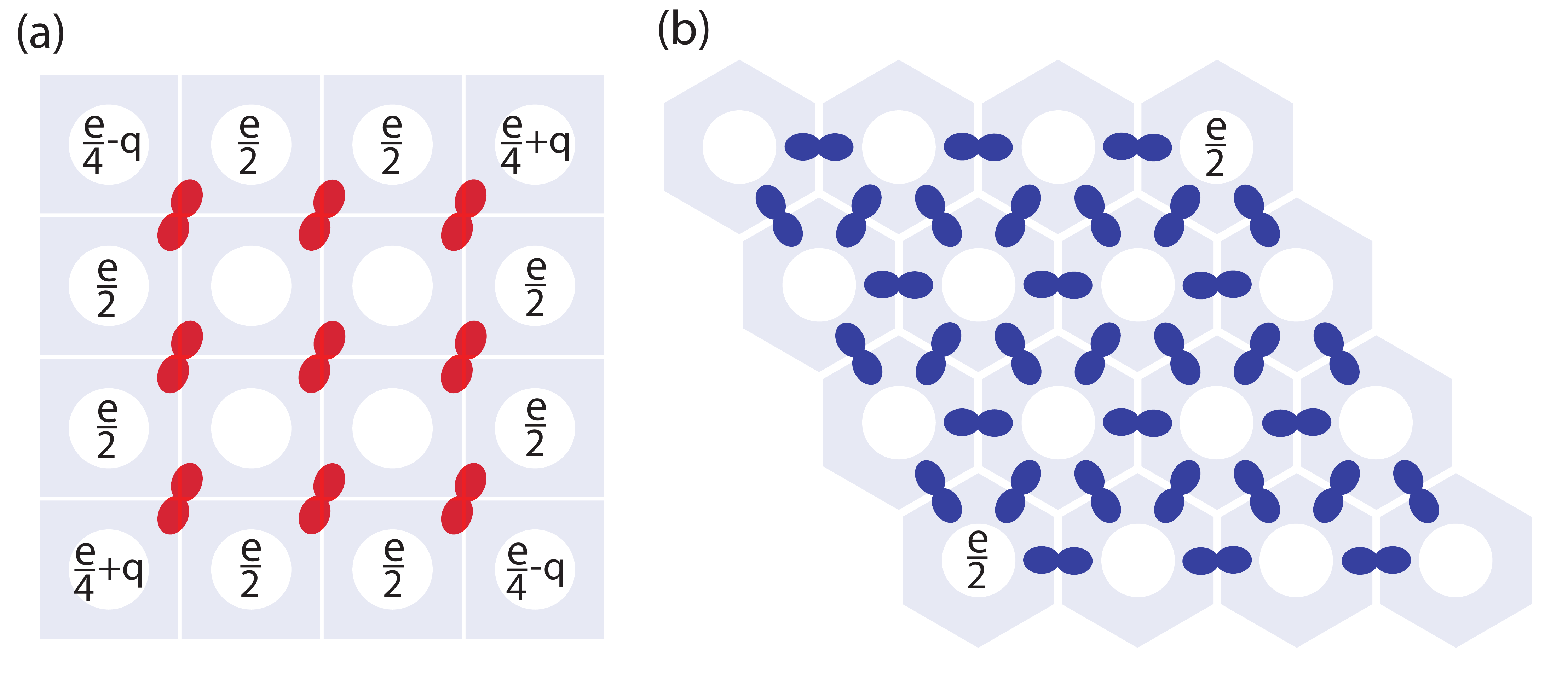}
	\caption{Orbitals in $C_2$-symmetric lattices having (a) one electron at Wyckoff position $b$, (b) three electrons, at Wyckoff positions $c$, $c''$, and $c'''$. Case (a) leads to non-quantized nominal corner charge. Case (b) leads to quantized corner charge at the two $120^\circ$ corners~\cite{noh2018}.}
	\label{fig:C2Orbitals}
\end{figure}
\begin{figure}[t]
	\centering
    \includegraphics[width=\columnwidth]{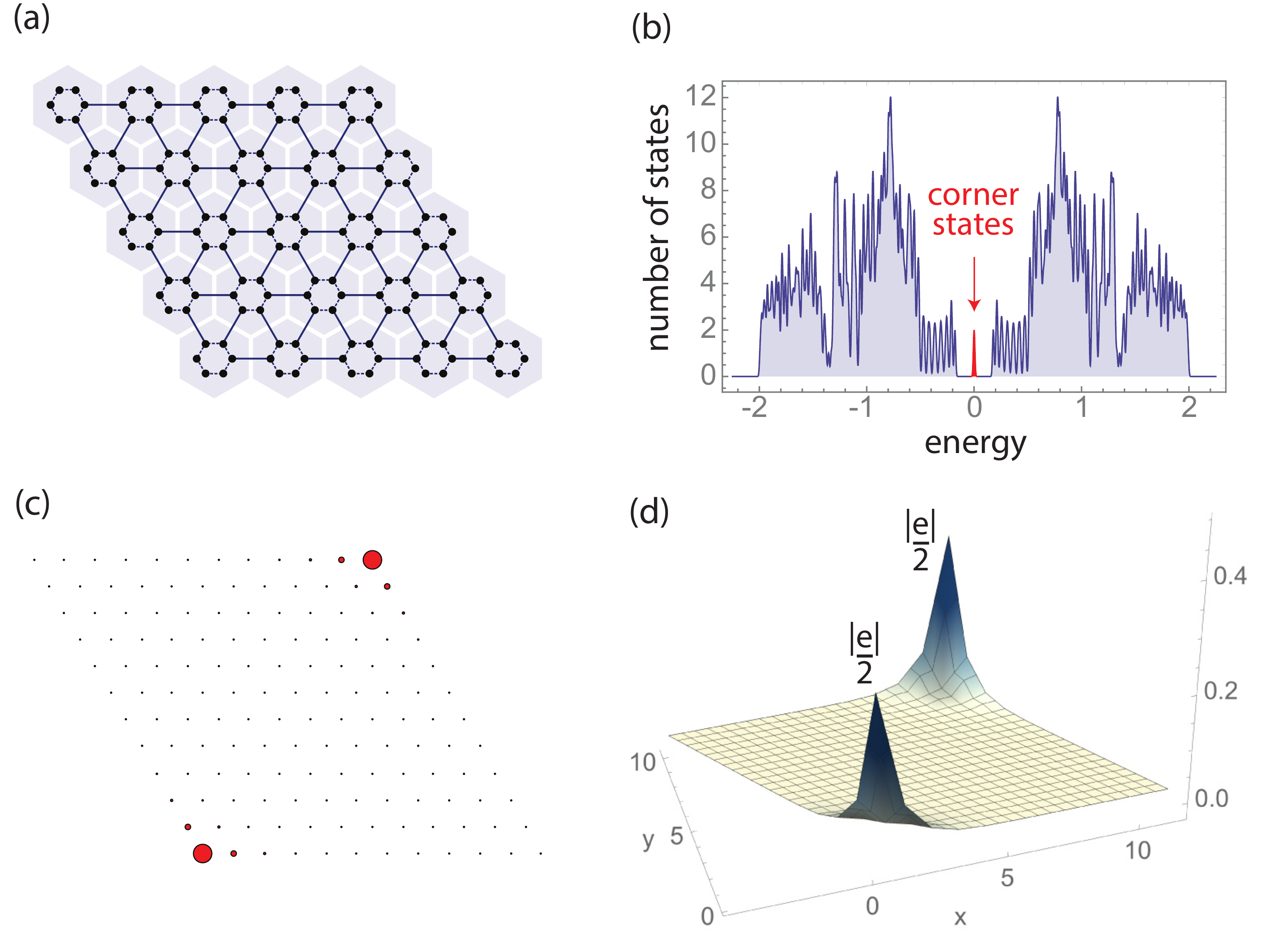}
	\caption{An example of charge quantization due to $C_2$ symmetry. (a) Lattice of $H^{(6)}_2({\bf k},t)$ [Eq.~\eqref{H_c6_2}] in a parallelogram. (b) Density of states for a simulation of the lattice in (a) with $t_0=0.5$. There are two states at zero energy. (c) Probability density of the two zero energy states. The two states exponentially localize at two opposite corners. (d) The charge density for this configuration. Each corner has a charge of $\frac{e}{2}$ when the Fermi level fills the lowest three bands but excludes the zero energy.}
	\label{fig:C2Quantization}
\end{figure}
We start by studying the case of $C_2$-symmetry. Fig.~\ref{fig:C2Orbitals} shows two cases: one in which $C_2$ symmetry does not quantize the charge at individual corners and one in which it does. Notice that in both cases, the lattice has $C_2$ symmetry, and thus each half of the lattice will have charge quantization in multiples of $\frac{e}{2}$.

In Fig.~\ref{fig:C2Orbitals}(a) we have one electron at Wyckoff position $b$. We have drawn Wannier orbitals that reflect its $C_2$ symmetry. To count charge at a unit cell, we count the fraction of the electronic charge that falls inside that unit cell. Since the unit cells cut the Wannier orbitals in 4 quadrants, two opposite quadrants will have a charge of $\frac{e}{4}-q$ and the other a charge of $\frac{e}{4}+q$, for any value of $q \in [-\frac{1}{2},\frac{1}{2}]$. Correspondingly, corner unit cells will have the same unquantized charge. Edge charge, on the other hand, remains quantized, because the charge contribution to each edge unit cell comes from two quadrants of the Wannier orbitals, one having charge $\frac{e}{4}+q$ and the other one having charge $\frac{e}{4}-q$.

In Fig~\ref{fig:C2Orbitals}(b) we show the case of $H^{(6)}_2({\bf k},t)$ [Eq.~\eqref{H_c6_2}] in the lattice of a parallelogram. In this lattice, two corners have fractional charges of $\frac{e}{2}$ while the other two have vanishing corner charge. The quantization of corner charge in this case occurs because the unit cells only cut the Wannier orbitals in two. Furthermore, this model has chiral symmetry, and the corner charges are associated with zero energy corner localized states~\cite{noh2018}. 

A short analysis of the features of this lattice is shown in Fig.~\ref{fig:C2Quantization}. A more detailed description of the states and their protection for this paralelogram lattice can be found in the Supplementary Information of Ref.~\onlinecite{noh2018}.

The two examples in Fig.~\ref{fig:C2Orbitals} lead to the conclusion that, in lattices with 4 corners, $C_2$ symmetry \emph{can} quantize the charge at individual corners, but the quantization is not guaranteed by its mere presence. Microscopically, the lack of quantization can occur when unit cells in the lattice cut the Wannier orbitals in more parts than the number of symmetry-related sectors of the orbital.

\begin{figure}[t]
	\centering
   \includegraphics[width=\columnwidth]{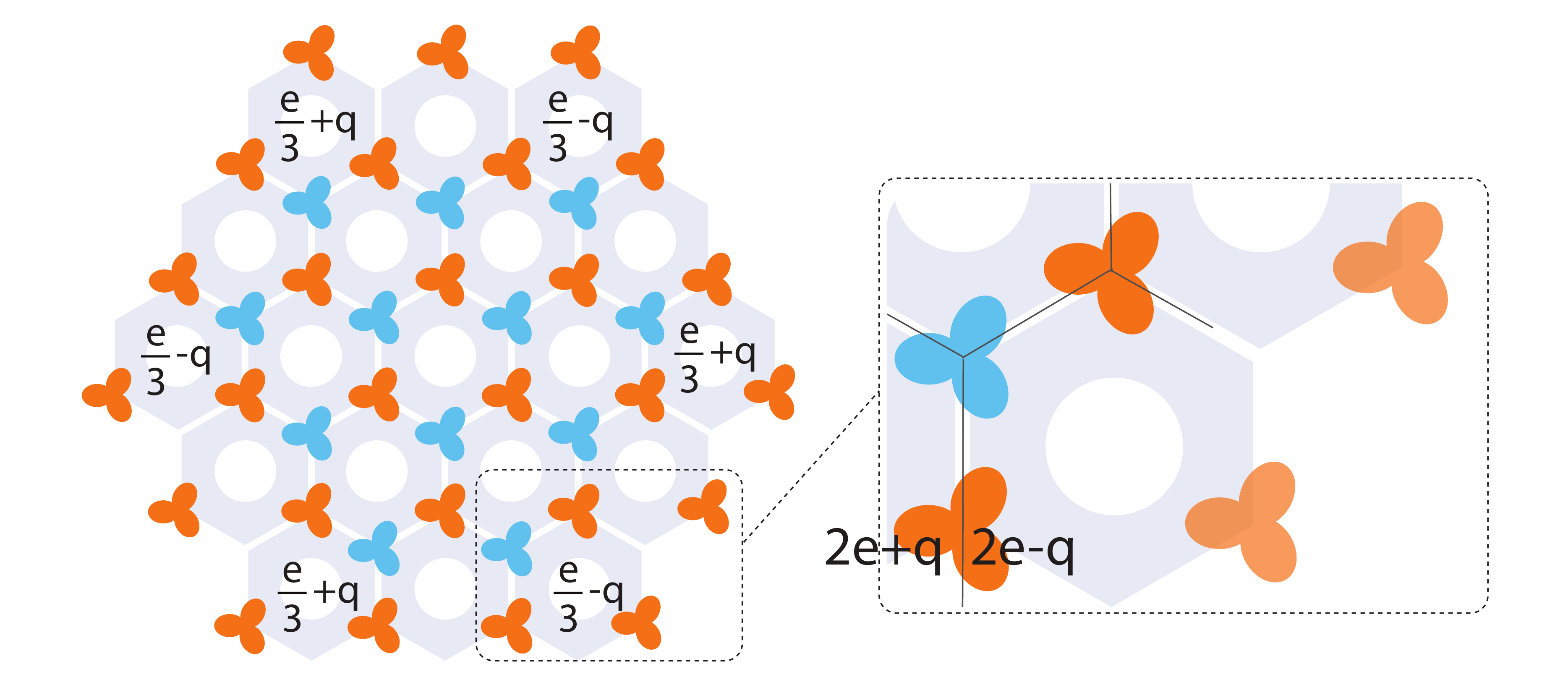}
	\caption{Orbitals in a hexagonal lattice having two electrons at Wyckoff position $b$ (orange orbitals) and two electrons at Wyckoff position $c$ (cyan orbitals). Extra terms are added to the Hamiltonians to break $C_6$ symmetry down to $C_3$ symmetry.}
	\label{orbital_c3}
\end{figure}

We now apply the same criteria to $C_3$ symmetric TCIs. Consider stacking generators $h_{2b}^{(3)}$ and $h_{2c}^{(3)}$ (either of these generators separately do not have non-vanishing polarization). The bulk Hamiltonian is originally $C_6$ symmetric, but terms are added to break this symmetry down to only $C_3$ symmetry [For example, adding the intra-cell hopping terms as shown in the inset of Fig.~\ref{simulations:corner_c3}(a)]. Fig.~\ref{orbital_c3}(a) shows the Wannier orbitals for such a model on a hexagonal lattice.  
It has two electrons located at Wyckoff position $b$ and two electrons located at Wyckoff position $c$. In the bulk, they are located at both Wyckoff positions $b$ and $c$, while at the edge they prefer to occupy either Wyckoff position $b$ or $c$.
Without loss of generality, we choose the Wyckoff position $b$ to illustrate how the breakdown of quantization at individual corners comes about. The contribution to corner charge comes from both bulk and edge Wannier orbitals. While bulk orbitals are cut in three parts, edge orbitals are cut only in two parts. Since only one edge Wannier orbital contributes to corner charge, these edge orbitals un-quantize the corner charge. The charge over two adjacent corners is quantized to $\frac{2e}{3}$, however, in agreement with the secondary indices in Eq~10 of the Main Text. 

\section{Localization of the fractional \\corner charge}
\label{sec:LocalizationOfCornerCharge}
In this section, we will take the model $H_1^{(6)}(\bd{k},t_0)$ [Eq.~\eqref{eq:H_61}] as an example to show that the fractional charge is exponentially localized at the corner. We put the model on a finite hexagon with $7$ unit cells on each edge. The decay of the fractional corner charge $Q(\bd{r})$ should follow an exponential law $Q(\bd{r})\propto e^{-\alpha \abs{\bd{r}}}$, where $\alpha$ is a constant that depends on the ratio between intra-cell hopping amplitude and inter-cell hopping amplitude $t_0$ (we have set the inter-cell hopping strength to be 1).

\begin{figure}[b]
\centering
\includegraphics[width=\columnwidth]{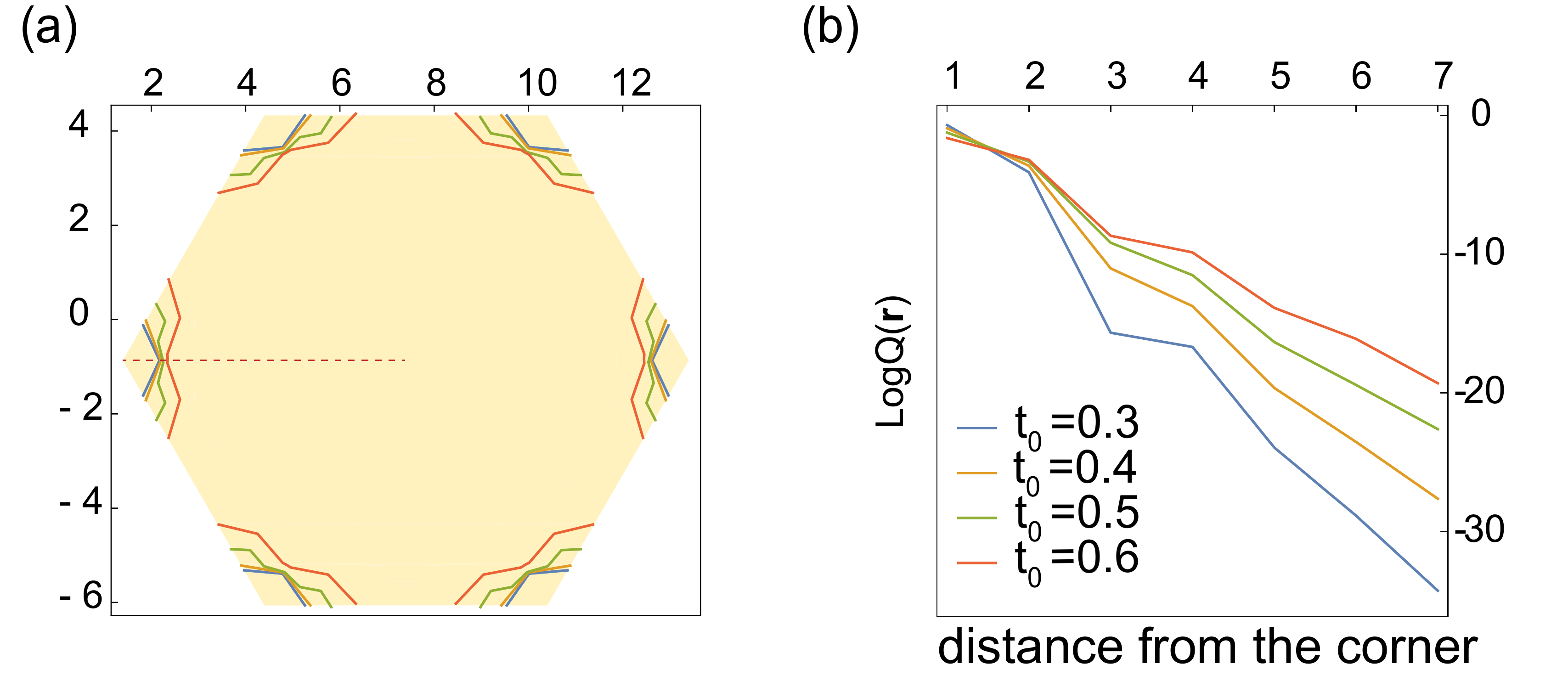}
\caption{(a) Contour plot of charge density $Q(\bd{r})$ for $H_1^{(6)}(\kk)$ [Eq.~\eqref{eq:H_61}] on a hexagon with $7$ unit cells on each edge. Blue, orange, green and red lines identify points where $Q(\bd{r})=Q_{max}/4$ for the intra- and inter-cell hopping amplitude ratio $t_0= 0.3, 0.4, 0.5, 0.6$, respectively. The legends of each line are shown in plot (b). (b) Plot of $\log Q(\bd{r})$ along the red dashed line in (a) as a function of distance from the corner (in unit cells).}
\label{fig:localization}
\end{figure}

In the extremely dimerized limit, $t_0=0$, the fractional charge will be fully localized at the corner unit cell, hence $\alpha\rightarrow \infty$. When turning on the intra-cell hoppings, the electrons penetrate into the bulk from the corner unit cell due to the tunneling and $\alpha$ is a finite constant, i.e. as one increases $t_0$, $\alpha$ decreases. In Fig.~\ref{fig:localization}~(a) we show the contour plot of the charge density at $Q=\frac{1}{4}Q^{max}$ for intra-cell and inter-cell hopping amplitude ratio $t_0=0.3,0.4,0.5,0.6$.   
It is clear that as $t_0$ increases, the charge density spreads more and more into the bulk. In Fig.~\ref{fig:localization}~(b), we show the charge density along the dashed red line in Fig.~\ref{fig:localization}~(a) in a log scale, $\log Q(\bd{r})\propto -\alpha \abs{\bd{r}}$. The charge density of different hopping strength ratios linearly depends on the distance from the corner unit cell. Therefore, the fractional charge is indeed exponentially localized at the corner unit cell.

\section{Fractional corner charge\\ in fragile TCIs}
\label{sec:Fragile}
In this section, we show a concrete example of a fragile topological TCI to verify that our indices indeed predict the correct corner charges. In the recent work of Ref.~\onlinecite{liu2018shift}, Liu $et\,\,al$ proposed a $C_6$-symmetric model that generates a series of fragile topological phases. It is constructed by stacking two Haldane models~\cite{haldane1988} with opposite Chern number $\pm 1$ and $p_x\pm i p_y$ orbitals on the lattice sites,
\begin{align}
H^{(6)}_s(t,\lambda)&=
\left(
\begin{array}{cc}
H_{h}(t,\lambda)& \gamma\\
\gamma^\dagger & H_{h}(t,-\lambda)
\end{array}
\right),
\label{eq:Hshift}
\end{align}
where $H_{h}(t,\lambda)$ represents the Haldane model with Chern number $-\textrm{sgn}(\lambda)$ and the $\gamma$ term corresponds to $C_6$ symmetry-preserving coupling terms between the two Haldane models~\cite{haldane1988}. In order to get a finite $C_6$-symmetric hexagonal configuration without cutting in between unit cells, we use an hexagonal unit cell that contains $6$ sites in the honeycomb lattice. Our block Hamiltonian is then the $12$-band model of Eq.~\eqref{eq:Hshift}, where each copy of the Haldane model has Bloch Hamiltonian
\begin{align}
\label{H_haldane}
H_{h}(t,\lambda)=&
\setlength{\arraycolsep}{0.pt}
-t\begin{pmatrix}
0  & 1 &  0 & e^{i\bd{k}\cdot \bd{a}_1} &0 & 1 \\
1 & 0 & 1  &  0 & e^{i\bd{k}\cdot \bd{a}_2} & 0\\
0 & 1 & 0 & 1 & 0 &  e^{-i\bd{k}\cdot \bd{a}_3}\\
e^{-i\bd{k}\cdot \bd{a}_1} & 0 & 1 & 0 & 1 &  0\\
0 & e^{-i\bd{k}\cdot \bd{a}_2} & 0 & 1 & 0 & 1\\
1 & 0 & e^{i\bd{k}\cdot \bd{a}_3} & 0 & 1 & 0
\end{pmatrix}\nonumber\\
&+\begin{pmatrix}
0  & 0 &  Q_1 & 0 &-Q_2 & 0 \\
0 & 0 & 0   &  Q_2 & 0 & -Q_3\\
Q_1^* & 0 & 0 & 0 & Q_3 & 0\\
0 & Q_2^* & 0 & 0 & 0 &  -Q_1^*\\
-Q_2^* & 0 & Q_3^* & 0 & 0 & 0\\\
0 & -Q_3^* & 0 & -Q_1 & 0 & 0
\end{pmatrix},\nonumber\\
Q_1=&i\lambda(1+e^{i\kk\cdot \bd{a}_1}+e^{i\kk\cdot \bd{a}_3})\nonumber\\
Q_2=&i\lambda(1+e^{i\kk\cdot \bd{a}_1}+e^{i\kk\cdot \bd{a}_2})\nonumber\\
Q_3=&i\lambda(1+e^{-i\kk\cdot \bd{a}_3}+e^{i\kk\cdot \bd{a}_2}),
\end{align}
and the coupling matrix $\gamma$ is
\begin{align}
\gamma=J
\begin{pmatrix}
0  & e^{-\frac{2i\pi}{3}} &  0 & 0 & 0 & e^{\frac{2i\pi}{3}} \\
e^{-\frac{2i\pi}{3}} & 0 & 1  &  0 & 0 & 0\\
0 & 1 & 0 & e^{\frac{2i\pi}{3}} & 0 &  0\\
0 & 0 & e^{\frac{2i\pi}{3}} & 0 & e^{-\frac{2i\pi}{3}} &  0\\
0 & 0 & 0 & e^{-\frac{2i\pi}{3}} & 0 & 1\\
e^{\frac{2i\pi}{3}} & 0 & 0 & 0 & 1 & 0
\end{pmatrix}
\end{align}
The $C_6$ rotation operator is
\begin{align}
\hat{r}_{6,s}=
\begin{pmatrix}
e^{-i\frac{\pi}{3}}\hat{r}_6 & 0\\
0 & e^{i\frac{\pi}{3}}\hat{r}_6
\end{pmatrix},
\end{align}
where $\hat{r}_6$ is defined in Eq.~\ref{rot_c6} and the phase factors are due to the opposite orbital types for each Haldane model. The band structure is gapped at $\frac{1}{2}$-filling. We chose the phase with both $t$ and $\lambda$ to be positive and calculate the rotation eigenvalues for the $6$ bands below the gap. The results are summarized in Table~\ref{Tab:shift}. 

\begin{table}[th]
\centering
\begin{tabular}{c|ccc}
\hline\hline
HSPs & $\bd{\Gamma}$ & $\bd{K}$ & $\bd{M}$\\\hline
\multirow{2}{*}{$ (t=1,\lambda=0.2)$}  
& $\pm1,\pm1$ 
                     & $1,e^{\pm\frac{i2\pi}{3}}$ 
              		 & $1,\pm1$\\
& $e^{\pm\frac{i\pi}{3}}$ 
                     & $1,e^{\pm\frac{i2\pi}{3}}$ 
              		 & $1,\pm1$\\
\hline\hline
\end{tabular}
\caption{rotation eigenvalues of the occupied energy bands at $\bd{\Gamma}$ ($C_6$ eigenvalues), $\bd{K}$ ($C_3$ eigenvalues) and $\bd{M}$ ($C_2$ eigenvalues) points of the BZ for the fragile TCI with Bloch Hamiltonian in Eq.~\eqref{eq:Hshift}.}
\label{Tab:shift}
\end{table}
The rotation representation in Table~\ref{Tab:shift} is incompatible with those from any band representations induced by atomic insulators (see Table~\ref{Tab:Wy_C6} for all possible such representations). Thus, there is an obstruction to construct Wannier representations in this phase. However, if we add to this fragile phase the bands of an atomic insulator $3c_0+2c_1$ (here, we denote the atomic insulators by the locations of their Wannier orbitals subindexed with their corresponding angular momentum, as in Section~\ref{sec:BandRep}) -- which has $3$ Wannier orbitals with angular momentum $0$ and $2$ Wannier orbitals with angular momentum $1$, located at Wyckoff positions $c,c'$ and $c''$ respectively -- the combined TCI is deformable to the atomic insulator $4b_0+3b_1+3b_2+a_0$.  Hence, the fragile TCI in Eq.~\ref{eq:Hshift} can be expressed as
\begin{align}
FT_{s}\sim4b_0+3b_1+3b_2-3c_0-2c_1+a_0.
\end{align}
\begin{figure}[th]
\centering
\includegraphics[width=\columnwidth]{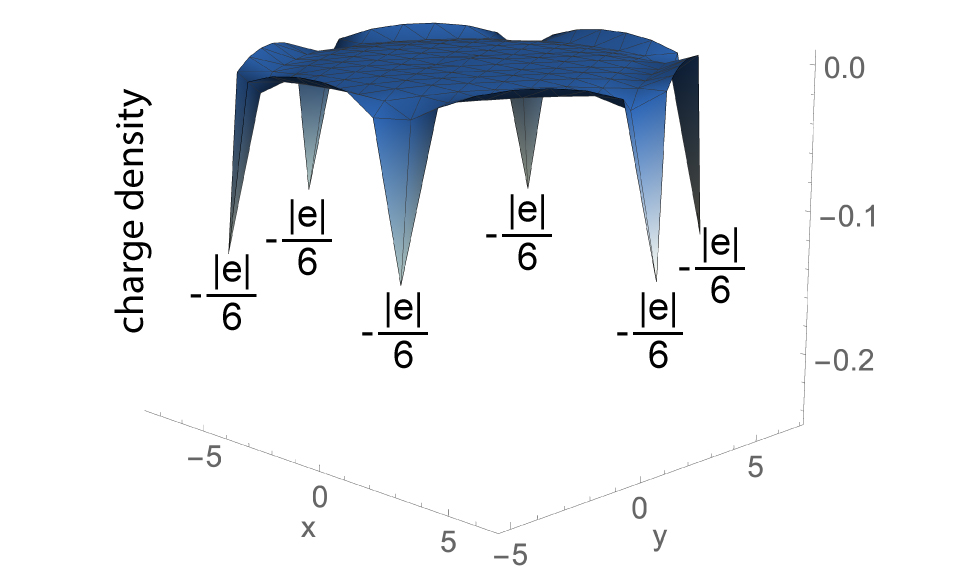}
\caption{The total charge density for the fragile TCI with Bloch Hamiltonian in Eq.~\eqref{eq:Hshift} on a finite hexagonal lattice at a filling of $6N_{uc}+1$.}
\label{fig:shift}
\end{figure} 
According to the pictorial counting method of the corner charge described in Section~IV of the Main Text, Wannier orbitals at Wyckoff positions $b$ and $b'$contribute fractional electronic corner charge of $\frac{2e}{3}$ [Fig.~6(c) in the Main Text] and Wannier orbitals at Wyckoff positions $c,c'$ and $c''$ contribute fractional electronic corner charge of $\frac{e}{2}$ [Fig.~6(d) in the Main Text]. Therefore we expect an electronic corner charge of $\frac{e}{6}$ for this fragile phase.
From the rotation representations in Table~\ref{Tab:shift}, the fragile phase is in class
\begin{align}
\chi^{(6)}=(2,-2). 
\end{align}
For these values of rotation invariants, the secondary indices (Eq.~11 in the Main Text) indeed predict a corner charge of $\frac{e}{6}$. We then numerically simulate this model on a finite hexagonal lattice. The resulting energy spectrum for the system with $N_{uc}$ unit cells is gapped at a filling of $6N_{uc}+1$. In Fig.~\ref{fig:shift}, we show the total (ionic and electronic) charge density at that filling. Since the filling anomaly is $-1$, the integral of total charge density over each of the symmetry-related sectors is equal to $-\frac{e}{6}$, matching the prediction of the secondary index.

\section{Charge density in neutral insulators}
\label{sec:neutrality}
\begin{figure}[b]
\centering
\includegraphics[width=\columnwidth]{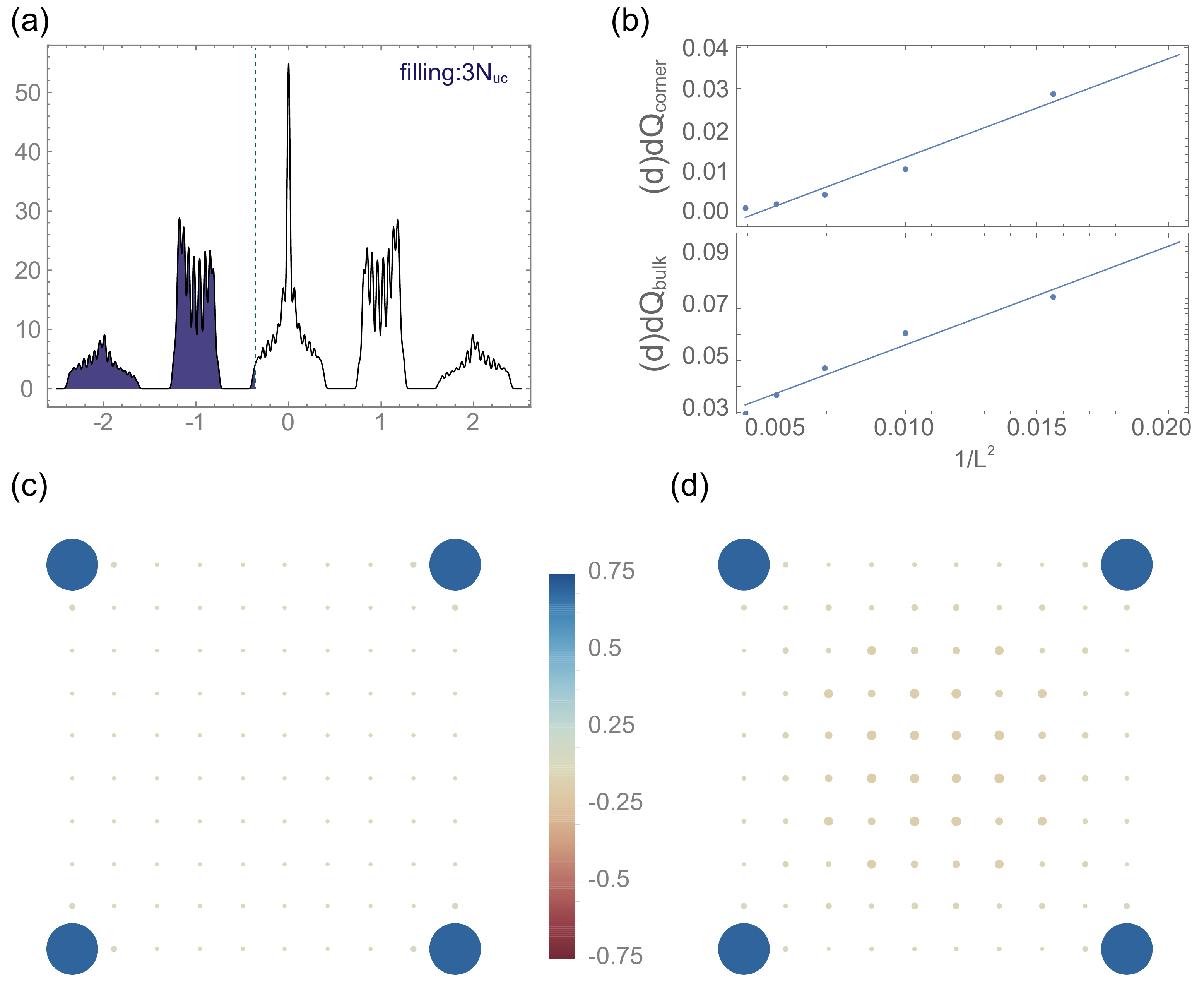}
\caption{(a) Density of states for Hamiltonian in Eq.~\ref{eq:simulation} on a finite square lattice with $10$ unit cell on each side. The blue area indicates the filled states. (b) Scaling of deviation of charge density at the corner $d Q_{corner}$ and deep in the bulk $d Q_{bulk}$ (c), (d) Charge density for a filling of a complete band, with $3N_{uc}-3$ states (left) and for a filling of $3N_{uc}$ for a neutral TCI. The size of the dots indicates the absolute value of charge per unit cell. The color of each dot represents the charge density as indicated by the color bar.}
\label{fig:charge_scale}
\end{figure}

In this section, we discuss how charge distributes in a TCI with nontrivial corner charges if we enforce charge neutrality. We take the model in Eq.~\ref{eq:simulation} as an example to show that the correction in charge density caused by enforcing neutrality is in the order of $\frac{1}{L^2}$, where $L$ is the length of the lattice (in units of unit cells). As mentioned in Sec.~\ref{sec:NumericSimulation}, the filling of the TCI in Eq.~\ref{eq:simulation} in a square lattice is $3N_{uc}-3$. This results in a charge imbalance due to a corner-induced filling anomaly. A neutral crystal, however, will fill the next $3$ states in the valence band [see Fig.~\ref{fig:charge_scale}(a)]. Comparing the charge density for the two different fillings [Fig.~\ref{fig:charge_scale}(c,d)], we verify that the corner charge persists, and is compensated by an overall charge of the opposite sign distributed across the bulk. As a result, the integral of charge density over a distance $r$ -that is smaller than half of the length of one edge but larger than the correlation length- away from the corner or from the center of the bulk deviates from the same integral of charge at filling of $3N_{uc}-3$ by a small amount, $\Delta Q$. We then define the change in charge density at the corner $d Q_{corner}$ (or in the bulk, $d Q_{corner}$), by dividing the integrated charge deviation with the number of unit cells included. In Fig.~\ref{fig:charge_scale}(b), we show the plot of $d Q_{corner}$ and $d Q_{bulk}$ as a function of $\frac{1}{L^2}$. Indeed, the deviation in charge density, both at the corner and in the bulk scales linearly with $\frac{1}{L^2}$, which is negligible in the thermodynamic limit.

\end{appendices}

\bibliographystyle{apsrev4-1}
\bibliography{references}